\documentclass{article} 
\usepackage{iclr2027_conference,times}

\usepackage{amsmath,amsfonts,bm}

\def\eqref#1{equation~\ref{#1}}

\def\1{\bm{1}}

\DeclareMathAlphabet{\mathsfit}{\encodingdefault}{\sfdefault}{m}{sl}
\SetMathAlphabet{\mathsfit}{bold}{\encodingdefault}{\sfdefault}{bx}{n}

\usepackage{hyperref}
\usepackage{url}
\usepackage{graphicx}
\usepackage{booktabs}
\usepackage{tabularx,array}
\usepackage{multirow}
\usepackage{makecell}
\usepackage{amsmath,amssymb}
\usepackage{xcolor}
\usepackage[labelformat=parens,labelfont=normalfont]{subcaption}
\usepackage{enumitem}
\usepackage{microtype}
\usepackage{pifont}
\usepackage{placeins}

\newif\ifTrainingDynamicTwoCases
\TrainingDynamicTwoCasestrue
\hypersetup{colorlinks=true,linkcolor=blue!60!black,citecolor=blue!60!black,urlcolor=blue!60!black}

\newif\ifshowtodos
\showtodostrue
\ifshowtodos
  \newcommand{\todo}[1]{{\color{red}\textbf{[TODO:} #1\textbf{]}}}
  
\else
  \newcommand{\todo}[1]{}
  
\fi

\newcommand{\ds}{\textsc{LLM-PowerTrace}}
\newcommand{\cmark}{\ding{51}}
\newcommand{\xmark}{\ding{55}}

\newcommand{\nTrainRuns}{1788}
\newcommand{\nInfRuns}{405}

\iclrfinalcopy
\setlist{itemsep=1pt,topsep=2pt,parsep=0pt,leftmargin=*}

\title{\ds: A Dataset of Time-Resolved GPU Power Traces for AI Compute}

\author{
\noindent Jiansheng Wang$^1$\thanks{Equal contribution;  \texttt{jw9461@nyu.edu, jaskeer1@ualberta.ca, hx2493@nyu.edu }} \quad
Jaskeerat Singh$^{2*}$ \quad
Haochen Xu$^{1*}$\quad \textbf{Xinyi Li}$^1$ \quad \textbf{Mingyuan Yan}$^1$ \\
\textbf{Shaohui Liu}$^3$ \quad 
\textbf{Yuzhuo Li}$^2$ \quad
\textbf{Yize Chen}$^2$\thanks{Equal advising; \texttt{yize.chen@ualberta.ca, wenqicui@nyu.edu}} \quad
\textbf{Wenqi Cui}$^{1\dagger}$ \\[0.6em]
{\small $^1$New York University, New York, NY, USA} \\
{\small $^2$University of Alberta, Edmonton, AB, Canada} \\
{\small $^3$Massachusetts Institute of Technology,
Cambridge, MA, USA} \\[0.4em]
}

\begin{document}

\maketitle

\begin{abstract}
AI compute is becoming large and fast-changing electrical loads.  Its impacts on data center infrastructure and power grids depend not only on total energy consumption, but also on how their power draw moves over sub-second to minute timescales, including peaks, ramps, and synchronized fluctuations across accelerators. Existing energy benchmarks for large language models (LLMs) report aggregate quantities such as joules per token and rarely release the underlying time-series measurements, leaving these power dynamics poorly characterized and difficult to model. We introduce \ds{}, an open dataset of time-resolved, per-GPU power traces for LLM fine-tuning and inference, sampled every 100\,ms on varying NVIDIA platforms  paired with complete workload and system metadata. For fine-tuning, \ds{} profiles open model families spanning a wide range of model sizes and fine-tuning methods, together with controlled sweeps over batch size, sequence length, precision, parallelism, and checkpointing. For inference, it covers variations in concurrency, GPU frequency, KV-cache and weight quantization, prompt and output lengths, and request-arrival patterns.  
Built upon power traces, we define power-grid-facing metrics as well as AI-compute power\&energy scaling law for extrapolating beyond the measured settings. Together, \ds{} provides a dataset foundation for modeling, forecasting, and managing the fast-changing electrical behavior of AI compute\footnote{Code and dataset are available at \url{https://systems-power-review.github.io/training-power-trace-explorer/}}.
\end{abstract}

\section{Introduction}
\label{sec:intro}
AI data centers are among the fastest-growing sources of electricity demand~\citep{shehabi2024datacenter,iea2025energyai}. Beyond their substantial energy consumption, their electrical behavior is shaped by the tightly coordinated execution of AI workloads. Modern AI clusters can synchronously shift thousands of accelerators among compute-bound, memory-bound, communication-bound, and idle states, producing rapid and possibly structured variations in power demand. These dynamics are already becoming operationally significant, where large-scale training jobs can generate synchronized power swings that require mitigation through software, hardware, and energy-storage mechanisms~\citep{choukse2025power}, while LLM workloads can induce irregular power transients impacting power grids~\citep{li2024unseen}. For instance, in July 2024, approximately 1{,}500\,MW of data center load in Northern Virginia was abruptly disconnected almost simultaneously following a single transmission-system fault~\citep{nerc2025incident}.  As power availability increasingly constrains the expansion and operation of AI data centers, understanding the magnitude, variability, and temporal structure of AI power demand is essential for both data center operation and energy systems planning.

Existing characterizations of AI power and energy use remain predominantly energy-centric~\citep{chung2025mlenergy, oviedo2026energy}. However, grid operators, data center owners, and AI servers are concerned not only about overall bulk energy or tokens per joule, but also about \emph{when} power is drawn and \emph{how fast} it changes. A workload with low total energy consumption can still create operational challenges through high power peaks, steep ramps, or unregulated fluctuations.  Such power dynamics can impose additional operational requirements both on data center electrical infrastructure such as feeders, transformers, power distribution and backup units, as well as on power systems such as frequency regulation, ramping capability, and other balancing resources~\citep{chen2025electricity}. 


Yet the data needed to study AI compute's power and energy dynamics remain limited. Benchmarks commonly report total energy consumption, energy per token, query, or run~\citep{samsi2023words,luccioni2024power,chung2025mlenergy,fernandez2025energy,tschand2025mlperfpower}. LLM serving datasets capture request arrivals and workload statistics but not corresponding power measurements~\citep{wang2025burstgpt}, while cluster telemetry datasets are not organized around controlled, configuration-labeled LLM workloads~\citep{samsi2021supercloud}. As a result, researchers seeking to characterize~\citep{maliakel2026characterizing}, model and scale~\citep{hoffmann2022training}, forecast, or control AI power dynamics~\citep{colangelo2026ai} lack open data that directly links workload configurations and execution events to high-resolution power trajectories.


To address this gap, we introduce \ds{} to connect time-resolved GPU power measurements with fine-tuning and inference workload characteristics, execution configurations, and hardware properties. It contains \nTrainRuns{} fine-tuning runs and \nInfRuns{} full serving runs on variations of NVIDIA GPU platforms and datasets, each with 100\,ms per-GPU telemetry, the full run configuration, and timestamped execution events. 
Our contributions are threefold. First, \ds{} provides a systematic, configuration-labeled collection of high-resolution power traces for both LLM fine-tuning and inference, enabling characterization of rich power behaviors beyond aggregate or average energy metrics. Second, building on these traces, we define grid-relevant metrics and demonstrate both model-based analysis and empirical scaling laws for identifying key factors and extrapolating power and energy behaviors beyond the measured configurations. Third, we release the data, metric-computation code, and an interactive explorer, together with validation pipeline that supports reviewed community extensions of the dataset. 

\section{Related Work}
\label{sec:related}

\begin{table}[t]
  \centering
  \small
  \caption{Comparison with related resources. ``Power series'': time-resolved power traces are released. ``Controlled'': runs form controlled configuration sweeps.}
  \label{tab:comparison}
  \footnotesize
  \setlength{\tabcolsep}{3.5pt}
  \begin{tabular}{lcccccc}
    \toprule
    Resource & Train & Infer & Power series & Per-GPU & Controlled & Resolution \\
    \midrule
    Words to Watts~\citep{samsi2023words} & \xmark & \cmark & \xmark & \xmark & \cmark & aggregate \\
    ML.ENERGY~\citep{chung2025mlenergy} & \xmark & \cmark & \xmark & \xmark & \cmark & aggregate \\
    BurstGPT~\citep{wang2025burstgpt} & \xmark & \cmark & \xmark & \xmark & \xmark & request log \\
    MIT Supercloud~\citep{samsi2021supercloud} & \cmark & \cmark & \cmark & \cmark & \xmark & 100\,ms \\
    \midrule
    \ds{} (ours) & \cmark & \cmark & \cmark & \cmark & \cmark & 100\,ms \\
    \bottomrule
  \end{tabular}
  \vspace{-8pt}
\end{table}

\textbf{Energy of LLM training and inference.} As AI models' parameters increase sharply over the past decade, AI training and inference's energy and emission impacts are drawing attention~\citep{strubell2019energy,patterson2021carbon}. For inference, \citet{samsi2023words} benchmark LLaMA across GPU generations and sharding choices, \citet{luccioni2024power} compare tasks and architectures, \citet{fernandez2025energy} study efficiency optimizations, ML.ENERGY automates inference's energy measurement~\citep{chung2025mlenergy}, and MLPerf Power unifies system-level reporting~\citep{tschand2025mlperfpower}. Note these works mostly report aggregate \emph{energy} with varying benchmarks and system-level control knobs; while \ds{} targets the \emph{power waveform} and releases raw time series.

\textbf{Grid impact of AI data centers.} Industry reports have documented synchronized training power swings and their mitigation~\citep{choukse2025power, li2026powerslider}; \citet{li2024unseen} analyzes LLM-induced transients from a power systems perspective, and reliability organizations now treat large computational loads as a distinct and urgent reliability  risk~\citep{nerc2025incident}. Together this calls for large AI load modeling and configuration-resolved data.

\textbf{Power management for LLM systems.} System work identifies GPU frequency (via DVFS), power caps, KV cache, and instance scheduling to save energy in training~\citep{you2023zeus,chung2024perseus} and serving~\citep{patel2024polca,stojkovic2025dynamollm, stojkovic2024towards, he2025freesh}, and separates or interleaves compute-bound prefill and memory-bound decode~\citep{patel2024splitwise,zhong2024distserve,agrawal2024sarathi}. Traces behind these studies are usually isolated or unreleased; \ds{} recovers the same knobs with the power trajectory as one of the targets.

\textbf{AI Workload and energy datasets.} Serving traces such as BurstGPT~\citep{wang2025burstgpt} capture requests but not power, and the MIT Supercloud dataset~\citep{samsi2021supercloud} releases telemetry from heterogeneous user jobs rather than controlled LLM sweeps. BuildingsBench~\citep{emami2023buildingsbench} showed how an open load dataset with standard tasks can mobilize ML research on building energy, which motivates us to adapt  design principles for profiling AI power demands  (Table~\ref{tab:comparison}).

\section{The \ds{} Dataset}
\label{sec:dataset}

\ds{} records GPU power together with workload configuration and execution
metadata. The main fine-tuning collection includes Gemma 1--4,
Llama 2 and 3.x,
Qwen2 and Qwen3,
DeepSeek LLM,
and Mistral/Mixtral. 
The collection includes both dense and mixture-of-experts (MoE) models, with full fine-tuning, LoRA~\citep{hu2022lora}, and
QLoRA~\citep{dettmers2023qlora}. Both data parallelism (DP) and layer-wise model parallelism (MP) are involved. 
In DP, each worker keeps a model replica. In MP, consecutive model layers are
placed across multiple GPUs.
 The main models, fine-tuning/inference configuration, hardware specifications
are summarized in Table~\ref{tab:hardware}. 

Inference traces are collected with vLLM~\citep{kwon2023vllm} on A100, H100, and H200 GPUs,
serving dense Qwen3 (0.6B--32B) and Llama-3.1 (8B, 70B) models and the MoE Qwen3-30B-A3B at
tensor-parallel degrees 1--4. Each sweep varies one factor at a time: maximum concurrency, SM clock, KV-cache and weight quantization, input/output length, and the arrival pattern. Arrival patterns are among closed-loop (a fixed number of requests in flight), Poisson, or replayed BurstGPT
traffic. Runs also include a request timeline aligned to the power trace.
The main configurations are summarized in Table~\ref{tab:stats}, and the full sweep values
are listed in Appendix~\ref{app:inf}.
\begin{figure}
    \centering
    \includegraphics[width=1\linewidth]{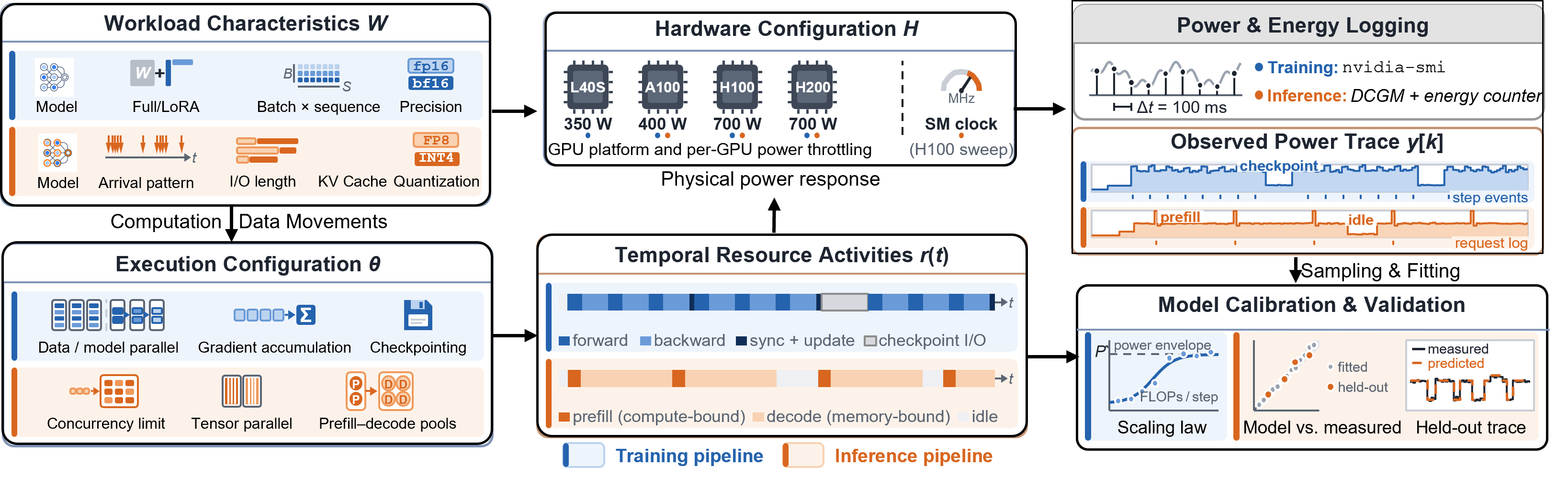}
    \caption{Schematic of power and energy data collection based on
    fine-tuning and inference profiling. To characterize power consumption, we identify levers including workload $W$ and execution
    configuration $\theta$ both determine hardware-dependent resource
    activities $\mathbf{r}(t)$; the hardware configuration $H$ and
    measurement process produce the observed time series trace $y[k]$, which
    supports scaling law analysis of LLM's power demand.}
    \label{fig:framework}
\end{figure}

\begin{table}[t]
\centering
\caption{Overview of the fine-tuning and inference data in \ds{}. }
\label{tab:stats}
\small
\setlength{\tabcolsep}{4pt}
\renewcommand{\arraystretch}{1.08}
\begin{tabularx}{\linewidth}{@{}>{\raggedright\arraybackslash}p{0.16\linewidth}
>{\raggedright\arraybackslash}X >{\raggedright\arraybackslash}X@{}}
\toprule
\textbf{Dimension} & \textbf{Fine-tuning subset} & \textbf{Inference subset} \\
\midrule
Observation unit &
A fine-tuning execution with per-GPU traces and aggregate GPU power. &
A serving execution with device telemetry and available request records. \\
GPU deployment &
L40S, A100, H100, H200; four homogeneous GPUs per main-collection run. &
A100, H100, H200; TP 1, 2, and 4 in designated configurations. \\
Models &
Gemma, Llama, Qwen, DeepSeek, Mistral/Mixtral; dense and MoE models. Model inventory in Table~\ref{tab:models}. &
Qwen3 dense models (0.6B--32B), Qwen3-30B-A3B MoE, and Llama-3.1-8B/70B. \\
Execution stack &
Transformers and Accelerate; full fine-tuning, LoRA, QLoRA; DP or layer-wise MP. &
vLLM 0.28.0; version 0.21.0 for the concurrency sweep. \\
Input workload &
Alpaca instruction data, up to 52K examples, with common preprocessing in the main collection. &
Controlled input/output lengths; chat, code, and math prompts; Poisson arrivals and BurstGPT replay. \\
Workload sweeps &
Micro-batch $\{1,2,4,8,16,32\}$; sequence length $\{128,256,512,1024,2048\}$; model size and type. &
Concurrency $\{1,2,4,\ldots,256\}$; input/output lengths 128/128, 512/512, 2048/128, 512/2048, 2048/2048; thinking on/off in free generation. \\
Execution sweeps &
Accumulation and checkpoint interval $\{1,2,4,8,16\}$; DP/MP and GPU platform. Main save intervals are in optimizer steps. &
TP degree; H100 SM clocks $\{345,600,900,1200,1605,1980\}$ MHz; Poisson rates $\{0.25,0.5,1,2,4,8\}$ requests/s. \\
Precision and quantization &
Recorded \texttt{bf16}, \texttt{fp16}, and \texttt{no} settings; QLoRA uses 4-bit NF4 and double quantization in the main implementation. &
KV cache: BF16 or FP8 (E4M3). Weights: BF16, FP8, or INT4 (GPTQ). \\
Power measurement &
Nominal 100-ms \texttt{nvidia-smi} polling; recorded timestamps, device power, SM clock, and available utilization, memory, and temperature fields. &
Nominal 100-ms DCGM polling; device power (field 155), cumulative energy, clocks, thermal/throttle and activity counters (no activity counters in the concurrency sweep or on A100). \\
\bottomrule
\end{tabularx}
\vspace{2pt}
\begin{minipage}{\linewidth}
\footnotesize
\textit{Notes:} 
DP-data parallelism; MP-layer-wise model parallelism; TP-tensor
parallelism. Details of the inference
patterns and designated multi-GPU configurations are in Appendix.~\ref{app:inf}.
The inference column summarizes the main vLLM collection; the SGLang
prefill--decode runs used in Section~\ref{sec:tasks} form a separate
calibration collection.
\end{minipage}
\end{table}

\paragraph{Hosting and Maintenance}
\label{sec:hosting}

The datasets together with an interactive explorer that filters by model family, GPU, workload type, and precision are available in our online website repository. The explorer also accepts community traces in the same format: submissions are validated automatically, stay private until reviewed, and are published only with the contributor's explicit consent. The complete dataset will be released upon acceptance. The code used to generate the dataset is open sourced\footnote{\url{https://anonymous.4open.science/r/data-collection-0866}}.

\section{Metrics and Empirical Characterization}
\label{sec:empirical_characterization}

\subsection{Grid-facing Metrics}
\label{sec:metrics}

Let $P_i(t)$ denote the power of GPU $i$, and let
$P_\Sigma(t)=\sum_{i=1}^{G}P_i(t)$ denote the aggregate GPU power of
the $G$ participating GPUs. We use $P(t)$ to denote either a per-GPU
or aggregate GPU power trace over an interval $\mathcal T=[t_0,t_1]$.
Let $P[n]=P(t_n)$ denote the power at sample time $t_n$. We characterize power dynamics using the following metrics.
\textbf{Energy:} $E=\int_{\mathcal{T}}P(t)\,dt$ and energy per token $e_{\mathrm{tok}}=E/N_{\mathrm{tok}}$ , where $N_{\rm tok}$ is the number of training tokens for fine-tuning or generated tokens for inference over $\mathcal T$. We also report energy per iteration for fine-tuning and per request for inference.
\textbf{Power level:} mean $\bar P=\frac{1}{N}\sum_{n=1}^{N}P[n]$
and quantiles $P^{(q)}=Q_q(\{P[n]\})$, $q\in\{0.05,0.95\}$, where $Q_q$ denotes the empirical $q$-quantile operator. Upper quantiles capture peaks while resisting single outliers.
\textbf{Ramps:} with $R_{\Delta}(t)=(P(t)-P(t-\Delta))/\Delta$, where $\Delta$ is the time lag, we report the 99th-percentile absolute ramp rate 
$R_\Delta^{(q)}[P]=Q_{q}(\{|R_\Delta(t_n)|\})$ at $q=0.99$.
\textbf{Regularity:} for the mean-centered power sequence 
$\tilde P[n]=P[n]-\bar{P}$, we compute the FFT power spectrum $S_k=|\mathcal{F}\{\tilde{P}[n]\}_k|^2$, where $\mathcal{F}$ denotes the discrete Fourier transform, exclude the DC component, and normalize the $K$ retained nonzero-frequency bins as $p_k=S_k/\sum_{j=1}^{K}S_j$. The normalized spectral entropy is $H_{\mathrm{spec}}=-\sum_{k=1}^{K}p_k\log p_k/\log K$; lower values indicate power concentrated at fewer frequencies and hence a more regular periodic pattern.
\textbf{Synchronization:} for the per-device traces $P_i$ of one node, $\Gamma_\Delta=R_\Delta^{(q)}(\sum_i P_i)/\sum_i R_\Delta^{(q)}(P_i)$ is the node's ramp as a fraction of the summed per-device ramps; values near one indicate that device ramps coincide and add fully, lower values that they partly cancel. 

\subsection{Characteristics of Fine-tuning and Serving Power Traces}
\label{sec:anatomy}

\begin{figure*}[t]
    \centering

    \begin{subfigure}[t]{0.49\linewidth}
        \centering
        \includegraphics[width=\linewidth]
        {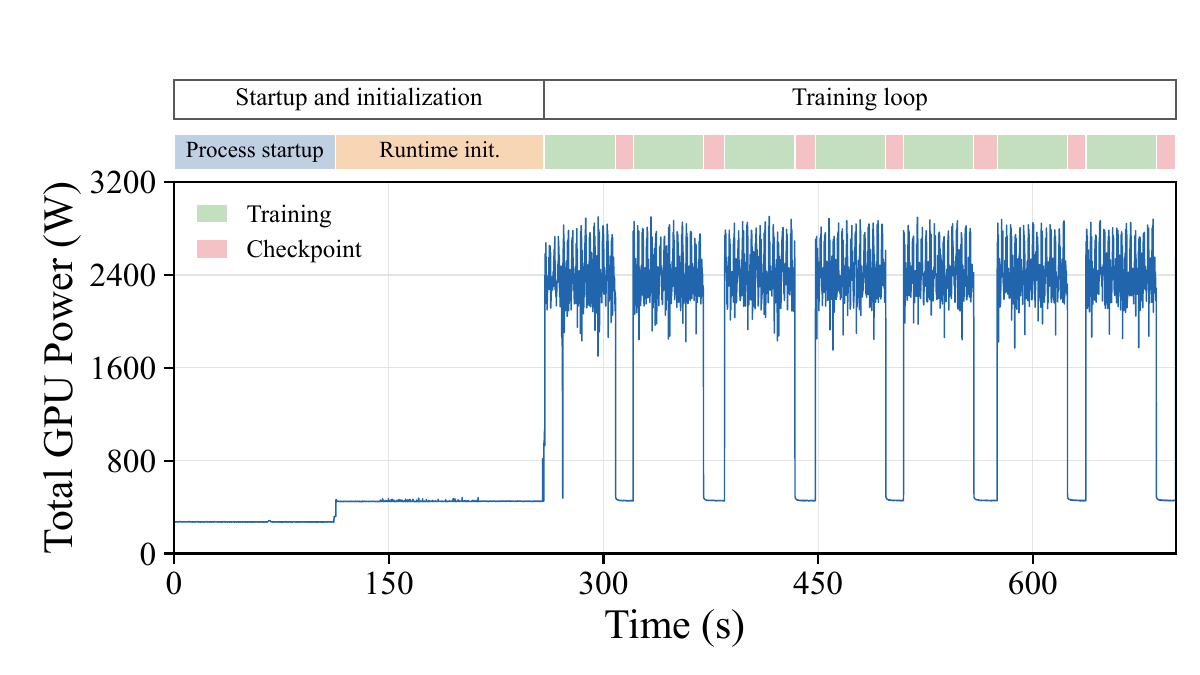}
        \caption{Fine-tuning: full run.}
        \label{fig:training_overview}
    \end{subfigure}
    \hfill
    \begin{subfigure}[t]{0.49\linewidth}
        \centering
        \includegraphics[width=\linewidth]
        {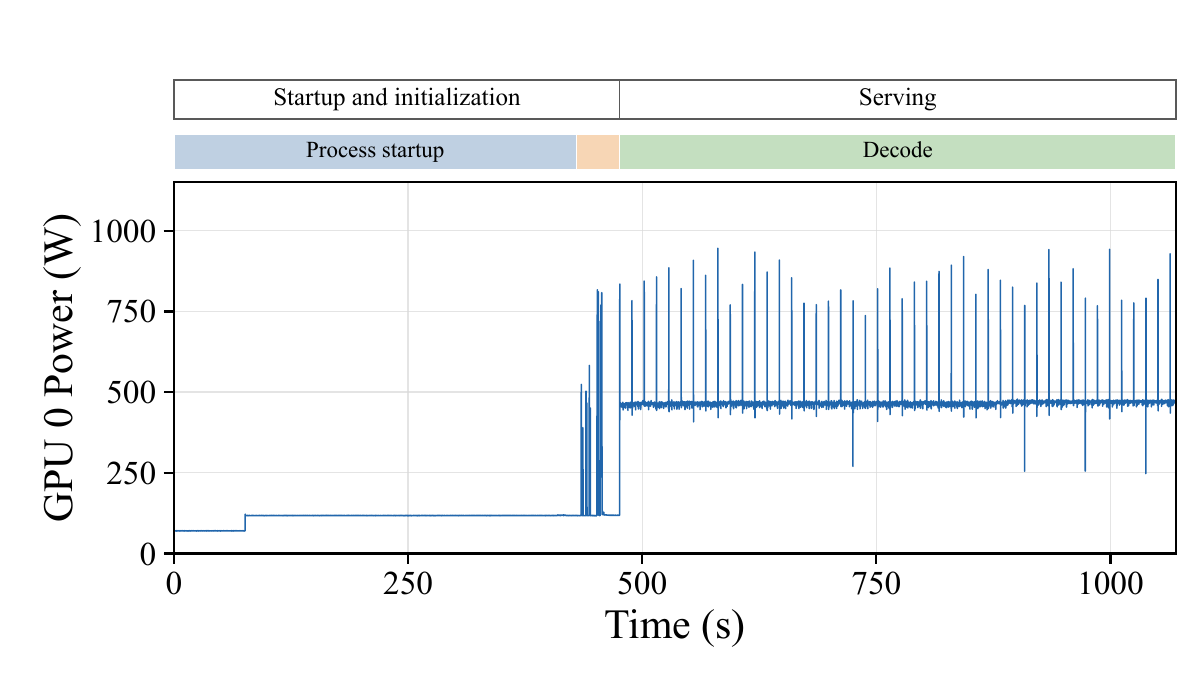}
        \caption{Inference: full run.}
        \label{fig:inference_overview}
    \end{subfigure}

    \caption{Representative LLM-POWERTRACE examples for fine-tuning and inference.
    (a) Fine-tuning power trace for QLoRA fine-tuning of Llama-2-13B on four H100 GPUs with data parallelism.
    (b) Inference power trace for Qwen3-32B served by vLLM on one H100 GPU measured using field 157 on DCGM with 20ms polling.}
    \label{fig:representative_training_inference}
\end{figure*}

\paragraph{Fine-tuning power trace.} In this work, we use fine-tuning workloads to represent the training-side workloads in our dataset. Fine-tuning starts from a pretrained model and retains the standard forward--backward--optimization structure of model training \cite{narayanan2021megatron,hu2022lora}, while the set of parameters being updated depends on the fine-tuning method \cite{hu2022lora,dettmers2023qlora}. Detailed method-specific computation is discussed in Appendix~\ref{sec:train_scale_summary}. LLM fine-tuning is an iterative workload with a structured execution pattern. Each run consists of a startup phase followed by a fine-tuning loop (Fig.~\ref{fig:training_overview}). The startup phase exhibits two relatively stable power plateaus, corresponding to process startup and runtime initialization. The fine-tuning loop then follows a nested periodic structure determined by gradient accumulation and checkpoint frequency. We refer to one optimizer step as one fine-tuning iteration. Each iteration consists of several gradient-accumulation micro-steps (a forward and backward pass on one micro-batch each), followed by gradient processing and the parameter update. During checkpoint saving, computation largely pauses and power drops to a plateau near the active-GPU baseline, producing a regular power swing at each checkpoint. Within each micro-step, the forward pass produces a short high-power interval, whereas the backward pass is longer and exhibits greater power variation. This difference reflects the additional computation required for backpropagation: for a typical linear layer, the forward pass requires one matrix multiplication to compute the output, while the backward pass requires two to compute the input and weight gradients. 
\paragraph{Inference-serving power trace.} LLM inference is also iterative, but its execution structure is organized around requests. Every run splits into startup and a serving loop (Fig.~\ref{fig:inference_overview}). Startup again shows two stable power plateaus, corresponding to process startup and runtime initialization, with brief bursts from graph compilation and CUDA-graph capture. The serving loop consists of repeated request cycles, each divided into \textit{prefill} and \textit{decode}. Prefill processes prompt tokens in parallel, producing a short compute-bound interval with high tensor-core activity and power. Decode generates one token per sequence per step while repeatedly accessing model weights and the growing KV cache, resulting in a longer memory-bound interval with lower tensor-core activity and power. Under closed-loop load, the requests in flight start together, reach their output length at similar times, and are replaced together, so power repeats with a \emph{batch-turnover period}: the time between these replacements, and the inference counterpart of the fine-tuning iterations. With a fixed number of requests in flight, this period is close to the mean request latency, so it is set largely by the output length and by how fast each decode step runs. Unlike checkpoint saving in fine-tuning, inference cycles do not return power to the GPU idle level for continuous serving, so steady serving appears as repeated prefill--decode power swings rather than large periodic drops.

\subsection{Factors Shaping Power Traces}
\label{sec:train-factors}

Despite sharing the same hierarchical structure, the traces exhibit substantial differences in power level, time scale, waveform characteristics, and cross-GPU power distribution. 
Fine-tuning factors fall into three layers (Fig.~\ref{fig:framework}, Table~\ref{tab:factor-effects}): the \emph{workload} (model, method, batch, sequence length, precision) determines the computation and data movement per iteration. The \emph{execution configuration} (parallelism, gradient accumulation, checkpointing) determines how computation, communication, and I/O events
are ordered, synchronized, and repeated. Finally, the \emph{hardware} (compute
throughput, data-transfer capability, power envelope) determines the dynamic response characteristics of the power trace.

Inference factors fall into the same three layers (Table~\ref{tab:inf-factors}): the \emph{workload} (model, concurrency, input and output length, weight and KV-cache precision, arrival process) determines the computation and data movement per decode step. The \emph{execution configuration} (tensor parallelism) determines how computation and communication are split and synchronized across GPUs. Finally, the \emph{hardware} (GPU platform, SM clock) determines the dynamic response characteristics of the power trace.

\begin{figure}[t]
    \centering
    \includegraphics[width=0.9\textwidth]{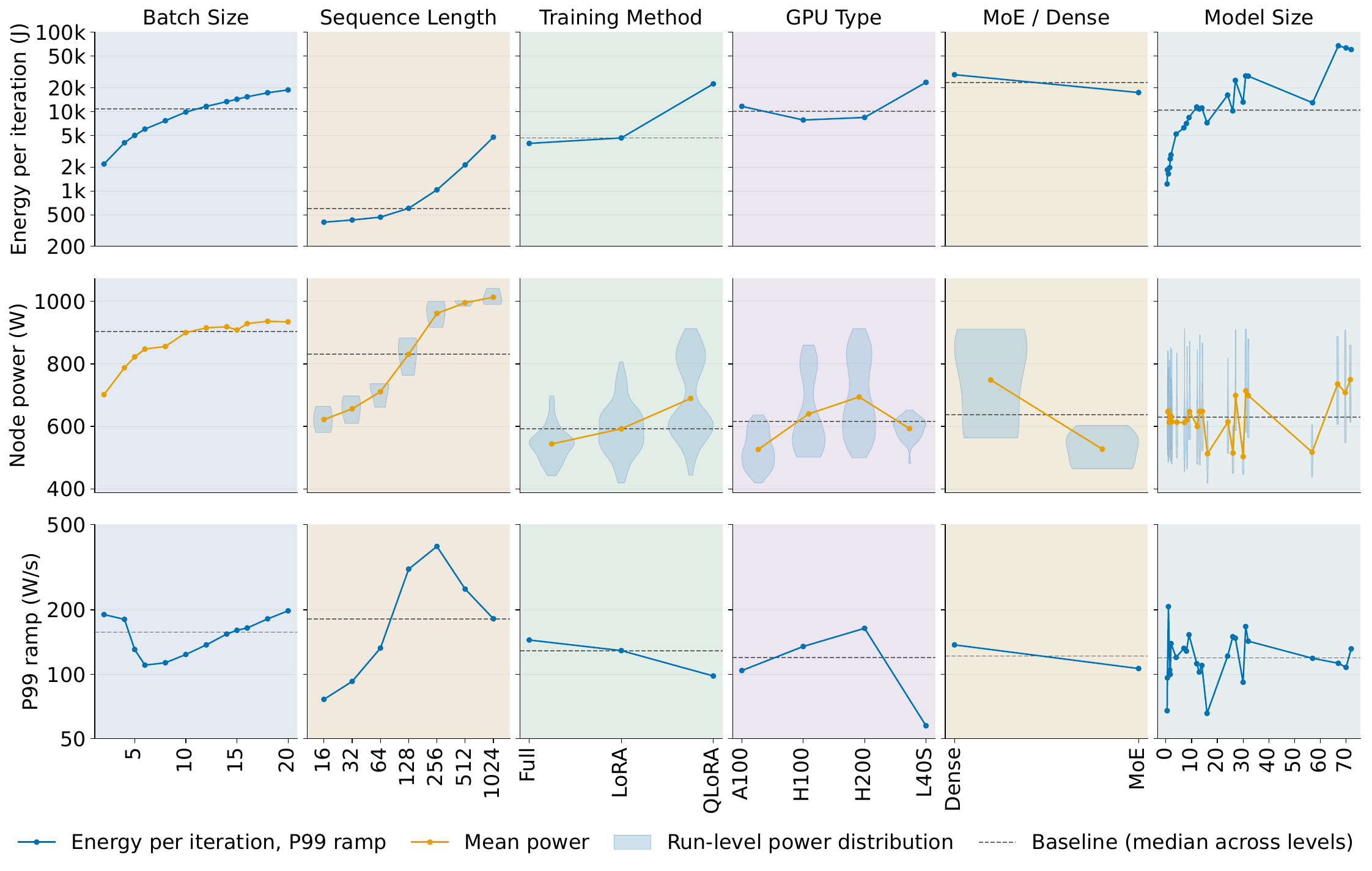}
    \caption{Effects of workload, fine-tuning configuration, and hardware on LLM fine-tuning power. Energy and ramp metrics are shown on logarithmic scales. \vspace{-10pt}}
    \label{fig:training_knobs_overview}
    \vspace{-4pt}
\end{figure}

Fig.~\ref{fig:training_knobs_overview} reveals several consistent patterns for LLM fine-tuning. We treat the fine-tuning method as a workload characteristic because it changes the set of trainable parameters and
the associated gradient and optimizer states.  First, increasing model size, batch size, or sequence length generally increases power, but the increase becomes smaller as power approaches a hardware-dependent envelope. Beyond this point, additional workload is reflected more in longer execution time than in higher power. This limited power headroom also helps explain why ramping does not necessarily increase monotonically with workload size. Second, different workload configurations affect energy efficiency in distinct ways. Larger batch sizes reduce energy per token by amortizing fixed overheads, while MoE models require less effective computation than dense counterparts. Despite lower memory footprint of QLoRA, its fine-tuning is slower and consumes more energy than Full and LoRA. Finally, hardware changes both magnitude and temporal structure of power traces. Compared with earlier GPU models, H100 and H200 achieve lower energy per token and shorter, more regular fine-tuning iterations, but operate at higher power levels with steeper ramps. See more comparisons in Appendix~\ref{app:more-train}.

\begin{figure}[t]
  \centering
  \includegraphics[width=0.9\linewidth]{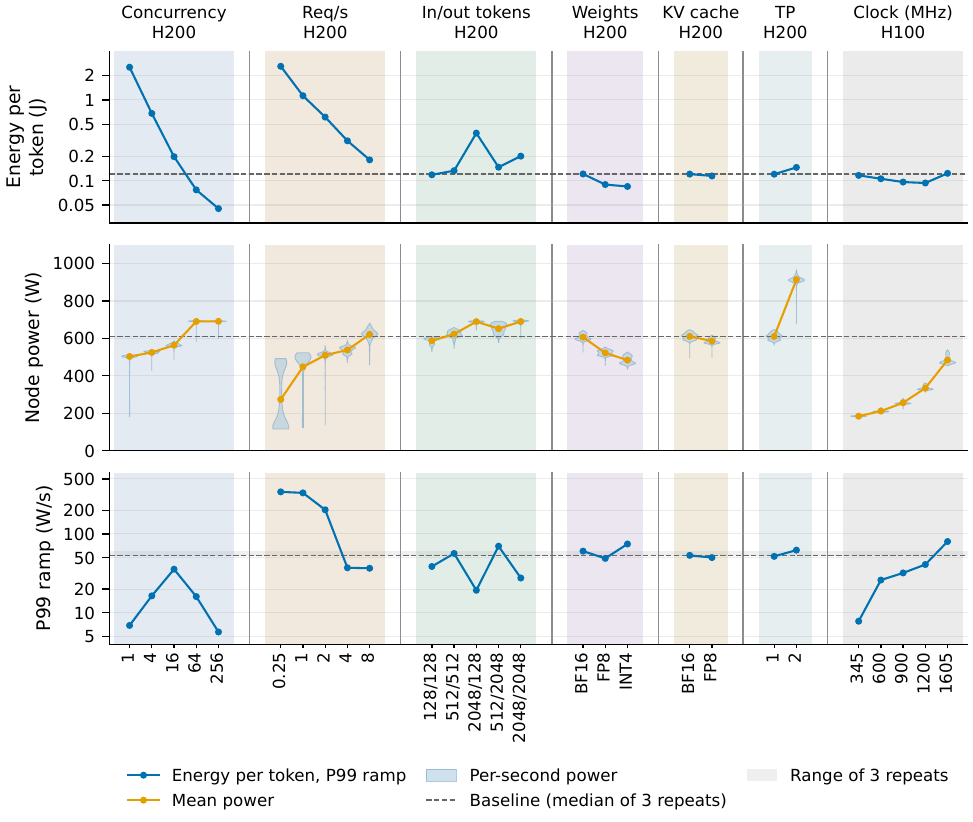}
  \caption{Effects of main serving knobs on energy per output token, aggregate node GPU power, and the 99th-percentile 1-s ramp for Qwen3-8B. Each shaded band changes one knob from the default serving configuration (dashed line; H200, 32 requests in flight, BF16, TP~1); the SM-clock band
  is on H100, and the concurrency band comes from the separate concurrency sweep (vLLM 0.21.0). Energy and ramp axes are logarithmic.
  \vspace{-10pt}}
  \label{fig:knob-effects}
\end{figure}

Figure~\ref{fig:knob-effects} shows several consistent patterns for LLM serving. First, heavier AI load, whether more requests in flight, longer prompts, or larger models, raises power only until the GPU reaches its power limit; beyond it, additional work appears as longer request latency rather than higher power. As in fine-tuning, this saturation compresses the power envelope, so ramps peak at intermediate concurrency and are smallest for long prompts that hold the GPU at its limit. Second, serving configurations change energy efficiency in distinct ways. Batching more requests amortizes each decode step and cuts energy per output token by 17--72$\times$ from 1 to 256 requests in flight; FP8 and INT4 weights cut it by a median 30\% and 39\% by shortening each step, whereas an FP8 KV cache saves only about 3\%. Third, the arrival process shapes the waveform more than any other factor: sparse or bursty arrivals let the GPU drain to idle between requests, widening the power band from about 20 to 1{,}100~W and raising the 99th-percentile ramp about twentyfold, and because tensor-parallel GPUs ramp together, their ramps add at the node. Finally, hardware sets the power range and the timescale. Newer GPU models lower energy per token and turn batches over faster and, unlike in fine-tuning, also run smoother; on a single GPU, the SM clock that minimizes energy per token (1200~MHz) is not the one that minimizes ramps. Detailed comparisons are provided in Appendices~\ref{app:inf-factors} and~\ref{app:inf-quantitative}.

\section{Models and Benchmark Results}
\label{sec:benchmark}


To help identify the key levers shaping power and energy behaviors, as well as generalizing beyond measured configurations to unseen AI model sizes and fine-tuning/inference settings, we propose to demonstrate the utilization of \ds{} for both model-based analysis and empirical scaling laws. For fine-tuning, optimizer-step energy closely follows analytical FLOPs, while model size and sequence length increase computational work and larger micro-batches reduce energy per token by amortizing step-level overhead. For inference, node power follows a hardware-dependent saturation relationship driven by model size, concurrency, and context length, while request-arrival patterns have the strongest effect on short-term power ramps. 


\subsection{Fine-tuning Power and Energy Scaling Law}
\label{sec:training_scaling}


We first validate a FLOPs-based scaling law for fine-tuning energy, summarized here with additional details provided in Appendix~\ref{sec:train_scale_summary}. In the scaling-law notation below, the subscript "step" refers to the same iteration-level interval defined in Sec.~\ref{sec:metrics}.
The computational workload of an optimizer step can be characterized as $
F_{\mathrm{step}}
\propto
A B_\mu\left(N_{\mathrm{act}}S+\kappa_{\mathrm{att}}Ld_{\mathrm{model}}S^2\right),
$
where $A$, $B_\mu$, and $S$ are the gradient accumulation, micro-batch size, and sequence length; and $N_{\mathrm{act}}$, $L$, and $d_{\mathrm{model}}$ are the active parameter count, layer count, and hidden dimension respectively; and $\kappa_{\mathrm{att}}$ is an architecture-dependent attention factor. 
We then relate the analytical computation to measured optimizer-step energy and normalize by the number of processed tokens (approximated by $N_{\mathrm{tok,step}}=AB_\mu S$), yielding a compact model characterizing the energy per token 
\begin{equation}
e_{\mathrm{tok}}
=
\frac{E_{\mathrm{step}}}{AB_\mu S}
\sim
c_1 N_{\mathrm{act}}+c_2 S+\frac{c_3}{AB_\mu S},
\label{eq:training_etok_scaling}
\end{equation}
where the coefficients absorb hardware, execution, and architecture-dependent factors. Architecture- and method-specific FLOP calculations, the derivation of the energy model, and additional validation details are provided in Appendix~\ref{sec:train_scale_summary}. 

\begin{figure*}[!t]
    \centering

    \begin{minipage}{0.32\linewidth}
        \centering
        \includegraphics[width=\linewidth, trim=0 0 0 11, clip]
        {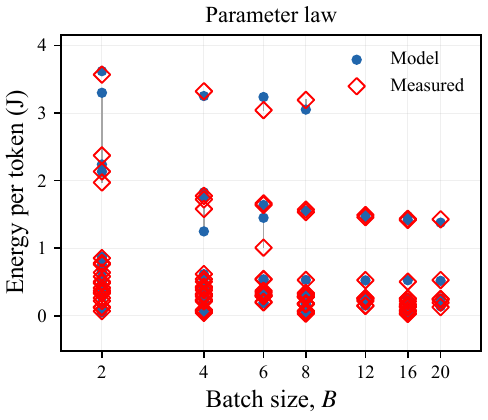}
        \subcaption{Batch size}
        \label{fig:train_pred_bs}
    \end{minipage}
    \hfill
    \begin{minipage}{0.32\linewidth}
        \centering
        \includegraphics[width=\linewidth, trim=0 0 0 11, clip]
        {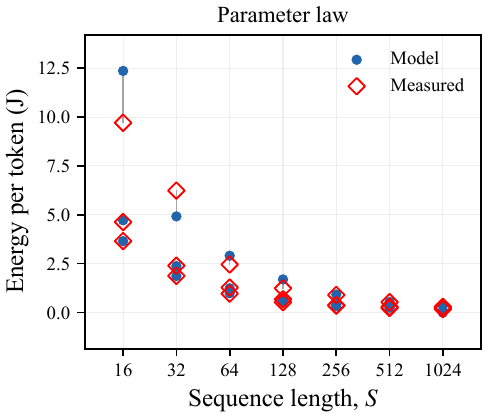}
        \subcaption{Sequence length}
        \label{fig:train_pred_seq}
    \end{minipage}
    \hfill
    \begin{minipage}{0.32\linewidth}
        \centering
        \includegraphics[width=\linewidth, trim=0 0 0 11, clip]
        {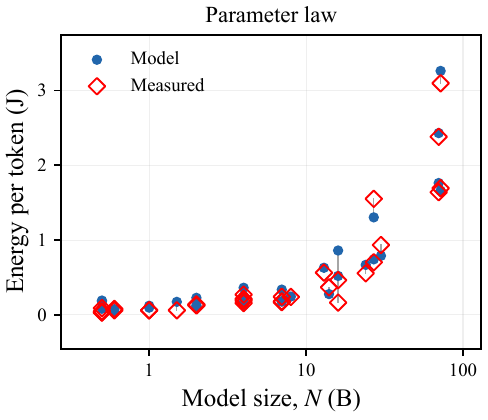}
        \subcaption{Model size}
        \label{fig:train_pred_model}
    \end{minipage}

    \caption{
    Out-of-sample validation of the fine-tuning-energy scaling law.
    Measured and leave-one-level-out predictions as batch size,
    sequence length, and model size vary.
    Model-size points are subsampled for visualization only; all samples
    are retained for fitting and evaluation.
    }
    \label{fig:training_scaling_validation}
\end{figure*}

The measured data closely follow the predicted scaling forms, with fitted coefficients and goodness-of-fit statistics reported in Appendix~\ref{app:training_scaling_result}.
 To evaluate generalization beyond the calibration data, we further perform level-held-out validation: each distinct batch size, sequence length, or model size is excluded in turn, and its energy is predicted using a model fitted to the remaining levels. Fig.~\ref{fig:training_scaling_validation} compares these out-of-sample predictions with measurements.  The close agreement indicates that the analytical structure Eq.~\ref{eq:training_etok_scaling} provides a compact surrogate for how fine-tuning energy scales with key workload dimensions, rather than simply interpolating individual measured configurations.

\subsection{Inference Power and Energy Scaling Law}
We next demonstrate how the dataset can be used to fit an empirical scaling law for LLM inference, which is strongly related to input sequence length (ISL) and output sequence length (OSL).
An inference server continuously processes the requests currently in flight. We therefore model instantaneous node power as a function of the number of active requests $n(t)$ obtained from the serving log. 
The law has three kinds of inputs (Table~\ref{tab:inf-inputs}). \emph{Discrete} settings select
constants: GPU config $H$ sets loaded idle power $P_{0,H}$, power scale $S_H$ and the
intercept $\theta_H$; the tensor-parallel (TP) degree $G$ multiplies the per-GPU terms; quantized weights shift the load by $b_w$; an FP8 KV cache scales the power by $m_{\mathrm{KV}}$; and a locked SM clock frequency 
$f$ scales it by $\phi(f)$. Three continuous inputs enter the load index $z$ \emph{linearly} on a log
scale: the active parameters per GPU ($\ell_N$), the running requests ($\ell_n$) and the mean
context held during decoding ($\ell_L$). The identified \emph{nonlinear} parts are a quadratic term in $\ell_n$, the clock
factor $\phi(f)$, and the logistic function $\sigma(z)=1/(1+e^{-z})$, which maps the load index $z(t)$ to the fraction of the power scale in use as the following relationship:
\refstepcounter{equation}\label{eq:inf-law}
\begin{center}
\makebox[\linewidth]{\hfill$\displaystyle
  \begin{gathered}
    P(t) = G\,P_{0,H} + \sum_{i=1}^{G}\Delta P_{i,H}(t), \qquad
    \Delta P_{i,H}(t) = \mathbf{1}[n(t)>0]\; S_H\,\phi(f)\,m_{\mathrm{KV}}\,\sigma\big(z(t)\big),
    \\
    z(t) = \theta_H + \delta\,\ell_N + \gamma_1\,\ell_n(t) + \gamma_2\,\ell_n(t)^2
    + b_{\mathrm{len}}\,\ell_L + b_w , \qquad \phi(f)=e^{\beta\ell_f+\beta_2\ell_f^2}.
  \end{gathered}$\hfill\llap{(\theequation)}}
\end{center}
Every workload factor moves the GPU along the same saturating curve: larger models per GPU, more
running requests and longer requests push it toward the envelope, where further work changes power
little. Note mean power, peak power and ramps can all be treated as statistics of the predicted trace, so the arrival pattern enters only through requests $n(t)$. The 13 continuous parameters are fitted jointly to measured mean
power, with the precision terms fixed from paired runs. Fitted values, the evidence for each term and the limitations are in Appendix~\ref{app:inf-law}.

\begin{table}[t]
\centering\small
\caption{Inputs of the inference scaling law (Eq.~\ref{eq:inf-law}). Logarithms are taken relative
to the reference values shown. Fitted values are in Table~\ref{tab:inf-vars}.}
\label{tab:inf-inputs}
\begin{tabular}{@{}llp{0.55\linewidth}@{}}
\toprule
Type & Input & Enters inference scaling law Eq. \ref{eq:inf-law} as \\
\midrule
Discrete & platform $H$ & $P_{0,H}$ (loaded idle power per GPU, measured), $S_H$ (power scale per GPU), $\theta_H$ (intercept) \\
 & TP $G$ & multiplies the per-GPU terms; divides $N_{\mathrm{act}}$ in $\ell_N$ \\
 & weight precision & shift $b_w$ in $z$ (FP8, INT4; 0 for BF16 and on A100) \\
 & KV-cache precision & factor $m_{\mathrm{KV}}$ on $S_H$ (FP8; 1 for BF16) \\
\midrule
Linear in $z$ & model size & $\ell_N=\log(N_{\mathrm{act}}/(8G))$, $N_{\mathrm{act}}$ in billions \\
 & running requests & $\ell_n=\log(n(t)/32)$, from the serving log \\
 & request length & $\ell_L=\log((\mathrm{ISL}+\mathrm{OSL}/2)/512)$, realized tokens (mean context during decoding) \\
\midrule
Nonlinear & load & $\gamma_2\,\ell_n^2$ in $z$ \\
 & locked SM clock $f$ & $\phi(f)$ on $S_H$, $\ell_f=\log(f/1980\,\mathrm{MHz})$; $\phi=1$ when unlocked \\
 & saturation & $\sigma(z)$, fraction of $S_H$ in use; no incremental power when $n(t)=0$ \\
\bottomrule
\end{tabular}
\end{table}

\begin{figure}[t]
  \centering
  \includegraphics[width=\linewidth]{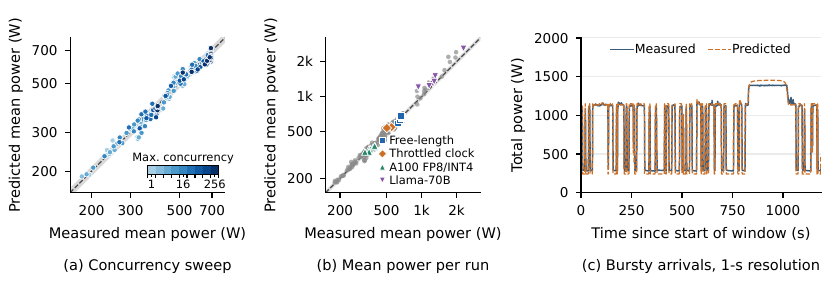}\vspace{-10pt}
  \caption{Fit of the inference scaling law (Eq.~\ref{eq:inf-law}).
  (a) Predicted versus measured mean node power for the 162 concurrency-sweep runs (six dense Qwen3 models on A100, H100, and H200), shaded by the maximum concurrency from 1 (light) to 256 (dark); median error 3.0\%, 67\% of runs within $\pm5\%$.
  (b) Predicted versus measured mean node power per run. Grey points are the 253 fitted runs; colored points were not used to fit the coefficients. The shaded band marks $\pm5\%$.
  (c) Measured and predicted total GPU power at 1-s resolution for a replayed bursty BurstGPT
  segment never used in fitting (Qwen3-32B, 2$\times$H200,
  predicted from per-request log).}
  \label{fig:inf-law}
\end{figure}

Beyond mean power, power and energy scaling law we propose is generalizable to modeling the ramp rates (Appendix \ref{app:inf-law}). Fig.~\ref{fig:inf-law} summarizes the model fit and validation. Across the 253 fitted runs the median error in mean
node power is 2.7\%: 1.7--4.2\% for every sweep and GPU except tensor parallelism, which the law
overpredicts (median error 5.2--7.9\%). On sets never used to fit a coefficient or to choose a term it is 1.3--6.1\% (Table~\ref{tab:inf-law-acc}). Over 27 BurstGPT replays, we are able to fit a short-term power load prediction, where the predicted 1-s trace strongly correlates with the measured one at a median of 0.83. Eq.~\ref{eq:inf-law} is fitted directly to measured traces, illustrating effective use of \textsc{LLM-PowerTrace}.

\subsection{Model Calibration and Validation for Disaggregated-Serving Inference}
\label{sec:tasks}

\begin{figure}[h]
  \centering
  \includegraphics[width=\linewidth]{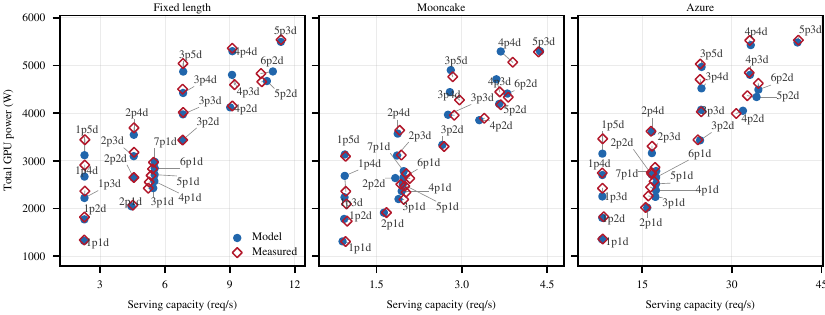}
  \caption{Serving capacity and total GPU power for the evaluated
  prefill--decode deployments.
  Labels follow the form $x\mathrm{p} \, y\mathrm{d}$, where $\mathrm{p}$ denotes
  prefill, $\mathrm{d}$ denotes decode, and $x$ and $y$ give the corresponding
  numbers of instances. \vspace{-10pt}
  }
  \label{fig:pd-validation-planes}
\end{figure}

Finally, we demonstrate how the dataset can be used to calibrate and validate the analytical model for inference serving
under prefill--decode disaggregation~\citep{yan2026analytical},
where each request is processed first by a pool of prefill instances and then
by a pool of decode instances.
The model maps workload characteristics, hardware limits, and the numbers of
prefill and decode instances to serving capacity and total GPU power.
For this evaluation, each coefficient set is fitted once from its designated
traces and then applied to every deployment without deployment-specific adjustment.
The traces were collected while serving Qwen3-32B with SGLang on H200 GPUs,
with one GPU assigned to each prefill or decode instance,
under a fixed-length workload and under the inference request length distributions of the Mooncake~\citep{qin2025mooncake} and Azure~\citep{patel2024splitwise} conversation traces.
We report MAPE for serving capacity and total GPU power,
with the model definition, the calibration protocol and the per-deployment values
in Appendix~\ref{app:pd-validation}.
Deployment-level comparisons of modeled and measured serving capacity and total
GPU power are provided in Fig.~\ref{fig:pd-validation-planes}. This shows that the model-based characterization of power and serving capacity closely matches the measured data and generalizes across different inference request distributions and workload types.

\section{Conclusion and Discussion}
\label{sec:conclusion}

In this work, we present \ds{}, an open dataset and benchmark of time-resolved GPU power traces for LLM fine-tuning and inference, with grid-facing metrics, and power and energy scaling laws. Beyond aggregated energy and joules per token metrics, our profiled traces show that the power waveform of AI workloads  exhibit structured relationships with choices of AI model, hardware, execution, and AI serving parameters. \ds{} provides a novel and necessary connection of fine-grained AI compute power modeling with data center and energy system tasks. In the future work, we will maintain and update the dataset through the community pipeline, and add multi-node training, server- and facility-level measurements, practical serving policies  (e.g., disabling or tuning chunked prefill, speculative decoding), and incoming generation of accelerators.

\bibliography{references}
\bibliographystyle{iclr2027_conference}

\newpage
\appendix
\section*{Appendix}
\section{Fine-Tuning Subset Details}
\label{app:train}
\subsection{Hardware, models, and fine-tuning configurations}
\paragraph{Hardware.} The LLM fine-tuning power traces were collected on a university High Performance Computing (HPC) cluster. 
We selected NVIDIA L40S, A100, H100, and H200 to cover a broad range of accelerator generations and hardware capabilities, including differences in compute performance, memory capacity, and memory bandwidth. Detailed specifications of the GPU platforms evaluated are summarized in Table~\ref{tab:hardware}.
\begin{table}[h]
  \centering
  \small
  \caption{GPU platforms in \ds{}.}
  \label{tab:hardware}
  \setlength{\tabcolsep}{5pt}
  \begin{tabular}{lcccc}
    \toprule
    & L40S & A100 & H100 & H200 \\
    \midrule
    Architecture & Ada Lovelace & Ampere & Hopper & Hopper \\
    Memory & 48\,GB GDDR6 & 80\,GB HBM2e & 80\,GB HBM3 & 141\,GB HBM3e \\
    Memory bandwidth & 864\,GB/s & 2.04\,TB/s & 3.35\,TB/s & 4.8\,TB/s \\
    BF16 Tensor Core & 362\,TFLOPS & 312\,TFLOPS & 989\,TFLOPS & 989\,TFLOPS \\
    Max.\ power limit & 350\,W & 400\,W & 700\,W & 700\,W \\
    Inter-GPU link & PCIe (no NVLink) & NVLink 600\,GB/s & NVLink 900\,GB/s & NVLink 900\,GB/s \\
    Subsets & Train & Train, Infer & Train, Infer & Train, Infer \\
    \bottomrule
  \end{tabular}

  \vspace{1mm}
  \begin{minipage}{0.98\linewidth}
  \footnotesize
  \textit{Notes:} BF16 throughput is dense (no structured sparsity). The power limit is the configurable per-GPU limit, not the measured power. A100, H100, and H200 are SXM variants.
  \end{minipage}
\end{table}

\textit{Logger.} We use \texttt{nvidia-smi} for fine-tuning because it provides a common lightweight interface across the evaluated GPU platforms for collecting the per-GPU power, clock, and utilization fields required by our measurements. 
At each 100\,ms poll the logger queries \texttt{nvidia-smi} for \texttt{timestamp}, \texttt{index}, \texttt{power.draw.instant}, \texttt{clocks.sm}, \texttt{utilization.gpu}, and \texttt{utilization.memory}. Node-level power is the sum of the participating GPUs' power at each aligned timestamp. The observed median and 95th-percentile sampling intervals of every run are stored in its metadata. For serving runs, the logger queries DCGM every 100\,ms for power (field 155), the cumulative energy counter (field 156), SM and memory clocks, temperatures, utilization, memory use, and throttle reasons, plus profiling counters for SM, tensor-core, and DRAM activity and PCIe/NVLink traffic on H100 and H200. We use DCGM for serving because it samples every GPU of a node together and exposes the energy counter and these activity counters, which \texttt{nvidia-smi} does not report. Field 155 is a smoothed average that attenuates sub-second peaks; future releases will log the instantaneous power (field 157), as in Fig.~\ref{fig:inference_overview}, to capture them.

\textit{Caveats.} On-board power sensors on A100 and H100 sample only part of the runtime and may average over a window~\citep{yang2024parttime}. Sub-100\,ms dynamics are therefore not resolved, and short spikes may be attenuated.


\paragraph{Models.} Table~\ref{tab:models} shows the LLMs that we characterize. Our selection represents some of the most representative open-source LLM families. Default configuration for the fine-tuning experiments is shown in Table~\ref{tab:train-defaults}.
\begin{table}[h]
\centering
\caption{Models included in the main fine-tuning collection and additional
fine-tuning runs.}
\label{tab:models}
\small
\setlength{\tabcolsep}{5pt}
\renewcommand{\arraystretch}{1.03}
\begin{tabularx}{\linewidth}{@{}ll>{\raggedright\arraybackslash}X@{}}
\toprule
\textbf{Family} & \textbf{Generation} & \textbf{Model sizes (B)} \\
\midrule
\multicolumn{3}{@{}l}{\textit{Main fine-tuning collection}} \\
Gemma & Gemma & 2, 7 \\
 & Gemma 2 & 2, 9, 27 \\
 & Gemma 3 & 0.27, 1, 4, 12, 27 \\
 & Gemma 4 & E2, E4, 26(4), 31 \\
\midrule
Llama & Llama 2 & 7, 13, 70 \\
 & Llama 3/3.1/3.3 & 8, 70 \\
\midrule
Qwen & Qwen2 & 0.5, 1.5, 7, 57(14), 72 \\
 & Qwen3 & 0.6, 1.7, 4, 8, 14, 32, 30(3), 235(22) \\
\midrule
DeepSeek & DeepSeek LLM & 7, 67 \\
 & DeepSeek-MoE & 16(2.8) \\
\midrule
Mistral/Mixtral & Mistral 7B & 7 \\
 & Mistral Small & 22, 24 \\
 & Ministral & 8 \\
 & Mistral 3 & 3, 8, 14 \\
 & Mixtral & $8\times7$B: 47(13); $8\times22$B: 141(39) \\
\midrule
\multicolumn{3}{@{}l}{\textit{Additional fine-tuning runs}} \\
GPT-J & GPT-J & 6 \\
GPT-NeoX & GPT-NeoX & 20 \\
Pythia & Pythia & 6.9 \\
\bottomrule
\end{tabularx}
\vspace{1mm}
\begin{minipage}{0.98\linewidth}
\footnotesize
\textit{Notes:} Model sizes are in billions of parameters. For MoE models, $x(y)$ denotes $x$B total and $y$B activated parameters per token. Mixtral names give nominal expert sizes. Gemma 4 E2/E4 are effective-size labels for dense models with 5.1B/8B total parameters, as recorded in the manuscript inventory.
\end{minipage}
\end{table}

\begin{table}[h]
  \centering
  \small
  \caption{Default configuration for the controlled LLM fine-tuning experiments.}
  \label{tab:train-defaults}
  \footnotesize
  \begin{tabular}{ll@{\hskip 12pt}ll}
    \toprule
    Parameter & Default & Parameter & Default \\
    \midrule
    Node & 1 node, 4 identical GPUs & Learning rate & $2\times10^{-5}$ \\
    Framework & Transformers + Accelerate & Warmup & 100 optimizer steps \\
    Dataset & Alpaca, $\le$52K samples & Gradient clipping & max norm 1.0 \\
    Sequence length & 1{,}024 tokens & Attention & eager \\
    Micro-batch size & 8 per process/sharded run & LoRA rank/scale/drop. & 64 / 16 / 0.05 \\
    Grad.\ accumulation & 4 micro-steps & QLoRA quantization & 4-bit NF4, double \\
    Optimizer & AdamW, $\beta{=}(0.9, 0.95)$, wd 0 & Checkpoint interval & every 4 opt.\ steps \\
    Precision & BF16 & Power sampling & 100\,ms \\
    \bottomrule
  \end{tabular}
\end{table}

\begin{table}[h]
  \centering
  \small
  \caption{Primary effects of the investigated factors on fine-tuning power traces.}
  \label{tab:factor-effects}
  \begin{tabular}{lll}
    \toprule
    Layer & Factor & Primary trace manifestation \\
    \midrule
    \multirow{6}{*}{Workload} & Fine-tuning method & Overall power profile and phase characteristics \\
    & Model scale & Power level and iteration duration \\
    & Model architecture & Waveform morphology and temporal pattern \\
    & Batch size & Micro-step duration and power level \\
    & Sequence length & Forward/backward duration \\
    & Numerical precision & Forward/backward power contrast and duration \\
    \midrule
    \multirow{3}{*}{Execution} & Parallelism strategy & Inter-GPU power distribution; communication-related variation \\
    & Gradient accumulation & Number of micro-steps per optimizer step \\
    & Checkpoint interval & Spacing of checkpoint-saving segments \\
    \midrule
    Hardware & GPU platform & Power range and execution timescale \\
    \bottomrule
  \end{tabular}
\end{table}




\paragraph{Fine-tuning configuration and sweeps.}
All models in the main fine-tuning collection are fine-tuned on the same
Alpaca subset using a common preprocessing pipeline. Fine-tuning is implemented
with Hugging Face Transformers and Accelerate, with full fine-tuning, LoRA,
and QLoRA, and with either data parallelism (DP) or layer-wise model
parallelism (MP). Table~\ref{tab:train-defaults} summarizes the default
fine-tuning configuration, while Table~\ref{tab:factor-effects} summarizes how
the investigated workload, execution, and hardware factors manifest in the
power trace. The controlled experiments further sweep model size and
architecture, micro-batch size, sequence length, gradient accumulation,
checkpoint interval, parallelism strategy, and GPU platform, with the main
sweep ranges summarized in Table~\ref{tab:stats}.

\FloatBarrier
\newpage
\subsection{Fine-tuning Power and Energy Scaling Law derivations}\label{sec:train_scale_summary}
\subsubsection{FLOPs-based computational workload estimation}
\label{sec:flops_computation}
To obtain a hardware-independent representation of the computational workload, we analytically estimate the matrix-multiplication FLOPs of each optimizer step from the model architecture and fine-tuning configuration. We count one fused multiply--add (FMA) as two FLOPs. As shown in Sec.~\ref{sec:anatomy}, dominant power variation during fine-tuning comes from the forward and backward compute. We therefore use analytical GEMM FLOPs as the primary workload measure, treating normalization, activation, softmax, and quantization/dequantization as secondary contributions.

Let $B_{\mu}$ denote the microbatch size, $S$ the padded sequence length, $L$ the number of Transformer layers, $d_{\mathrm{model}}$ the hidden dimension, $I$ the MLP intermediate dimension, $N_h$ the number of query heads, $N_{kv}$ the number of key/value heads, $D_h$ the head dimension, and $V$ the vocabulary size. For a standard dense Transformer, we decompose the base-model forward computation as
\begin{equation}
F_{\mathrm{base}}^{\mathrm{fwd}}
=
F_{\mathrm{lin}}+F_{\mathrm{att}},
\label{eq:flops_base_fwd}
\end{equation}
where $F_{\mathrm{lin}}$ contains the attention projections, MLP layers, and LM head, while $F_{\mathrm{att}}$ contains the two attention matrix multiplications.

For the attention projections, the query projection maps $d_{\mathrm{model}}$ to $N_hD_h$, the key and value projections map $d_{\mathrm{model}}$ to $N_{kv}D_h$, and the output projection maps $N_hD_h$ back to $d_{\mathrm{model}}$. Their total cost is
\begin{equation}
F_{\mathrm{proj}}
=
2 B_{\mu} S L\left[d_{\mathrm{model}}(N_hD_h)+2d_{\mathrm{model}}(N_{kv}D_h)+(N_hD_h)d_{\mathrm{model}}\right].
\label{eq:flops_proj}
\end{equation}
Because the query and key/value projections use $N_h$ and $N_{kv}$ separately, the expression also covers grouped-query attention (GQA), where $N_{kv}\leq N_h$.

For a gated MLP with gate, up, and down projections, the MLP cost is
\begin{equation}
F_{\mathrm{MLP}}
=
c_{\mathrm{MLP}}B_{\mu}Sd_{\mathrm{model}}IL.
\label{eq:flops_mlp}
\end{equation}
where $c_{\mathrm{MLP}}=6$ for gated MLPs and $4$ for conventional
two-projection feed-forward blocks (GPT-J, GPT-NeoX, and Pythia). The final language-model head contributes
\begin{equation}
F_{\mathrm{LM}}
=
2 B_{\mu} S d_{\mathrm{model}} V.
\label{eq:flops_lm}
\end{equation}
Therefore,
\begin{equation}
F_{\mathrm{lin}}
=
F_{\mathrm{proj}}+F_{\mathrm{MLP}}+F_{\mathrm{LM}}.
\label{eq:flops_linear}
\end{equation}

The attention-score and value-aggregation matrix multiplications are evaluated layer by layer. Let $S_{\ell}$ denote the effective attention span of layer $\ell$. Their total cost is
\begin{equation}
F_{\mathrm{att}}
=
4B_{\mu}N_hD_h
\sum_{\ell=1}^{L}SS_{\ell},
\label{eq:flops_attention_general}
\end{equation}
where $S_{\ell}=S$ for full attention and
$S_{\ell}=\min(S,W_{\ell})$ for a sliding-window layer with window
$W_{\ell}$. When all layers use full attention,
\begin{equation}
F_{\mathrm{att}}
=
4B_{\mu}N_hD_hLS^2.
\label{eq:flops_attention_full}
\end{equation}
Hence, the dominant workload contains both approximately linear and quadratic
dependence on sequence length.

\paragraph{Fine-tuning method.}
For full fine-tuning, all base-model weights participate in gradient computation. We use
\begin{align}
F_{\mathrm{fwd}}^{\mathrm{full}}
&=
F_{\mathrm{lin}}+F_{\mathrm{att}},
\label{eq:full_fwd_flops}\\
F_{\mathrm{bwd}}^{\mathrm{full}}
&=
2\left(F_{\mathrm{lin}}+F_{\mathrm{att}}\right).
\label{eq:full_bwd_flops}
\end{align}
The backward approximation reflects the two matrix multiplications required to obtain input and weight gradients for a trainable linear operation.

For LoRA and QLoRA, the complete frozen backbone is still executed during the forward pass.
Let $N_{\mathrm{ad}}$ denote the number of trainable adapter parameters.
Since each trainable adapter parameter participates in one matrix-multiplication path per token, the adapter forward cost is
\begin{equation}
F_{\mathrm{ad}}
=
2B_{\mu}S N_{\mathrm{ad}},
\label{eq:adapter_flops}
\end{equation}
leading to the following
\begin{align}
F_{\mathrm{fwd}}^{\mathrm{LoRA/QLoRA}}
&=
F_{\mathrm{lin}}+F_{\mathrm{att}}+F_{\mathrm{ad}},
\label{eq:lora_qlora_fwd_flops}\\
F_{\mathrm{bwd}}^{\mathrm{LoRA/QLoRA}}
&=
F_{\mathrm{lin}}+2F_{\mathrm{att}}+2F_{\mathrm{ad}}.
\label{eq:lora_qlora_flops}
\end{align}



The backward cost is lower than for full fine-tuning because frozen base-model linear layers require the activation gradient but not a weight gradient, whereas the attention matrix multiplications and trainable adapter branches require both backward matrix-multiplication paths. For QLoRA, we use the same analytical GEMM graph as for LoRA and do not assign additional FLOPs to quantization or dequantization kernels.

The total computational workload of one microstep is then
\begin{equation}
F_{\mu}
=
F_{\mathrm{fwd}}+F_{\mathrm{bwd}}.
\label{eq:microstep_flops}
\end{equation}
For $A$ gradient-accumulation microsteps per optimizer step,
\begin{equation}
F_{\mathrm{step}}
=
A F_{\mu}
=
A\left(F_{\mathrm{fwd}}+F_{\mathrm{bwd}}
\right).
\label{eq:step_flops}
\end{equation}
Optimizer updates and checkpoint I/O are excluded from this GEMM count and
are modeled separately.

\paragraph{Architecture-specific extensions.}
For architectures that deviate from the standard dense Transformer, the same accounting is applied using the actual active matrix dimensions. In particular, MoE models count only the routed experts activated for each token, together with shared-expert and router computation. Grouped-query attention, sliding-window attention, KV sharing, and architecture-specific projections are likewise evaluated from their layer-specific dimensions.

For interpretation, the detailed architecture-aware workload can be summarized
by the dominant scaling form
\begin{equation}
F_{\mathrm{step}}
\propto
A B_{\mu}\left(N_{\mathrm{act}}S+\kappa_{\mathrm{att}}Ld_{\mathrm{model}}S^2\right),
\label{eq:flops_compact}
\end{equation}
where $N_{\mathrm{act}}$ is the active parameter count and
$\kappa_{\mathrm{att}}$ absorbs architecture-dependent attention constants.
The reported FLOP values use the detailed calculation above, while
Eq.~\ref{eq:flops_compact} provides a compact representation of the dominant
workload scaling for parameter-level analysis and for characterizing the
relationship between fine-tuning parameters and measured power and energy metrics.

\subsubsection{Fine-tuning Scaling Law}
\label{sec:training_scaling_details}

\paragraph{Compute--energy scaling.}
We next relate the computational workload to measured fine-tuning energy of the same optimizer step. In general, the effective workload includes both the analytical computation and additional non-GEMM operations, which can be written
conceptually as
\[
W_{\mathrm{step}}
=
F_{\mathrm{step}}
+
W_{\mathrm{non\text{-}FLOP}} .
\]
where \(W_{\rm non\text{-}FLOP}\) denotes memory-related and runtime work not represented by the analytical GEMM FLOP count. The latter includes memory-related and runtime operations and is particularly
relevant to QLoRA because of its additional quantization/dequantization operations. Since GEMM computation dominates the training workload, we use $F_{\mathrm{step}}$ as the primary workload proxy in the following analysis.

Let $E_{\mathrm{step}}$ denote the aggregate GPU energy over an optimizer step, and let $F_{\mathrm{step}}$ denote the corresponding aggregate analytical FLOPs across the participating GPUs (Appendix~\ref{sec:flops_computation}). For a fixed hardware and execution realization $(H,\boldsymbol{\theta})$, we observe an approximately
linear relationship
\begin{equation}
E_{\mathrm{step}}
=
\alpha_{H,\boldsymbol{\theta}}
F_{\mathrm{step}}
+
\beta_{H,\boldsymbol{\theta}}
+
\varepsilon_E ,
\label{eq:compute_energy_scaling}
\end{equation}
where $\alpha_{H,\boldsymbol{\theta}}$ captures the empirical energy cost per analytical FLOP, while $\beta_{H,\boldsymbol{\theta}}$ is a fitted energy offset that captures effects not explicitly represented by the analytical FLOP count, and $\varepsilon_E$ denotes the residual error.


\paragraph{Energy-per-token scaling.}
Let $N_{\mathrm{tok,step}}=A B_\mu S$ be the number of fine-tuning tokens processed within the same optimizer step. Dividing
Eq.~\ref{eq:compute_energy_scaling} by $N_{\mathrm{tok,step}}$ and
substituting the workload scaling in Eq.~\ref{eq:flops_compact} gives
\begin{equation}
\begin{aligned}
e_{\mathrm{tok}}
&=
\frac{E_{\mathrm{step}}}{N_{\mathrm{tok,step}}} \\
&\approx
\alpha_{H,\boldsymbol{\theta}}
\frac{F_{\mathrm{step}}}{N_{\mathrm{tok,step}}}
+
\frac{\beta_{H,\boldsymbol{\theta}}}
     {N_{\mathrm{tok,step}}} \\
&\approx
\alpha_{H,\boldsymbol{\theta}}
\left(
N_{\mathrm{act}}
+
\kappa_{\mathrm{att}}Ld_{\mathrm{model}}S
\right)
+
\frac{\beta_{H,\boldsymbol{\theta}}}
     {A B_\mu S}.
\end{aligned}
\label{eq:energy_token_scaling}
\end{equation}
Thus, energy per token separates naturally into computational work per token and an amortized step-level overhead. For parameter-level analysis, the above relation can be reduced to
\begin{equation}
e_{\mathrm{tok}}
\sim
c_1 N_{\mathrm{act}}+c_2 S+\frac{c_3}{A B_\mu S},
\label{eq:energy_token_parameter_scaling}
\end{equation}
where the coefficients absorb the remaining architecture- and
hardware-dependent factors.

\paragraph{Power--duration decomposition.}
Measured step energy satisfies
$E_{\mathrm{step}}=\overline{P}_{\mathrm{step}}T_{\mathrm{step}}$,
where $\overline{P}_{\mathrm{step}}$ and $T_{\mathrm{step}}$ denote the mean aggregate GPU power and duration, respectively, over the same optimizer-step interval.
Combining this identity with Eq.~\ref{eq:compute_energy_scaling} yields
\begin{equation}
\overline{P}_{\mathrm{step}}T_{\mathrm{step}}
\approx
\alpha_{H,\boldsymbol{\theta}}F_{\mathrm{step}}
+
\beta_{H,\boldsymbol{\theta}}.
\label{eq:power_duration_scaling}
\end{equation}

At low workload, Sec.~\ref{sec:empirical_characterization} shows that both $\overline{P}_{\mathrm{step}}$ and $T_{\mathrm{step}}$ may increase with $F_{\mathrm{step}}$. The available pre-saturation measurements, however, are not sufficient to justify a universal parametric form for $\overline{P}_{\mathrm{step}}(F_{\mathrm{step}})$, so we characterize this regime empirically rather than imposing a specific saturation function.

Once mean power approaches the hardware-dependent sustained envelope
$P_{\mathrm{sat},H}$,
Eq.~\ref{eq:power_duration_scaling} reduces to
\begin{equation}
T_{\mathrm{step}}
\approx
\frac{\alpha_{H,\boldsymbol{\theta}}F_{\mathrm{step}}+\beta_{H,\boldsymbol{\theta}}}{P_{\mathrm{sat},H}},
\label{eq:saturated_duration_scaling}
\end{equation}
so additional computational work appears primarily as longer execution rather than higher power.

\subsection{Empirical Fine-tuning Energy Scaling Law}
\label{app:training_scaling_result}

\begin{figure*}[ht]
    \centering

    \begin{minipage}{0.45\linewidth}
        \centering
        \includegraphics[width=\linewidth]
        {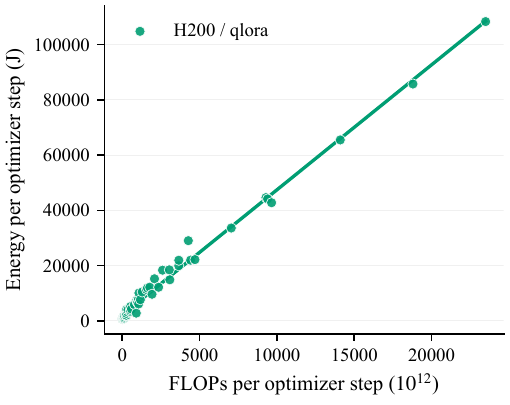}
        \subcaption{FLOPs}
        \label{fig:fit_flops}
    \end{minipage}
    \hfill
    \begin{minipage}{0.45\linewidth}
        \centering
        \includegraphics[width=\linewidth]
        {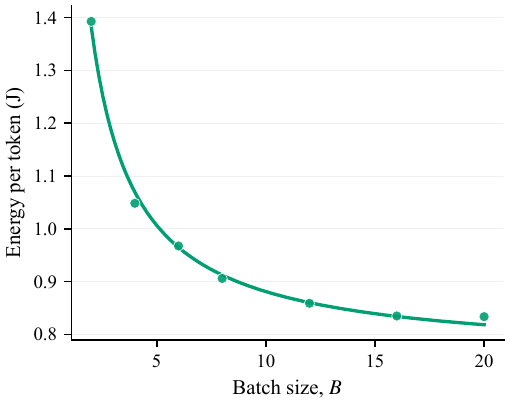}
        \subcaption{Batch size}
        \label{fig:fit_batch}
    \end{minipage}

    \vspace{2mm}

    \begin{minipage}{0.45\linewidth}
        \centering
        \includegraphics[width=\linewidth]
        {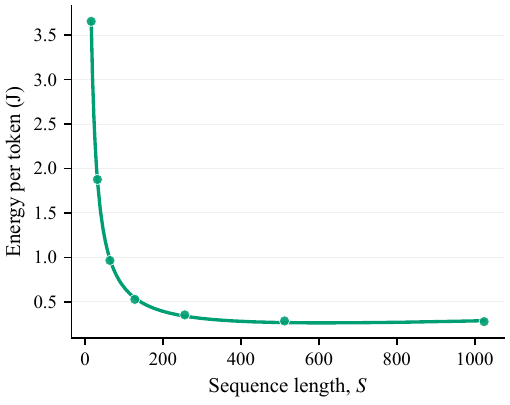}
        \subcaption{Sequence length}
        \label{fig:fit_seq}
    \end{minipage}
    \hfill
    \begin{minipage}{0.45\linewidth}
        \centering
        \includegraphics[width=\linewidth]
        {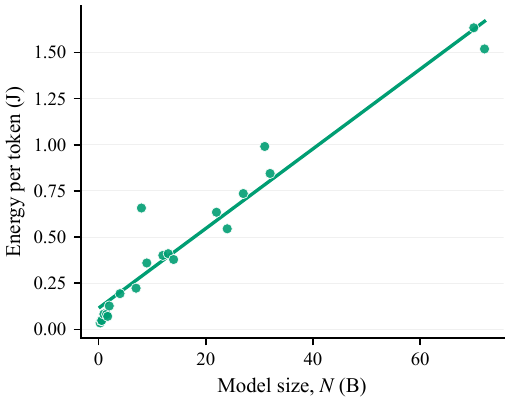}
        \subcaption{Model size}
        \label{fig:fit_model}
    \end{minipage}

    \caption{Calibration of the energy scaling law for QLoRA fine-tuning on H200. (a) Optimizer-step energy versus analytical FLOPs, excluding checkpoint intervals. (b--d) Energy per token as a function of batch size, sequence length, and model size. Lines show the fitted relations used to motivate Eq.~\ref{eq:energy_token_parameter_scaling}.}
    \label{fig:training_scaling_fit}
\end{figure*}

\paragraph{Fitting.}
We calibrate Eq.~\ref{eq:energy_token_parameter_scaling} using controlled experiments under a fixed hardware and fine-tuning configuration (H200 with QLoRA). For each sweep, we retain only configurations available across all values of the varied parameter and aggregate repeated measurements at the configuration level. The resulting fits are summarized in Table~\ref{tab:training_scaling_fit} and Fig.~\ref{fig:training_scaling_fit}.
As an independent check of the underlying assumption, checkpoint intervals
are removed and optimizer-step energy is regressed against analytical FLOPs;
the relation is approximately affine ($R^2=0.992$).

\begin{figure*}[!t]
    \centering

    \begin{minipage}{0.32\linewidth}
        \centering
        \includegraphics[width=\linewidth]
        {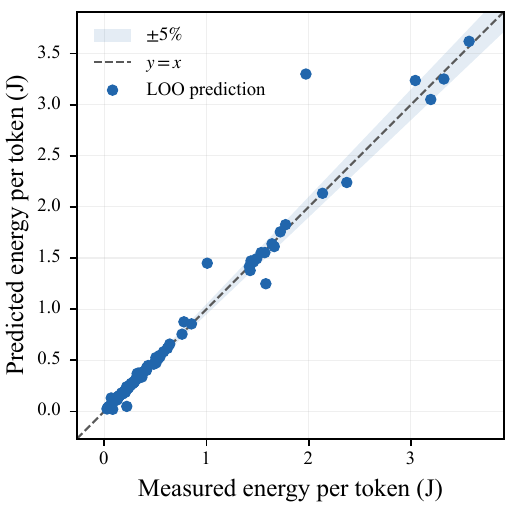}
        \subcaption{Batch size}
        \label{fig:train_mvp_bs}
    \end{minipage}
    \hfill
    \begin{minipage}{0.32\linewidth}
        \centering
        \includegraphics[width=\linewidth]
        {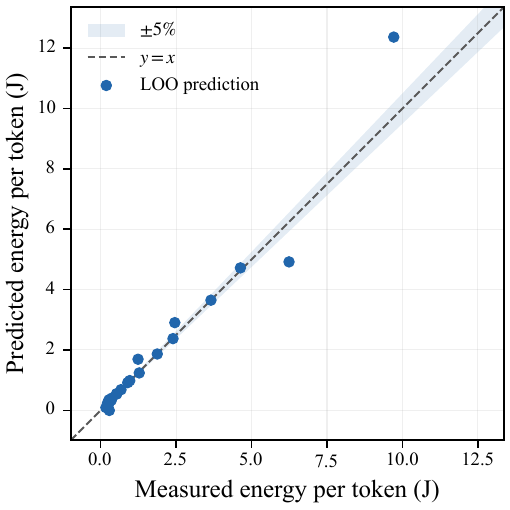}
        \subcaption{Sequence length}
        \label{fig:train_mvp_seq}
    \end{minipage}
    \hfill
    \begin{minipage}{0.32\linewidth}
        \centering
        \includegraphics[width=\linewidth]
        {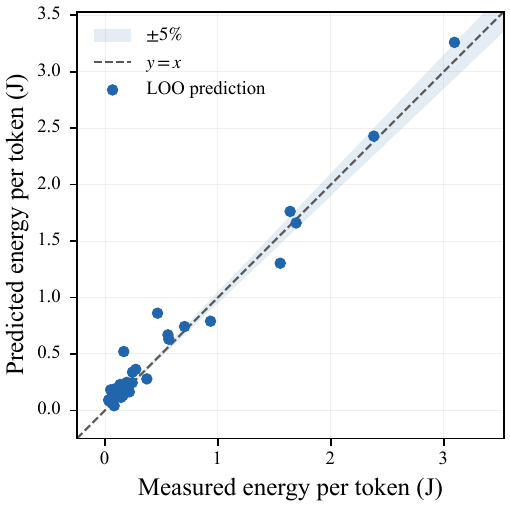}
        \subcaption{Model size}
        \label{fig:train_mvp_model}
    \end{minipage}

    \caption{
    Predicted versus measured energy per token for the held-out
    experiments in Fig.~\ref{fig:training_scaling_validation}.
    The dashed diagonal denotes perfect prediction, and the shaded region
    marks $\pm5\%$ relative error.
    Model-size points are subsampled for visualization only; all samples
    are retained for fitting and evaluation.
    }
    \label{fig:training_scaling_parity}
\end{figure*}

\begin{table}[h]
\centering
\caption{Calibration results for the energy scaling law for QLoRA fine-tuning on H200.}
\label{tab:training_scaling_fit}
\small
\begin{tabular}{lcccc}
\toprule
Relation & $x$ & $y$ & Fitted law & $R^2$ \\
\midrule
Compute--energy
& $F_{\mathrm{step}}$
& $E_{\mathrm{step}}$
& $2176.3 + 4.5241x$
& 0.992 \\

Batch size
& $B_\mu$
& $e_{\mathrm{tok}}$
& $0.7555 + 1.2552/x$
& 0.997 \\

Sequence length
& $S$
& $e_{\mathrm{tok}}$
& $0.0770 + 1.51{\times}10^{-4}x + 57.20/x$
& 1.000 \\

Model size
& $N$
& $e_{\mathrm{tok}}$
& $0.1147 + 0.0216x$
& 0.940 \\
\bottomrule
\end{tabular}
\vspace{1mm}
\begin{minipage}{0.98\linewidth}
\footnotesize
\textit{Notes:} $F_{\mathrm{step}}$ denotes analytical FLOPs per optimizer step, expressed in $10^{12}$ FLOPs, and $N$ denotes total model size in billions of parameters, distinct from the active parameter count $N_{\mathrm{act}}$ used in the analytical FLOP model.
\end{minipage}
\end{table}

\paragraph{Level-holdout validation.}
Prediction is evaluated without using the target parameter level during
calibration. For each distinct value of $B_\mu$, $S$, or $N$, all configurations
at that value are removed, the corresponding law is refitted on the remaining
levels, and the excluded level is predicted. This procedure tests whether the
scaling structure transfers across parameter values rather than only fitting
the observations used for calibration. Complete held-out results are shown in
Fig.~\ref{fig:training_scaling_validation}.


\newpage
\section{Inference Subset Details}
\label{app:inf}

Table~\ref{tab:inf-defaults} lists the default serving configuration, which we use as the baseline setting. Each sweep in Table~\ref{tab:stats} changes one of these settings. Table~\ref{tab:inf-factors} summarizes where each factor shows up in the power trace.

\begin{table}[h]
  \centering
  \small
  \caption{Default configuration for the controlled inference experiments.}
  \label{tab:inf-defaults}
  \footnotesize
  \begin{tabular}{ll@{\hskip 12pt}ll}
    \toprule
    Parameter & Default & Parameter & Default \\
    \midrule
    Node & 1 node, 1 GPU (TP 1) & Request pattern & closed loop (\texttt{b32}) \\
    Engine & vLLM 0.28.0 & Max Batch Size & 32 \\
    Batching & continuous & Output length & 512 tokens (EOS ignored) \\
    Chunked prefill & on & Maximum context & 2{,}048 tokens (closed loop) \\
    Prefix caching & on & Sampling & temperature 0.7, top-$p$ 0.9$^\ast$ \\
    Weights / KV cache & BF16 / BF16 & Warm-up & $\sim$1 min before the window \\
    SM clock & unlocked (default boost) & Measurement window & 600\,s (closed loop) \\
    Power sampling & DCGM, 100\,ms & Energy & hardware energy counter \\
    \bottomrule
  \end{tabular}
\end{table}

\begin{table}[h]
  \centering
  \small
  \caption{Primary effects of the investigated factors on inference power traces.}
  \label{tab:inf-factors}
  \begin{tabular}{lll}
    \toprule
    Layer & Factor & Primary trace manifestation \\
    \midrule
    \multirow{8}{*}{Workload} & Arrival process & Idle--busy switching; ramp magnitude \\
    & Concurrency limit & Power level until saturation; ramps peak mid-load \\
    & Model scale & Power level and batch-turnover period \\
    & Model architecture & Power level and cross-GPU balance \\
    & Input length & Prefill share and proximity to the power limit \\
    & Output length & Spacing of batch-turnover events \\
    & Weight precision & Execution speed (turnover period, energy per token) \\
    & KV-cache precision & Small power reduction; little change in waveform \\
    \midrule
    Execution & Tensor parallelism & Inter-GPU distribution; node power; turnover period \\
    \midrule
    \multirow{2}{*}{Hardware} & GPU platform & Power range and execution timescale \\
    & SM clock & Power level, energy per token, and ramp magnitude \\
    \bottomrule
  \end{tabular}
\end{table}

\textbf{Request patterns.} \texttt{b32} is the closed-loop default of Table~\ref{tab:inf-defaults}.
In the concurrency sweep the client caps requests in flight at 1–256. In the other closed-loop
experiments, the server's maximum batch is 32 and the client keeps 32 in flight, so the server
queue stays essentially empty. The
concurrency sweep is closed loop with the client capping requests in flight at 1--256, short
prompts (median 212 tokens), 512 max output tokens, and 20-minute runs on vLLM~0.21.0. $^\ast$This sweep used the models' default sampling (temperature 0.0, top-$k$ 20, top-$p$ 0.95) and drew prompts in a loop from a pool of 2{,}000, with a median prefix-cache hit rate of 6\%; its DCGM capture has no activity counters.
\texttt{work} and \texttt{burst} replay two 20-minute windows of the BurstGPT
trace~\citep{wang2025burstgpt} at native speed, taking arrival times and prompt lengths from the
trace and letting generation end naturally (at most 2{,}048 output tokens, maximum context
16{,}384): \texttt{work} carries steady traffic (6{,}171 requests, 5.1\,req/s, inter-arrival
coefficient of variation 2.1), and \texttt{burst} is sparser but spikier (1{,}921 requests,
1.6\,req/s, CV 5.4). The replay sweep adds a window at the trace's peak arrival rate
(13{,}998 requests) and replays of \texttt{work} rescaled to 0.25, 0.5, 1, and 2 times its
native rate. Poisson runs draw arrivals at a fixed rate with free-length generation. Open-loop windows are about 1{,}200\,s at native rate (up to about 4,800\,s for slowed replays), longer when a backlogged server drains after the trace ends, and use a maximum context of 16{,}384 tokens; the long input/output pairs use 4{,}096 or 8{,}192. Free
generation uses chat (ShareGPT), code (InstructCoder), and math (competition mathematics: AIME 1983–2024 and the NuminaMath-CoT competition
subset) prompts.

\textbf{Coverage.} The concurrency sweep covers the six dense Qwen3 models on all three GPUs
(162 runs). The TP, weight-precision, and KV-cache sweeps cover Qwen3-8B, 14B, 32B, and
30B-A3B on all three GPUs, plus Llama-3.1-70B at TP~4 on all three GPUs in the TP sweep and at TP~2 on H200 in the precision sweeps (36, 39, and 26 runs). The input/output-length
(30 runs), Poisson (21), and trace-replay (37, of which 35 are complete) sweeps run on H200. The SM-clock sweep (54 runs, 52 with complete artifacts, of which 15 did not hold their target clock)
runs Qwen3-8B, Qwen3-30B-A3B, and Llama-3.1-8B on rented H100 GPUs, each model's closed-loop sweep on a single GPU, under \texttt{b32},
\texttt{work}, and \texttt{burst}; clocks are locked with \texttt{nvidia-smi --lock-gpu-clocks}
and the achieved clock is read from the trace. Of the 15 runs that missed their target, nine are at 1980\,MHz, which throttles to a median of 1830\,MHz under load, five at 1605\,MHz, and one at 345\,MHz.

\textbf{Repeats and quality flags.} No cell is repeated by design, but the closed-loop default
appears on the weight-precision, KV-cache, and TP sheets, giving up to three runs of the same
configuration per model and GPU; we use these to estimate run-to-run variation: for Qwen3-8B on H200 the three runs agree within 0.7\% in energy per token and mean power, while their 99th-percentile ramps span 52--61\,W/s (17\%). Some rented H100 hosts ran with the GPU power limit lowered to 525--599\,W, and the only two-A100 host (all four A100 TP~2 runs) held its application clock at 1155\,MHz instead of 1410\,MHz; these runs are released with a flag and excluded from the scaling-law fit and the pooled comparisons.

\subsection{Empirical AI Serving Power and Energy Scaling Law}
\label{app:inf-law}

\paragraph{Overview.}
Eq.~\ref{eq:inf-law} was built one factor at a time. Steps~1--5 each add one factor, fitted on
the sweep that varies it, with the parameters of earlier steps held fixed. Step~6 refits every
continuous parameter jointly, Step~7 tests the remaining settings for a term, and Step~8 validates
the frozen law on runs never used before. The selection rules were fixed before any fitting.

\paragraph{Fitting objective.}
Each run $r$ contributes one data point. Its load is sampled by vLLM's running-request count,
logged every 10\,s ($K_r$ snapshots $n_k$, 60--484 per run). The predicted power above loaded
idle is the mean of the law over these snapshots, with no excess power when $n_k=0$. Parameters
minimize the squared log error against measured mean power above idle:
\begin{equation}
  \label{eq:inf-fit}
  \min \sum_r \Big[\log\Big(\tfrac{1}{K_r}\textstyle\sum_{k=1}^{K_r}\mathbf 1[n_k>0]\,G\,S_H\,\phi(f)\,m_{\mathrm{KV}}\,
  \sigma\big(z(n_k)\big)\Big) - \log\big(\bar P_r - G P_{0,H}\big)\Big]^2 .
\end{equation}
$\bar P_r$ is the mean of every power sample in the run's measurement window, including any
idle moments. $P_{0,H}$ is measured, not fitted: it is the median power between ``server ready''
and the first request (A100 72.9, H100 126.3, H200 120.9, and 77.8\,W on the locked-clock H100
host).

\paragraph{Run roles.}
Every run's role was declared before fitting. We exclude 21 runs on hosts whose GPU power limit
was lowered (525--599\,W on H100), the four runs on the 2$\times$A100 host whose application clock
was held at 1155\,MHz, 15 runs that failed or lack a summary, and 3 open-loop thinking runs. The
following sets were frozen until Step~8 and informed neither a coefficient nor the choice of a
term: free-length generation, the three locked-clock runs that targeted 1980\,MHz but throttled to
about 1830\,MHz, every Llama-3.1-70B run, and every Poisson and BurstGPT run.

\paragraph{Selection rules.}
Held-out error is the RMS of predicted minus measured $\log(\bar P-GP_{0,H})$ on held-out runs. A
candidate's primary score is its mean held-out error over the tests every candidate can be scored
on, and the lowest score wins, unless a candidate with fewer parameters lies within
$\sigma_{\mathrm{within}}=0.006$ of it (the run-to-run spread of repeated baseline runs on one
host). A categorical term (precision, thinking mode, architecture) is kept only if it lowers
held-out error and a paired $t$-test on the held-out absolute errors gives $p<0.05$. A kept term
goes inside $z$ (a shift along the curve) or outside $\sigma$ (a factor on $S_H$), whichever has
the lower held-out error.

\paragraph{Derivation.}
Table~\ref{tab:inf-terms} lists every candidate and its held-out scores.
\emph{Step~1} fits the 162 concurrency runs: six dense Qwen3 models at 1--256 requests on A100,
H100 and H200. A power law in model size fails, because its size slope falls toward zero as every
model approaches the power limit. The saturating form $S_H\sigma(z)$, which reduces to the power
law $S_He^{z}$ at light load, wins with a quadratic load term.
\emph{Step~2} fits 25 tensor-parallel runs, refitting only $\theta_H$. It keeps $\alpha=-\delta$:
each GPU behaves like one holding a model $G$ times smaller. This is the weakest choice, since a
free exponent is steeper ($-0.90$).
\emph{Step~3} fits 15 runs on the locked-clock H100 host and places the clock as a curved factor
$\phi(f)$ on the power scale.
\emph{Step~4} fits 15 H200 runs at five forced ISL/OSL pairs. The mean context held during
decoding, $\mathrm{ISL}+\mathrm{OSL}/2$, ties with total length, and the lower score is taken.
\emph{Step~5} estimates precision from BF16/quantized pairs on the same host
(Table~\ref{tab:inf-precision}). On Hopper, quantized weights act inside $z$, because their
saving shrinks near the ceiling as a lighter workload's would. An FP8 KV cache is a flat factor on
$S_H$. A100 behaves differently, because it has no FP8 tensor cores: vLLM serves FP8 weights
there weight-only through the Marlin kernel and converts them to 16-bit inside every matrix
multiply, while H200 runs them natively. INT4 weights also use a different kernel on each GPU
(Marlin on A100, Machete on Hopper). On A100, quantized weights draw up to 15\% more power than
BF16, neither term passes, and they get no term.
\emph{Step~6} refits the 13 continuous parameters jointly on 253 runs, with the precision terms
fixed (Table~\ref{tab:inf-vars}). Every step-wise value lies inside its 95\% bootstrap interval.
\emph{Step~7} finds no term for thinking mode or MoE, because a term raised held-out error.
\emph{Step~8} evaluates the frozen law on the held-out sets (Table~\ref{tab:inf-law-acc}). On
BurstGPT, the predicted 1-s trace correlates with the measured one at a median of 0.83.

\begin{table}[!htb]
\centering\footnotesize
\caption{Candidate forms and held-out error at each step. Errors are held-out RMS in
$\log(\bar P-GP_{0,H})$. T1--T4 are the step's tests, listed in its header row; $a\to b$ means fit on $a$ and predict $b$. The primary score
averages the tests every candidate can be scored on; -- marks a test on which a parameter is not
identified. $k$ is the number of free parameters, and the chosen form is in bold. Here
$\ell_N=\log(N_{\mathrm{act}}/8)$, $\ell_n=\log(n/32)$ and $\ell_f=\log(f/1980\,\mathrm{MHz})$.}
\label{tab:inf-terms}
\setlength{\tabcolsep}{2pt}
\begin{tabular*}{\linewidth}{@{\extracolsep{\fill}}llcccccc@{}}
\toprule
Candidate & Form & $k$ & T1 & T2 & T3 & T4 & Primary \\
\midrule
\multicolumn{8}{@{}p{\linewidth}@{}}{\emph{Step 1, saturation} (162 runs). T1: leave one model out. T2: 1--64 $\to$ 128--256 requests.} \\
Power law & $e^{\theta_H+\delta\ell_N+\gamma\ell_n}$ & 5 & 0.222 & 0.256 & & & 0.239 \\
Power law, quadratic load & $e^{\theta_H+\delta\ell_N+\gamma_1\ell_n+\gamma_2\ell_n^2}$ & 6 & 0.216 & 0.301 & & & 0.259 \\
Saturating & $S_H\,\sigma(\theta_H+\delta\ell_N+\gamma\ell_n)$ & 8 & 0.164 & 0.228 & & & 0.196 \\
\textbf{Saturating, quadratic load} & $S_H\,\sigma(\theta_H+\delta\ell_N+\gamma_1\ell_n+\gamma_2\ell_n^2)$ & 9 & \textbf{0.107} & \textbf{0.114} & & & \textbf{0.110} \\
Floor plus rise & $S_H\,[F+(1-F)\,R]$, two logistics & 12 & 0.110 & 0.120 & & & 0.115 \\
\midrule
\multicolumn{8}{@{}p{\linewidth}@{}}{\emph{Step 2, TP} (25 runs). T1: TP 1--2 $\to$ 4. T2: TP 2--4 $\to$ 1. T3: leave one model out.} \\
$\boldsymbol{\alpha=-\delta}$ & $-\delta\log G$ in $z$ & 3 & \textbf{0.155} & \textbf{0.107} & \textbf{0.170} & & \textbf{0.170} \\
Free $\alpha$ & $+\alpha\log G$ in $z$ & 4 & 0.329 & 0.069 & 0.178 & & 0.178 \\
Free $\alpha$ per GPU & $+\alpha_H\log G$ in $z$ & 6 & -- & 0.154 & 0.177 & & 0.177 \\
$\alpha=-\delta$, TP$>$1 step & $-\delta\log G+j\,\mathbf 1[G>1]$ in $z$ & 4 & 0.166 & -- & 0.171 & & 0.171 \\
One scale per node & $+\alpha\log G$ in $z$, node scale $S_H$ & 4 & 0.853 & 0.176 & 0.543 & & 0.543 \\
\midrule
\multicolumn{8}{@{}p{\linewidth}@{}}{\emph{Step 3, clock} (15 runs). T1: leave one clock out. T2: leave one model out. T3: 600--1200 $\to$ 1605\,MHz.} \\
In $z$, linear & $+\beta\ell_f$ in $z$ & 2 & 0.409 & 0.296 & 0.289 & & 0.332 \\
In $z$, quadratic & $+\beta\ell_f+\beta_2\ell_f^2$ in $z$ & 3 & 0.344 & 0.251 & 0.072 & & 0.222 \\
On $S_H$, linear & $S_H\,e^{\beta\ell_f}$ & 2 & 0.265 & 0.150 & 0.136 & & 0.184 \\
\textbf{On $\boldsymbol{S_H}$, quadratic} & $S_H\,e^{\beta\ell_f+\beta_2\ell_f^2}$ & 3 & \textbf{0.051} & \textbf{0.047} & \textbf{0.072} & & \textbf{0.057} \\
\midrule
\multicolumn{8}{@{}p{\linewidth}@{}}{\emph{Step 4, length} (15 runs). T1: leave one pair out. T2, T3: ISL, OSL $\le512\to2048$. T4: leave one model out.} \\
No length term & -- & 3 & 0.074 & 0.109 & 0.043 & 0.117 & 0.086 \\
ISL and OSL & $b_I\log(\mathrm{ISL}/256)+b_O\log(\mathrm{OSL}/512)$ & 5 & 0.086 & 0.123 & 0.086 & 0.035 & 0.083 \\
Total length & $b\log((\mathrm{ISL}+\mathrm{OSL})/768)$ & 4 & 0.080 & 0.072 & 0.035 & 0.046 & 0.0583 \\
\textbf{Decode context} & $b\log((\mathrm{ISL}+\mathrm{OSL}/2)/512)$ & 4 & \textbf{0.100} & \textbf{0.063} & \textbf{0.026} & \textbf{0.042} & \textbf{0.0577} \\
\bottomrule
\end{tabular*}
\end{table}

\begin{table}[!htb]
\centering\footnotesize
\caption{Precision terms from paired runs: held-out RMS (leave one pair out) with no term, a shift
inside $z$, or a factor on $S_H$. $p$ is the paired $t$-test of the better placement against no
term.}
\label{tab:inf-precision}
\begin{tabular}{@{}lcccccl@{}}
\toprule
Level & Pairs & No term & Inside $z$ & On $S_H$ & $p$ & Term kept \\
\midrule
KV cache FP8, all GPUs & 10 & 0.037 & 0.022 & \textbf{0.017} & 0.021 & $m_{\mathrm{KV}}=0.967$ \\
FP8 weights, H200 & 4 & 0.197 & \textbf{0.032} & 0.082 & 0.012 & $b_w=-0.81$ \\
INT4 weights, H200 & 4 & 0.232 & \textbf{0.025} & 0.089 & 0.015 & $b_w=-0.92$ \\
FP8 weights, A100 & 4 & 0.070 & 0.089 & 0.056 & 0.96 & none \\
INT4 weights, A100 & 4 & 0.097 & 0.074 & 0.068 & 0.60 & none \\
\bottomrule
\end{tabular}
\end{table}

\paragraph{Ramps and limitations.}
Ramps are read off the predicted trace $\hat p(t)$ and are not modeled separately. The
99th-percentile 1-s ramp is overpredicted by a median factor of 1.25 on Poisson traffic,
consistent with the sensor averaging each step over about a second, and underpredicted by 1.31 on
BurstGPT. On always-full servers, ramps come from batch turnover, which $n(t)$ does not capture.
Tensor parallelism is the weakest term: TP runs are overpredicted by 4.8--8.7\%. Llama-3.1-70B,
outside the fitted model family, is overpredicted by 10.4\% in closed loop. Because $S_H$ is
fitted rather than tied to the power limit, the H200 ceiling $P_{0,H}+S_H$ is 725\,W, above the
700\,W limit. $b_{\mathrm{len}}$ and the locked-clock $\theta$ are weakly identified, and the
clock terms come from one H100 host and three models. $P_{0,H}$ absorbs host effects: it is
77.8\,W on the locked-clock H100 host and 126.3\,W on the other H100 hosts. The coefficients are
specific to vLLM and the measured platforms.

\begin{table}[!htb]
\centering\footnotesize
\caption{Fitted values of the inference scaling law (Eq.~\ref{eq:inf-law}; inputs in
Table~\ref{tab:inf-inputs}) from the joint fit on 253 runs, with 95\% bootstrap intervals in
brackets. $P_{0,H}$ is measured; the locked-clock host shares $S_{H100}$. TP enters through
$\alpha=-\delta$. $b_w$ shifts $z$ (0 on A100) and $m_{\mathrm{KV}}$ multiplies $S_H$ (1 for BF16);
both come from paired runs and are held fixed. $\theta$ on the locked-clock host is weakly
identified: there $\sigma\approx1$, and $\phi(f)$ carries the clock effect.}
\label{tab:inf-vars}
\setlength{\tabcolsep}{4pt}
\begin{tabular}{@{}llllll@{}}
\toprule
Symbol & Meaning & A100 & H100 & H200 & H100, locked clock \\
\midrule
$P_{0,H}$ & idle power per GPU (W) & 72.9 & 126.3 & 120.9 & 77.8 \\
$S_H$ & power scale per GPU (W) & 329 {\scriptsize[318, 346]} & 537 {\scriptsize[509, 593]} & 604 {\scriptsize[583, 639]} & shared with H100 \\
$\theta_H$ & intercept of $z$ & 1.99 {\scriptsize[1.55, 2.21]} & 1.32 {\scriptsize[0.88, 1.50]} & 1.47 {\scriptsize[1.20, 1.68]} & 5.53 {\scriptsize[2.96, 238]} \\
\midrule
\multicolumn{2}{@{}l}{\emph{Shared across all platforms}} & & & & \\
\midrule
$\delta$ & model size per GPU & \multicolumn{4}{c}{0.598 {\scriptsize[0.465, 0.666]}} \\
$\gamma_1$ & load, linear & \multicolumn{4}{c}{0.749 {\scriptsize[0.649, 0.823]}} \\
$\gamma_2$ & load, quadratic & \multicolumn{4}{c}{0.137 {\scriptsize[0.117, 0.157]}} \\
$b_{\mathrm{len}}$ & request length & \multicolumn{4}{c}{0.440 {\scriptsize[0.016, 0.551]}} \\
$\beta$ & clock, linear & \multicolumn{4}{c}{1.665 {\scriptsize[1.550, 1.836]}} \\
$\beta_2$ & clock, quadratic & \multicolumn{4}{c}{0.430 {\scriptsize[0.380, 0.507]}} \\
$b_w$ & quantized weights (Hopper) & \multicolumn{4}{c}{FP8 $-0.814$, INT4 $-0.918$} \\
$m_{\mathrm{KV}}$ & FP8 KV cache & \multicolumn{4}{c}{0.967} \\
\bottomrule
\end{tabular}
\end{table}

\begin{table}[!htb]
\centering\small
\caption{Accuracy of the frozen law: median absolute error, share within 5\%, and bias
(positive is overprediction). Only the first row was fitted, and the A100 runs informed Step~5;
every other set was held out. ``10-s count'' is vLLM's running count; ``per-request log'' rebuilds
$n(t)$ from request times on a 50-ms grid.}
\label{tab:inf-law-acc}
\setlength{\tabcolsep}{5pt}
\begin{tabular}{@{}lrccr@{}}
\toprule
Set & $n$ & Median $|$err$|$ (\%) & $\le 5\%$ & Bias (\%) \\
\midrule
Mean power, fitted runs                  & 253 & 2.7 & 68\% & $+0.1$ \\
A100 FP8/INT4 weights (no term)          & 8   & 4.3 & 62\% & $-2.9$ \\
\midrule
Free-length generation                   & 9   & 4.6 & 78\% & $-3.5$ \\
Throttled clock (about 1830\,MHz)        & 3   & 1.3 & 67\% & $+1.6$ \\
Llama-3.1-70B, closed loop               & 7   & 6.0 & 43\% & $+10.4$ \\
Poisson, 2-min segments, per-request log & 120 & 2.3 & 100\% & -- \\
Poisson, 2-min segments, 10-s count      & 180 & 3.2 & 62\% & -- \\
BurstGPT, H200, mean power               & 25  & 2.3 & 68\% & $+2.3$ \\
BurstGPT, H100 locked clock, mean power  & 23  & 6.1 & 39\% & $-2.2$ \\
BurstGPT, Llama-3.1-70B (H200), mean power & 10 & 2.4 & 100\% & $+2.3$ \\
BurstGPT, per second (per-request log)   & 27  & 2.8 & -- & -- \\
BurstGPT, P95 power (per-request log)    & 27  & 2.3 & 78\% & -- \\
\bottomrule
\end{tabular}
\end{table}

\FloatBarrier

\subsection{Disaggregated-Serving Model Calibration and Validation}
\label{app:pd-validation}


The calibration traces provide the measurements needed to identify both the capacity and power models. 
Fixed-length prefill-limited runs provide service rates across input lengths,
which fit the achieved fraction of peak compute and an effective attention
coefficient.
Decode runs provide mean iteration times across batch sizes and context
lengths, which fit the achieved fraction of peak memory bandwidth together with
a fixed per-step cost and a per-request cost.
Power calibration uses per-instance mean GPU power and serving throughput from
3 fixed-length input-length settings, the request-rate sweep, and the Mooncake
and Azure workloads,
with each throughput normalized by the capacity limit of the same role under
the same workload.

Least-squares fits estimate the capacity coefficients from these timing
relationships.
The power measurements are pooled across workloads, and bounded nonlinear
least squares fits one capped ramp per role, using 905 per-GPU observations for
prefill and 781 for decode.
Only the request-rate sweep spans the rising part of the decode ramp and
identifies its slope and knee; the other workloads contribute one operating
point per deployment.
The fitted coefficients and the 2 ramps are applied to all evaluated deployments
without deployment-specific adjustment,
so the comparison below measures agreement under one shared calibration rather
than performance on measurements excluded from every fit.

\paragraph{Cross-deployment comparison.}

\begin{table}[h]
  \centering
  \caption{MAPE (\%) across the 24 evaluated deployments of each workload under
  the shared calibration.
  }
  \label{tab:pd-validation}
  \small
  \setlength{\tabcolsep}{8pt}
  \begin{tabular}{lcc}
    \toprule
    Workload & Capacity & Power  \\
    \midrule
    Fixed length & 1.25 & 2.28 \\
    Mooncake     & 3.15 & 2.35 \\
    Azure        & 1.68 & 2.23 \\
    \bottomrule
  \end{tabular}
\end{table}

Table~\ref{tab:pd-validation} reports errors across the evaluated deployments
for the 3 workloads.
Modeled serving capacity and total GPU power remain close to measurement in each case.
Fig.~\ref{fig:pd-validation-planes} shows the corresponding
deployment-level comparisons.

\section{Additional Fine-tuning Results}
\label{app:additional_training}

This section provides additional fine-tuning results complementing the analysis in Sec.~\ref{sec:empirical_characterization}. We first present representative power traces across a broader range of fine-tuning configurations, and then provide complete quantitative comparisons of the major factors considered in our experiments.

\subsection{Additional Fine-tuning Traces}
\label{app:more-train}

We provide additional representative fine-tuning power traces to illustrate how workload characteristics, execution configurations, and hardware platforms reshape the LLM fine-tuning power profile. These examples complement the representative cases in the main text and provide broader coverage of the configurations included in our dataset.

\begin{table}[h]
\centering
\caption{Guide to the figures in the appendix.}
\label{tab:training_figure_guide}
\footnotesize
\setlength{\tabcolsep}{3.5pt}
\renewcommand{\arraystretch}{1.08}
\begin{tabularx}{\linewidth}{
@{}p{0.13\linewidth}
p{0.19\linewidth}
p{0.25\linewidth}
X@{}}
\toprule
\textbf{Figure} & \textbf{Analysis} &
\textbf{Factor / comparison} & \textbf{Content} \\
\midrule

Fig.~\ref{fig:training_scaling_fit}
& Scaling-law calibration
& FLOPs, batch size, sequence length, model size
& Calibration of the fine-tuning-energy scaling law on H200 with QLoRA. \\

Fig.~\ref{fig:training_scaling_parity}
& Scaling-law validation
& Held-out parameter levels
& Predicted versus measured energy per token under level-held-out validation. \\

Fig.~\ref{fig:parallelism_power_traces}
& Representative traces
& MP vs.\ DP
& Per-GPU and aggregate power under layer-wise model and data parallelism. \\

Fig.~\ref{fig:training_power_trace_workload}
& Representative traces
& Batch size; sequence length
& Changes in fine-tuning waveform and power level as workload dimensions increase. \\

Fig.~\ref{fig:workload_power_traces}
& Representative traces
& Model size; dense vs.\ MoE
& Effects of model scale and architecture on cycle duration and power profile. \\

Fig.~\ref{fig:method_power_traces}
& Representative traces
& Full, LoRA, QLoRA
& Power-profile differences across fine-tuning methods. \\

Fig.~\ref{fig:gpu_power_traces}
& Representative traces
& GPU platform
& Comparison across L40S, A100, H100, and H200. \\

Fig.~\ref{fig:quant_model_size}
& Quantitative comparison
& Model size
& Duration, energy, power distribution, ramping rate, and spectral entropy. \\

Fig.~\ref{fig:quant_batch_size}
& Quantitative comparison
& Batch size
& Six fine-tuning metrics across batch-size levels. \\

Fig.~\ref{fig:quant_moe_dense}
& Quantitative comparison
& Dense vs.\ MoE
& Six fine-tuning metrics across model architectures. \\

Fig.~\ref{fig:quant_training_method}
& Quantitative comparison
& Fine-tuning method
& Six fine-tuning metrics for Full, LoRA, and QLoRA. \\

Fig.~\ref{fig:quant_gpu}
& Quantitative comparison
& GPU platform
& Six fine-tuning metrics across L40S, A100, H100, and H200. \\

Fig.~\ref{fig:quant_seq_length}
& Quantitative comparison
& Sequence length
& Six fine-tuning metrics across sequence-length settings. \\

\bottomrule
\end{tabularx}
\vspace{1mm}
\begin{minipage}{0.98\linewidth}
\footnotesize
\textit{Notes:} Representative-trace figures illustrate waveform structure under controlled configurations, while quantitative figures summarize how major fine-tuning factors affect energy, power, ramping, and temporal regularity.
\end{minipage}
\end{table}


\begin{figure}[!t]
  \centering

  \begin{subfigure}[t]{\linewidth}
    \centering
    \includegraphics[width=\linewidth]{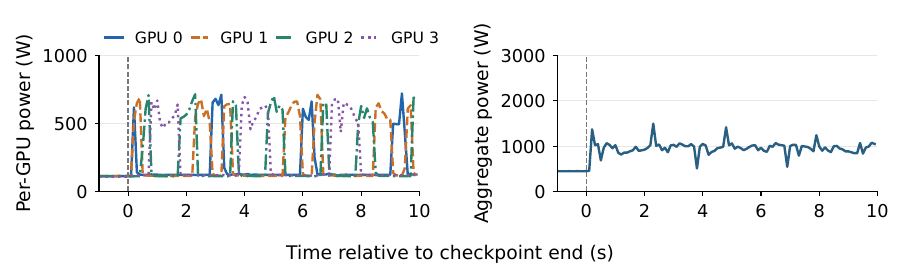}
    \caption{Model parallelism}
    \label{fig:trace_mp}
  \end{subfigure}

  \vspace{0.6em}

  \begin{subfigure}[t]{\linewidth}
    \centering
    \includegraphics[width=\linewidth]{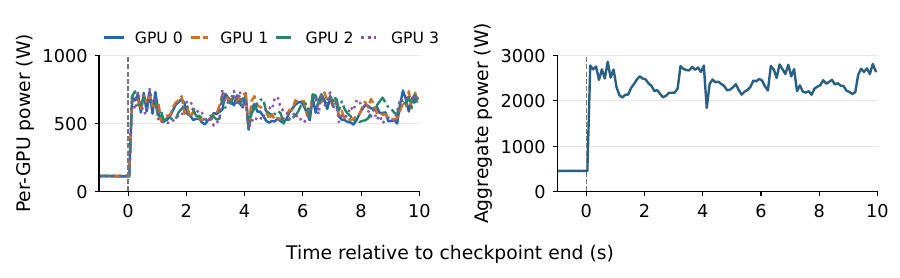}
    \caption{Data parallelism}
    \label{fig:trace_dp}
  \end{subfigure}

  \caption{[Fine-tuning] \textbf{Parallelism strategy.}
   Comparison of layer-wise model parallelism (MP) and data parallelism (DP) for Llama 2 13B under QLoRA on four H100 GPUs. MP exhibits a staggered and cyclic cross-GPU pattern, with recurring forward- and backward-related peak trains distributed across GPUs. The overlap between adjacent GPU activities produces pronounced surges in aggregate power. In contrast, DP executes replicated computation across GPUs, resulting in closely aligned per-GPU traces with higher per-GPU and aggregate power. The aggregate DP trace therefore largely preserves the individual waveform, while the iteration periods remain comparable between MP and DP.}
  \label{fig:parallelism_power_traces}
\end{figure}




\begin{figure*}[t]
    \centering
    \begin{subfigure}[t]{0.49\linewidth}
        \centering
        \includegraphics[width=\linewidth]{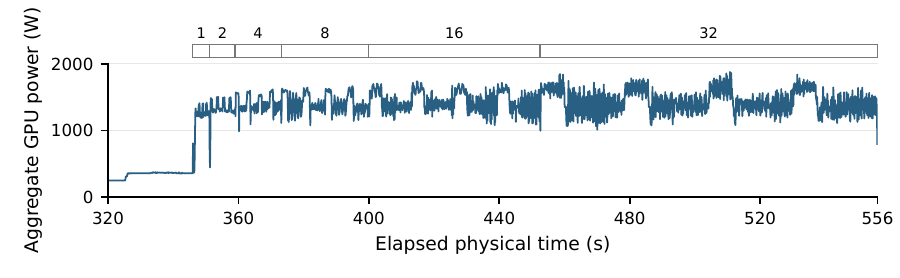}
        \caption{Batch size.}
        \label{fig:batch_size_trace}
    \end{subfigure}
    \hfill
    \begin{subfigure}[t]{0.49\linewidth}
        \centering
        \includegraphics[width=\linewidth]{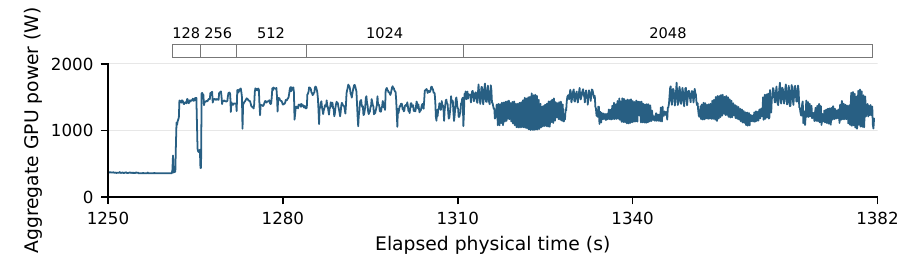}
        \caption{Sequence length.}
        \label{fig:sequence_length_trace}
    \end{subfigure}

    \caption{[Fine-tuning] \textbf{Batch size and sequence length.} Comparison of fine-tuning power traces across batch sizes and sequence lengths for Llama 2 13B under QLoRA on four H100 GPUs. Increasing either workload dimension increases the computation within each fine-tuning iteration, leading to longer execution intervals and generally higher power levels. These comparisons show that workload scale affects both the temporal structure and power level of fine-tuning.}
    \label{fig:training_power_trace_workload}
\end{figure*}

\begin{figure}[!t]
  \centering

  \begin{subfigure}[t]{\linewidth}
    \centering
    \includegraphics[width=\linewidth]{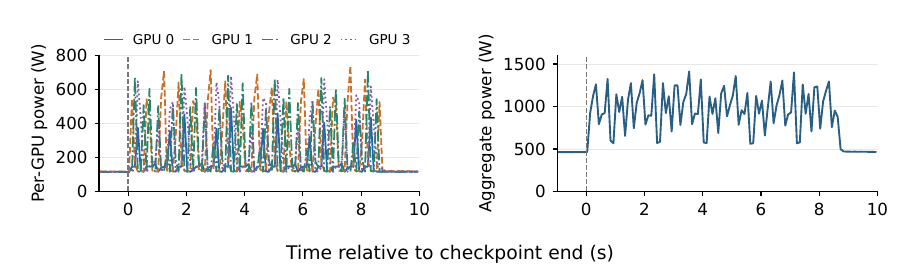}
    \caption{Gemma 3 4B}
    \label{fig:trace_4b}
  \end{subfigure}

  \vspace{0.35em}

  \begin{subfigure}[t]{\linewidth}
    \centering
    \includegraphics[width=\linewidth]{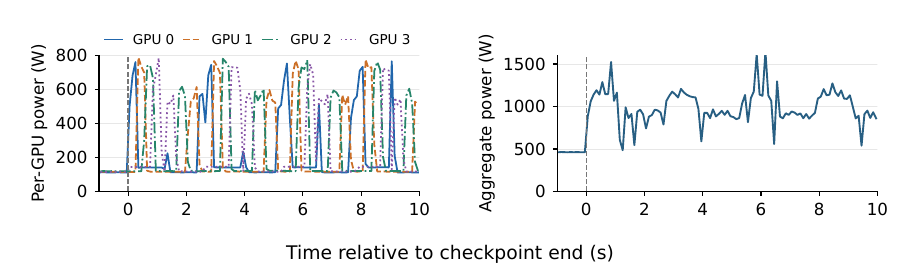}
    \caption{Gemma 3 27B}
    \label{fig:trace_27b}
  \end{subfigure}

  \vspace{0.35em}

  \begin{subfigure}[t]{\linewidth}
    \centering
    \includegraphics[width=\linewidth]{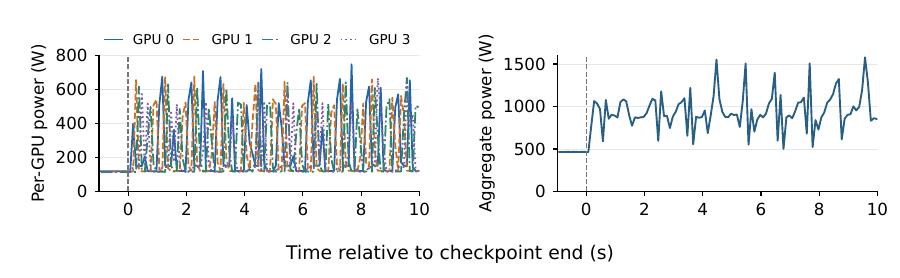}
    \caption{Gemma 4 26B-A4B}
    \label{fig:trace_moe}
  \end{subfigure}

  \vspace{0.35em}

  \begin{subfigure}[t]{\linewidth}
    \centering
    \includegraphics[width=\linewidth]{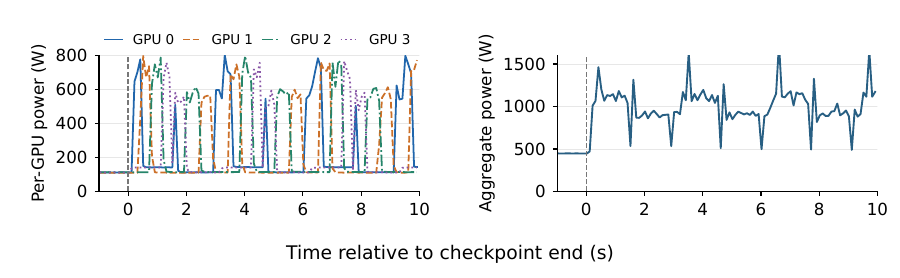}
    \caption{Gemma 4 31B}
    \label{fig:trace_dense}
  \end{subfigure}

  \caption{[Fine-tuning] \textbf{Model size.} Comparison of Gemma 3 4B and 27B shows that increasing model size increases the computational demand of each fine-tuning iteration, resulting in longer execution intervals and generally higher power levels.
  \textbf{Model architecture.} Gemma 4 26B-A4B and Gemma 4 31B have comparable total parameter counts, but the MoE model activates substantially fewer parameters per token. The resulting lower effective computation leads to shorter fine-tuning cycles and lower sustained power, while its temporal pattern is more irregular than that of the dense model.}
  \label{fig:workload_power_traces}
\end{figure}
\FloatBarrier


\begin{figure}[!t]
  \centering

  \begin{subfigure}[t]{\linewidth}
    \centering
    \includegraphics[width=\linewidth]{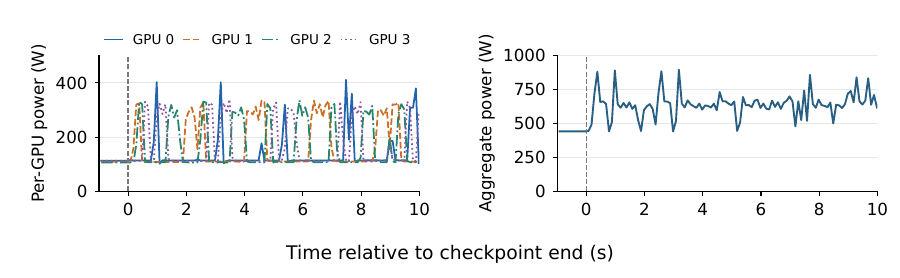}
    \caption{Full fine-tuning}
    \label{fig:trace_full}
  \end{subfigure}

  \vspace{0.5em}

  \begin{subfigure}[t]{\linewidth}
    \centering
    \includegraphics[width=\linewidth]{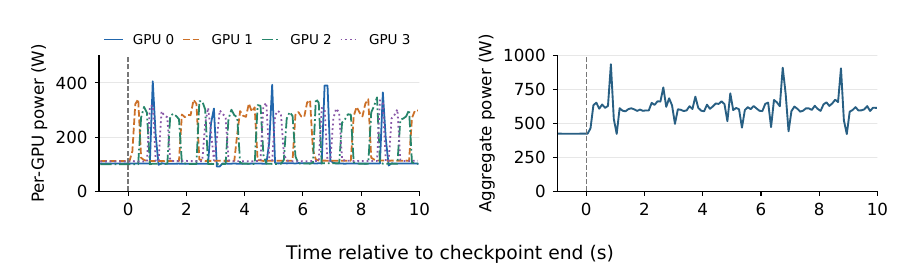}
    \caption{LoRA}
    \label{fig:trace_lora}
  \end{subfigure}

  \vspace{0.5em}

  \begin{subfigure}[t]{\linewidth}
    \centering
    \includegraphics[width=\linewidth]{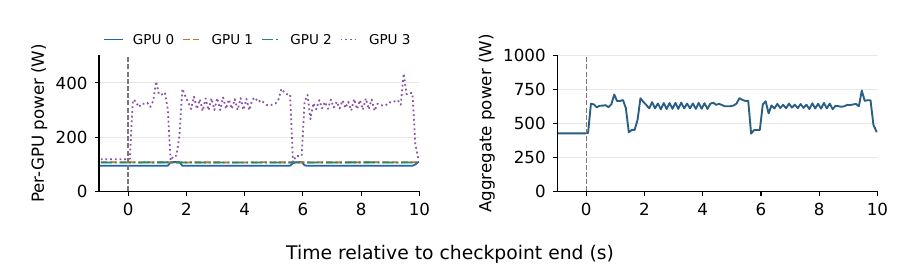}
    \caption{QLoRA}
    \label{fig:trace_qlora}
  \end{subfigure}

  \caption{[Fine-tuning]  \textbf{Fine-tuning method.} Comparison of per-GPU and aggregate power of full fine-tuning, LoRA, and QLoRA power traces for Gemma 3 4B on L40S GPUs. 
This figure illustrates the power profile under different fine-tuning methods. In our measured traces, the checkpoint-saving duration progressively decreases from full fine-tuning to LoRA and QLoRA. But as QLoRA further reduces the memory footprint, the resulting cross-GPU workload redistribution concentrates more computation on a single GPU in our experiments. As a result, the heavily utilized GPU exhibits a higher power level and longer execution intervals.
}
  \label{fig:method_power_traces}
\end{figure}


\begin{figure}[!t]
    \centering

    \begin{subfigure}[t]{\linewidth}
        \centering
        \includegraphics[width=\linewidth]{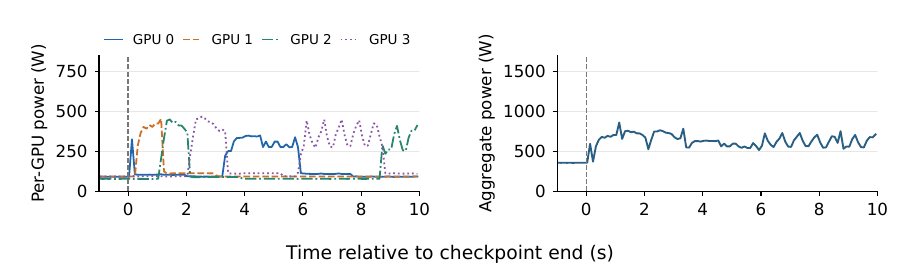}
        \caption{A100}
        \label{fig:trace_a100}
    \end{subfigure}

    \vspace{0.45em}

    \begin{subfigure}[t]{\linewidth}
        \centering
        \includegraphics[width=\linewidth]{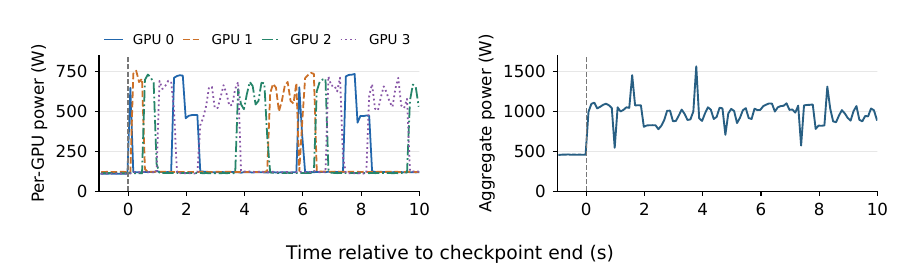}
        \caption{H100}
        \label{fig:trace_h100}
    \end{subfigure}

    \vspace{0.45em}

    \begin{subfigure}[t]{\linewidth}
        \centering
        \includegraphics[width=\linewidth]{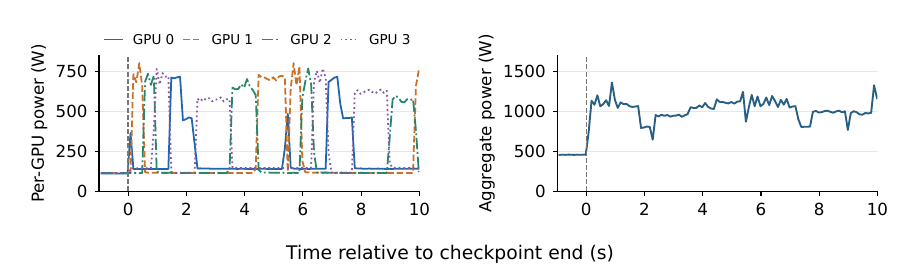}
        \caption{H200}
        \label{fig:trace_h200}
    \end{subfigure}

    \vspace{0.45em}

    \begin{subfigure}[t]{\linewidth}
        \centering
        \includegraphics[width=\linewidth]{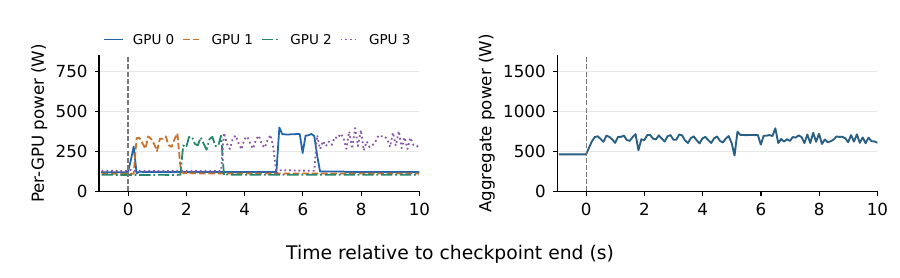}
        \caption{L40S}
        \label{fig:trace_l40s}
    \end{subfigure}

    \caption{[Fine-tuning] \textbf{GPU platform.} Comparison of Gemma 3 27B QLoRA fine-tuning power traces across A100, H100, H200, and L40S GPUs. The L40S and A100 traces exhibit weaker temporal regularity, whereas H100 and H200 show more clearly repeated fine-tuning patterns. The fine-tuning intervals are longest on L40S, followed by A100, while H100 and H200 exhibit similar and substantially shorter durations. In contrast, the sustained power level increases progressively from L40S to A100, H100, and H200.
}
    \label{fig:gpu_power_traces}
\end{figure}





\FloatBarrier


\subsection{Complete fine-tuning comparisons}
\label{app:training_quantitative}

We next provide the complete quantitative comparisons across the major fine-tuning factors considered in our experiments. These results complement the representative traces above by systematically quantifying how each factor affects iteration duration, energy per iteration, energy per token, power distribution, ramping behavior, and temporal regularity. We characterize these dimensions using the metrics defined in Sec.~\ref{sec:metrics}, including power percentiles, ramping rate percentiles, and spectral entropy.


\begin{figure*}[htbp]
    \centering

    \begin{subfigure}[t]{0.48\textwidth}
        \centering
        \includegraphics[width=\linewidth]
        {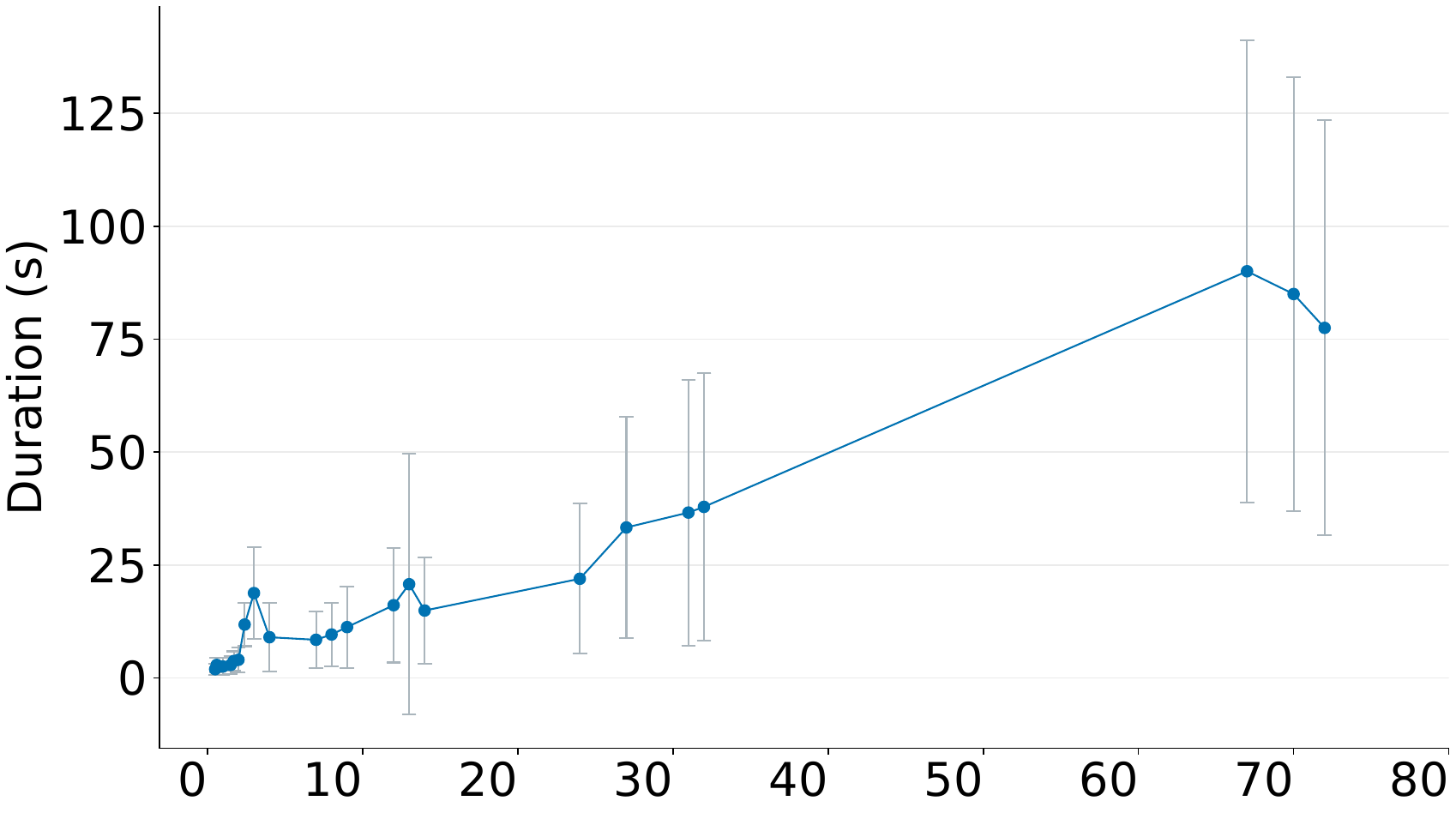}
        \caption{Iteration duration}
        \label{fig:quant_model_size_duration}
    \end{subfigure}
    \hfill
    \begin{subfigure}[t]{0.48\textwidth}
        \centering
        \includegraphics[width=\linewidth]
        {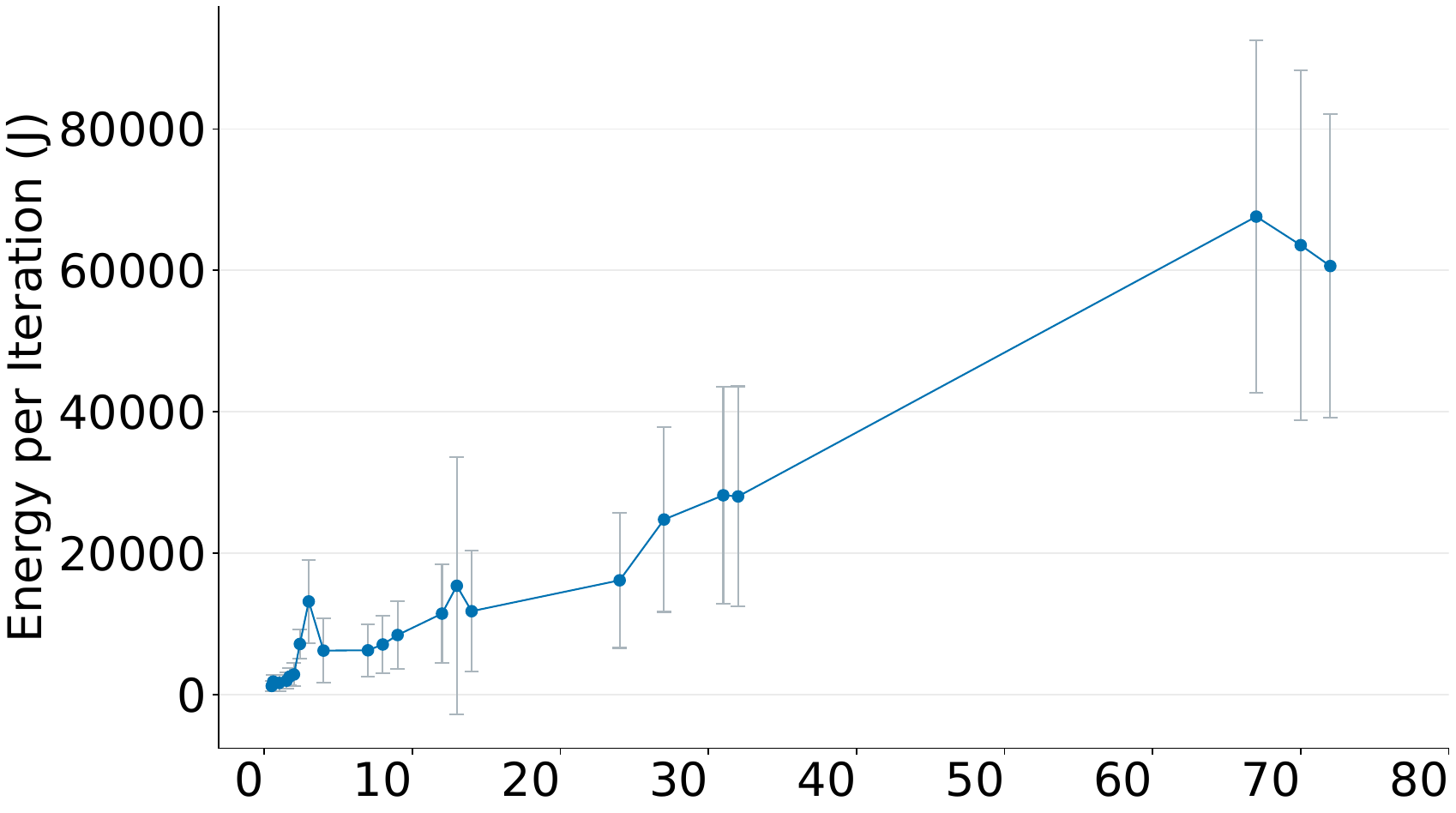}
        \caption{Energy per iteration}
        \label{fig:quant_model_size_energy_iteration}
    \end{subfigure}

    \vspace{0.5em}

    \begin{subfigure}[t]{0.48\textwidth}
        \centering
        \includegraphics[width=\linewidth]
        {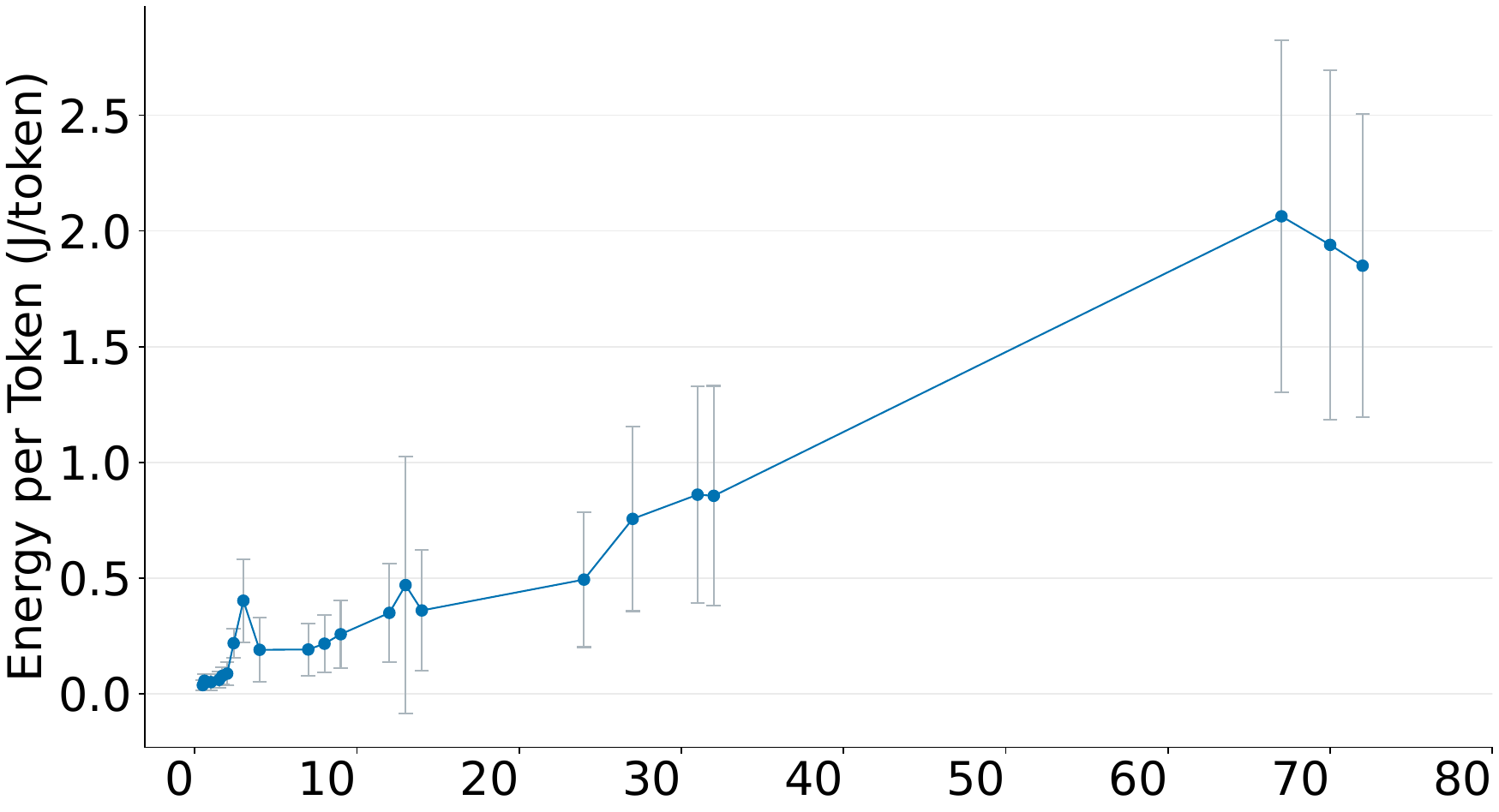}
        \caption{Energy per token}
        \label{fig:quant_model_size_energy_token}
    \end{subfigure}
    \hfill
    \begin{subfigure}[t]{0.48\textwidth}
        \centering
        \includegraphics[width=\linewidth]
        {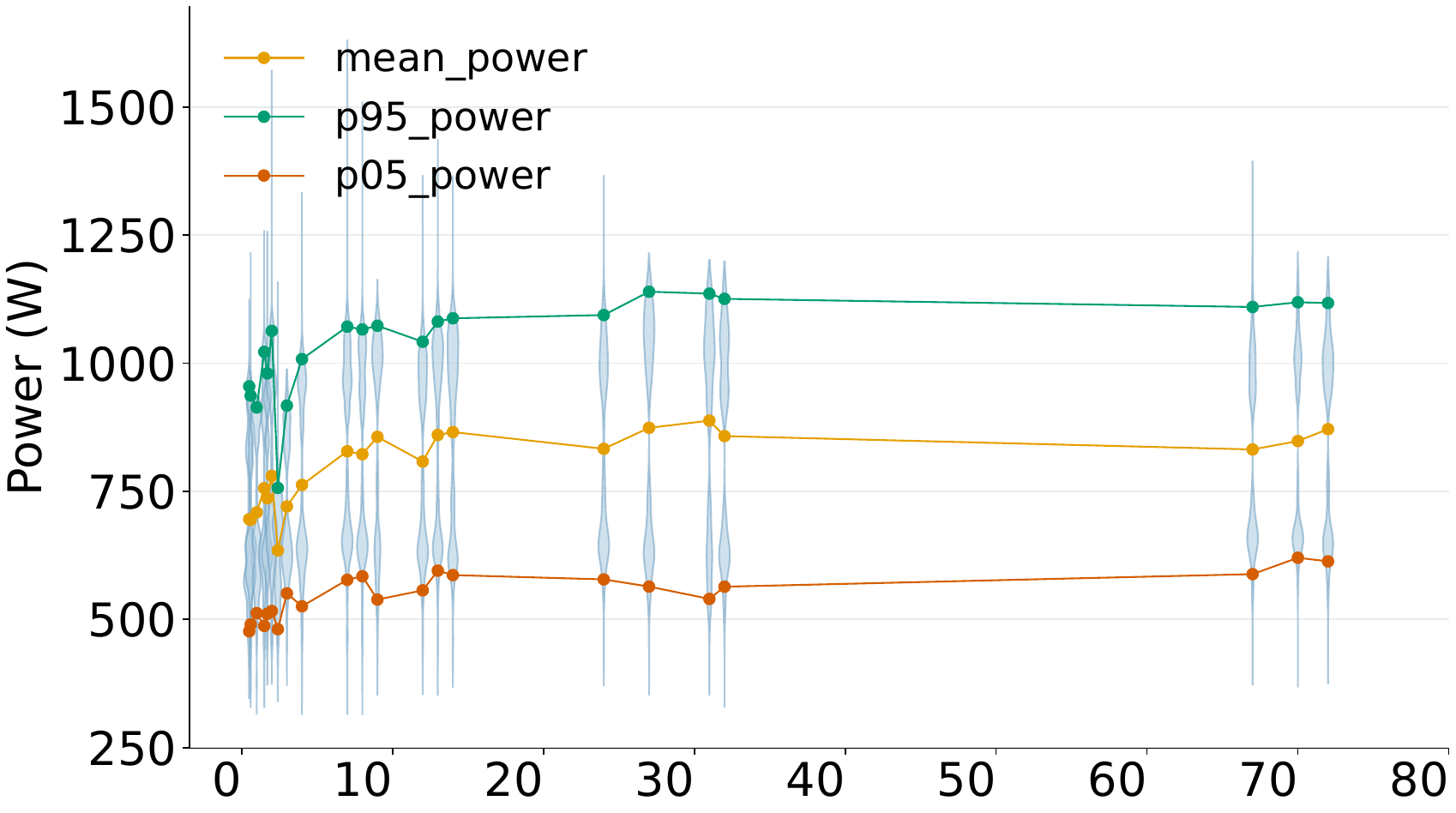}
        \caption{Power distribution}
        \label{fig:quant_model_size_power}
    \end{subfigure}

    \vspace{0.5em}

    \begin{subfigure}[t]{0.48\textwidth}
        \centering
        \includegraphics[width=\linewidth]
        {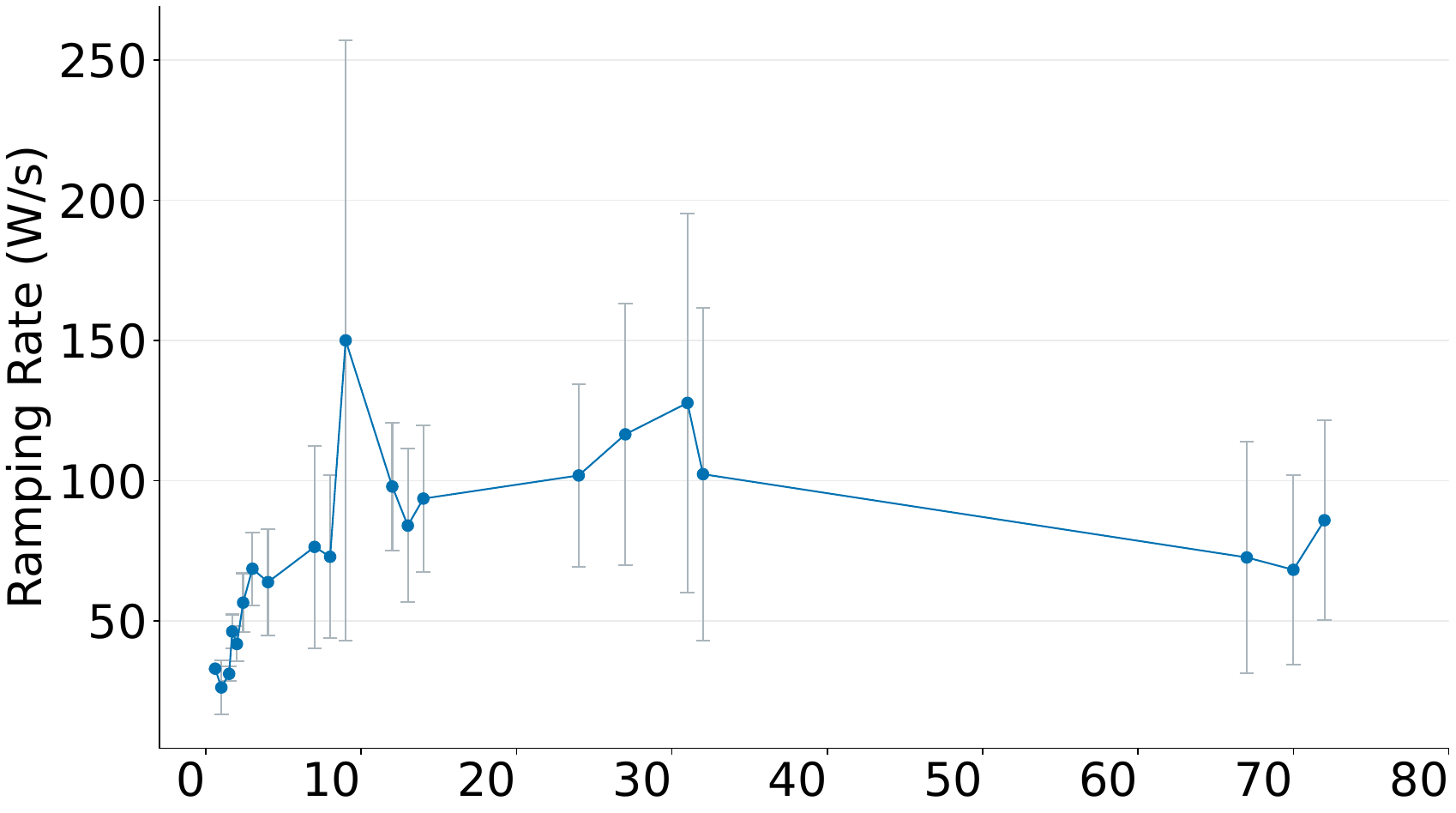}
        \caption{Ramping rate}
        \label{fig:quant_model_size_ramp}
    \end{subfigure}
    \hfill
    \begin{subfigure}[t]{0.48\textwidth}
        \centering
        \includegraphics[width=\linewidth]
        {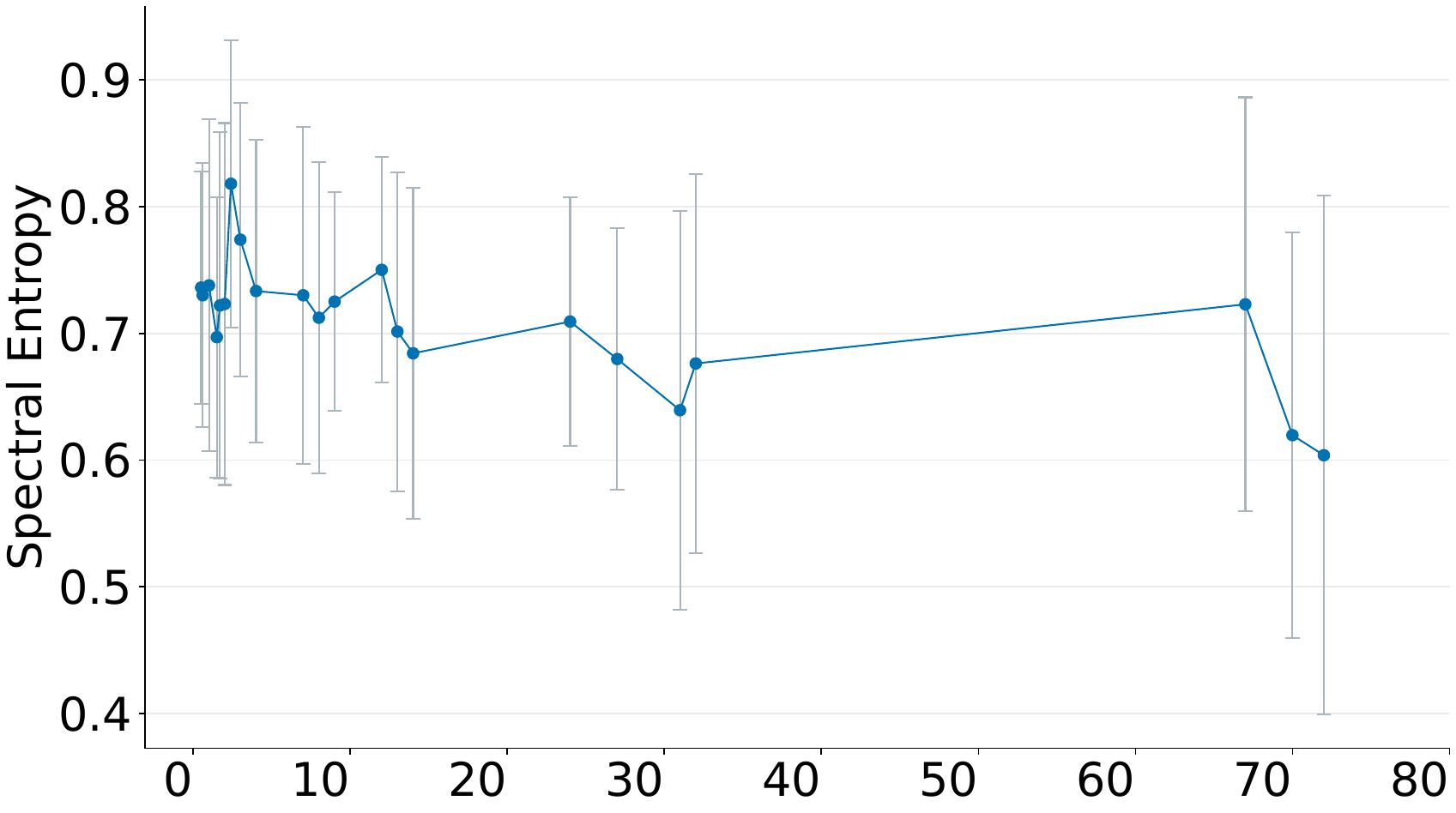}
        \caption{Spectral entropy}
        \label{fig:quant_model_size_entropy}
    \end{subfigure}

    \caption{[Fine-tuning] \textbf{Model size.} Complete quantitative comparison across different model sizes for fine-tuning tasks.
    The panels report iteration duration, energy per iteration, energy per token,
    power distribution, ramping rate, and spectral entropy. It summarizes how the six fine-tuning metrics vary
with active parameter count. Mean iteration duration and energy per iteration
increase with model size, with a particularly pronounced increase for the
largest models (Figs.~\ref{fig:quant_model_size_duration},
\ref{fig:quant_model_size_energy_iteration}). Energy per token exhibits a
similar upward trend at larger model sizes, while remaining non-monotonic
across intermediate configurations
(Fig.~\ref{fig:quant_model_size_energy_token}). Mean and upper-tail power
increase initially before approaching a broad plateau, whereas lower-tail
power varies more modestly (Fig.~\ref{fig:quant_model_size_power}). Ramping
rate increases toward medium--large models but remains non-monotonic across
configurations, while spectral entropy exhibits no consistent relationship
with model size (Figs.~\ref{fig:quant_model_size_ramp},
\ref{fig:quant_model_size_entropy}). }
    \label{fig:quant_model_size}
\end{figure*}


\begin{figure*}[!t]
    \centering

    \begin{subfigure}[t]{0.48\textwidth}
        \centering
        \includegraphics[width=\linewidth]
        {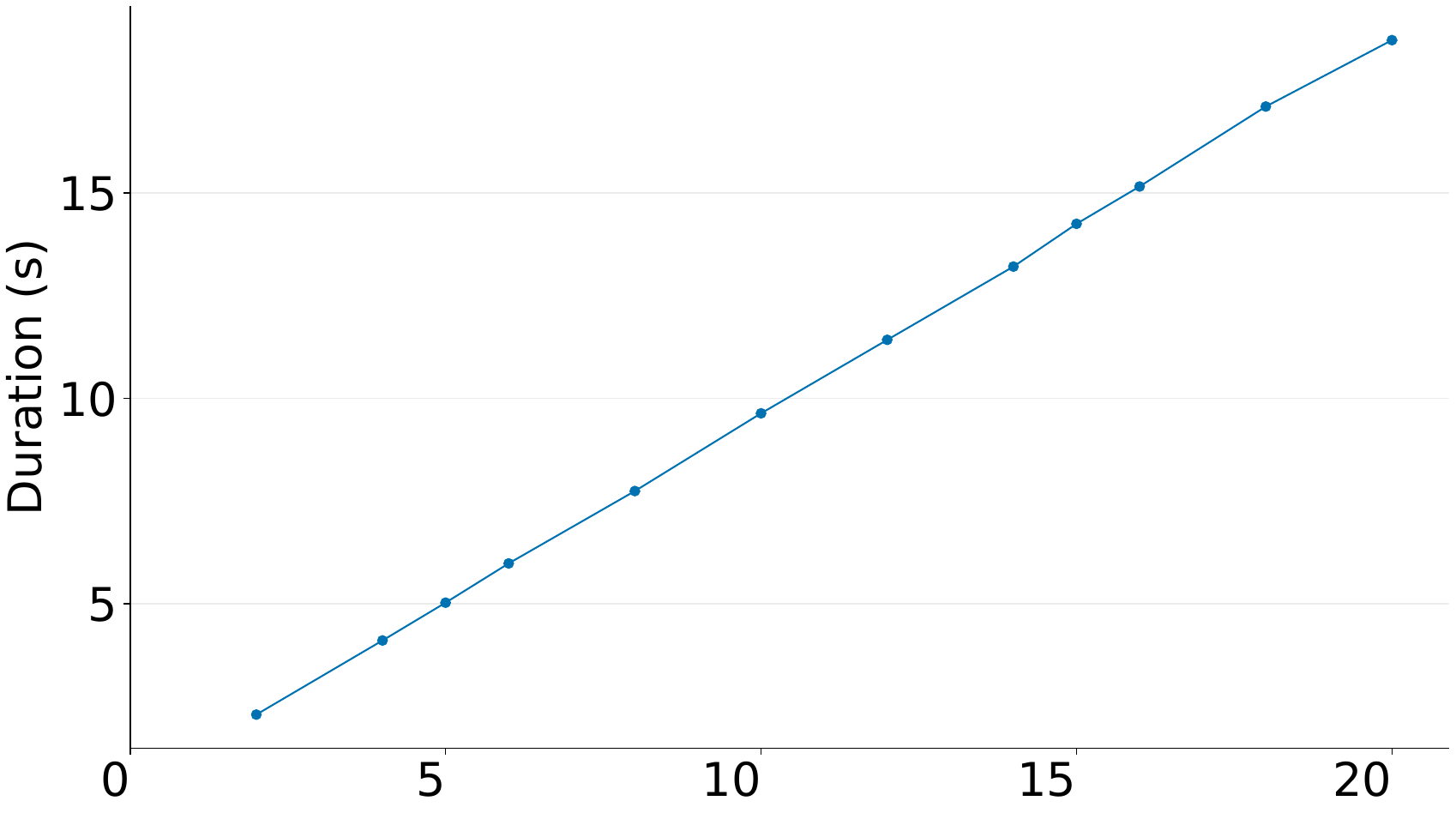}
        \caption{Iteration duration}
        \label{fig:quant_batch_size_duration}
    \end{subfigure}
    \hfill
    \begin{subfigure}[t]{0.48\textwidth}
        \centering
        \includegraphics[width=\linewidth]
        {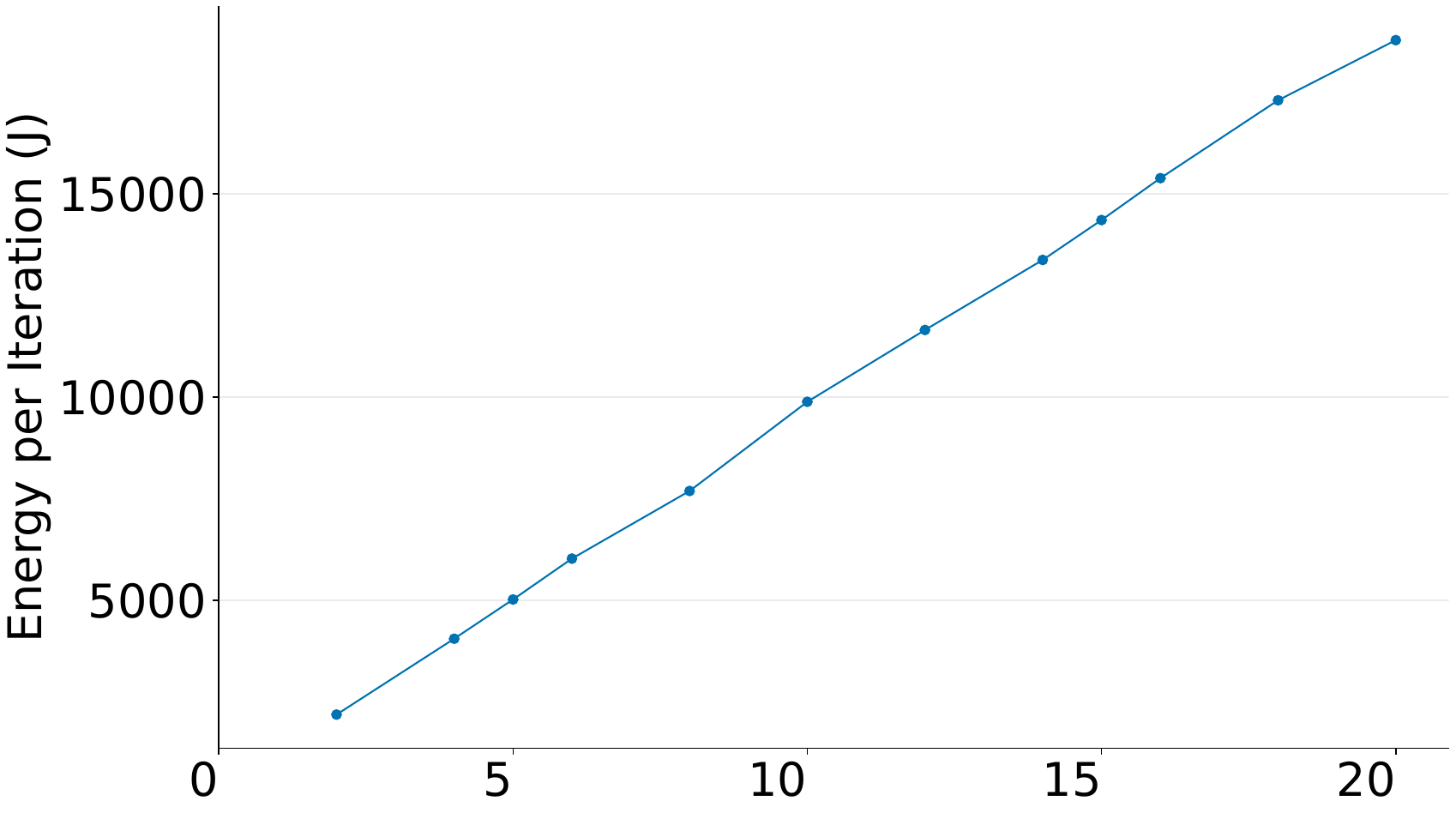}
        \caption{Energy per iteration}
        \label{fig:quant_batch_size_energy_iteration}
    \end{subfigure}

    \vspace{0.5em}

    \begin{subfigure}[t]{0.48\textwidth}
        \centering
        \includegraphics[width=\linewidth]
        {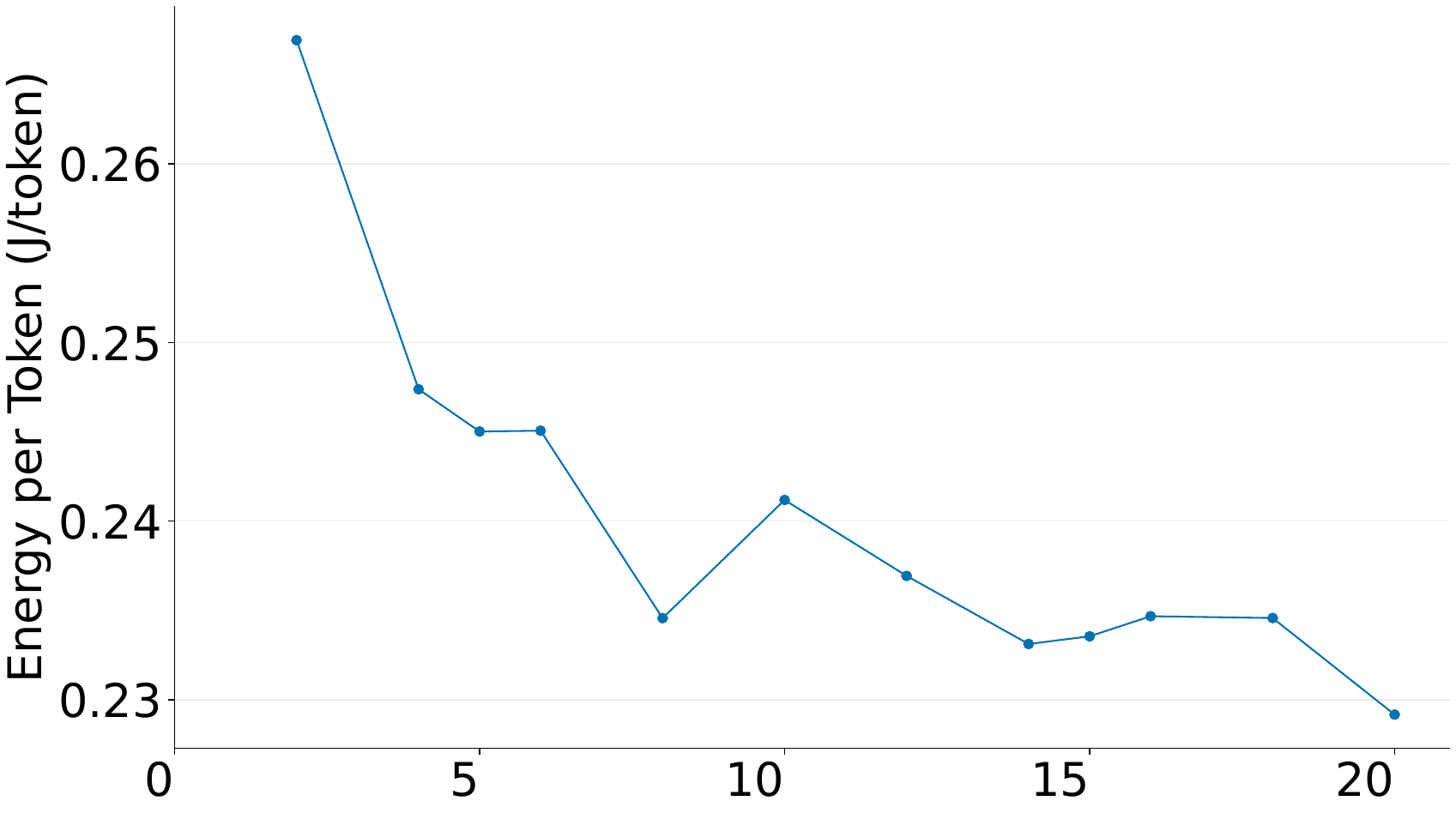}
        \caption{Energy per token}
        \label{fig:quant_batch_size_energy_token}
    \end{subfigure}
    \hfill
    \begin{subfigure}[t]{0.48\textwidth}
        \centering
        \includegraphics[width=\linewidth]
        {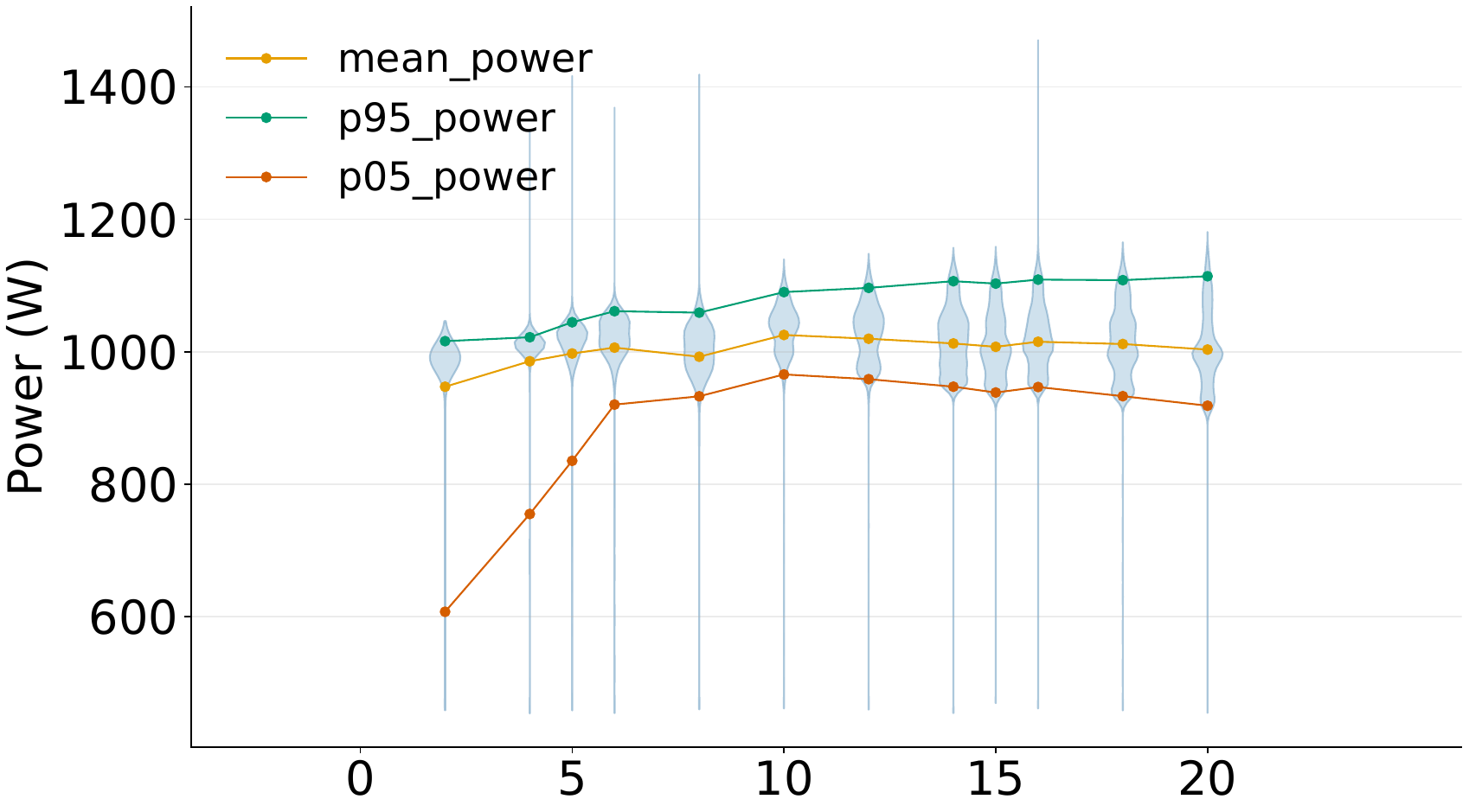}
        \caption{Power distribution}
        \label{fig:quant_batch_size_power}
    \end{subfigure}

    \vspace{0.5em}

    \begin{subfigure}[t]{0.48\textwidth}
        \centering
        \includegraphics[width=\linewidth]
        {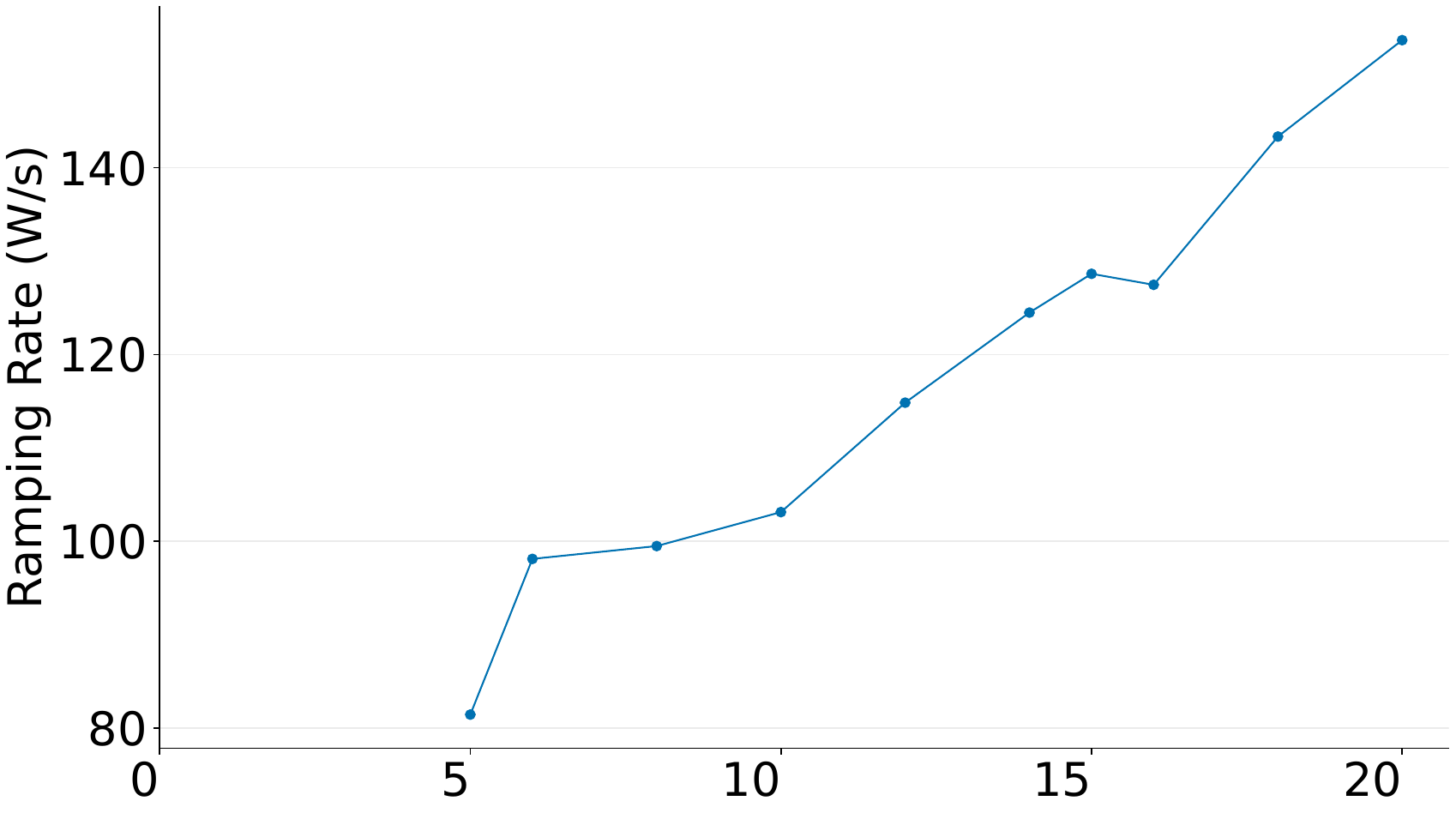}
        \caption{Ramping rate}
        \label{fig:quant_batch_size_ramp}
    \end{subfigure}
    \hfill
    \begin{subfigure}[t]{0.48\textwidth}
        \centering
        \includegraphics[width=\linewidth]
        {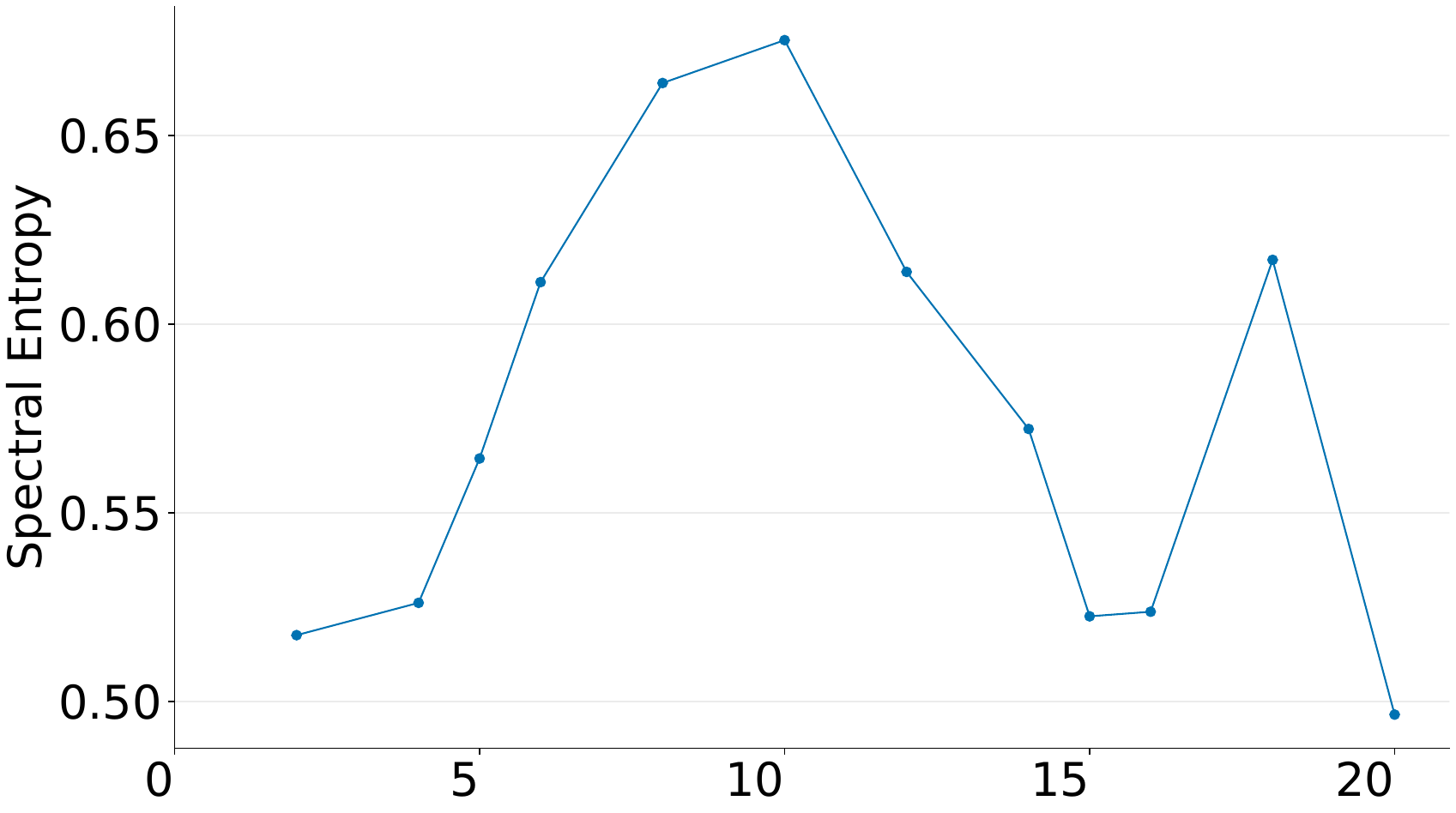}
        \caption{Spectral entropy}
        \label{fig:quant_batch_size_entropy}
    \end{subfigure}

    \caption{[Fine-tuning]  \textbf{Batch size.} Complete quantitative comparison across different batch sizes.
    The panels report iteration duration, energy per iteration, energy per token,
    power distribution, ramping rate, and spectral entropy. It summarizes the effects of batch size on the
six fine-tuning metrics. Iteration duration and energy per iteration increase
with batch size as more samples are processed per optimizer step
(Figs.~\ref{fig:quant_batch_size_duration},
\ref{fig:quant_batch_size_energy_iteration}). In contrast, energy per token
generally decreases, indicating that larger batches amortize fixed
computational and synchronization overheads over more tokens
(Fig.~\ref{fig:quant_batch_size_energy_token}). Mean and 95th-percentile
power increase only moderately, while 5th-percentile power rises sharply at
small batch sizes before leveling off (Fig.~\ref{fig:quant_batch_size_power}).
Ramping rate also increases with batch size, whereas spectral entropy varies
non-monotonically and peaks at intermediate batch sizes
(Figs.~\ref{fig:quant_batch_size_ramp},
\ref{fig:quant_batch_size_entropy}).}
    \label{fig:quant_batch_size}
\end{figure*}


\begin{figure*}[!t]
    \centering

    \begin{subfigure}[t]{0.48\textwidth}
        \centering
        \includegraphics[width=\linewidth]
        {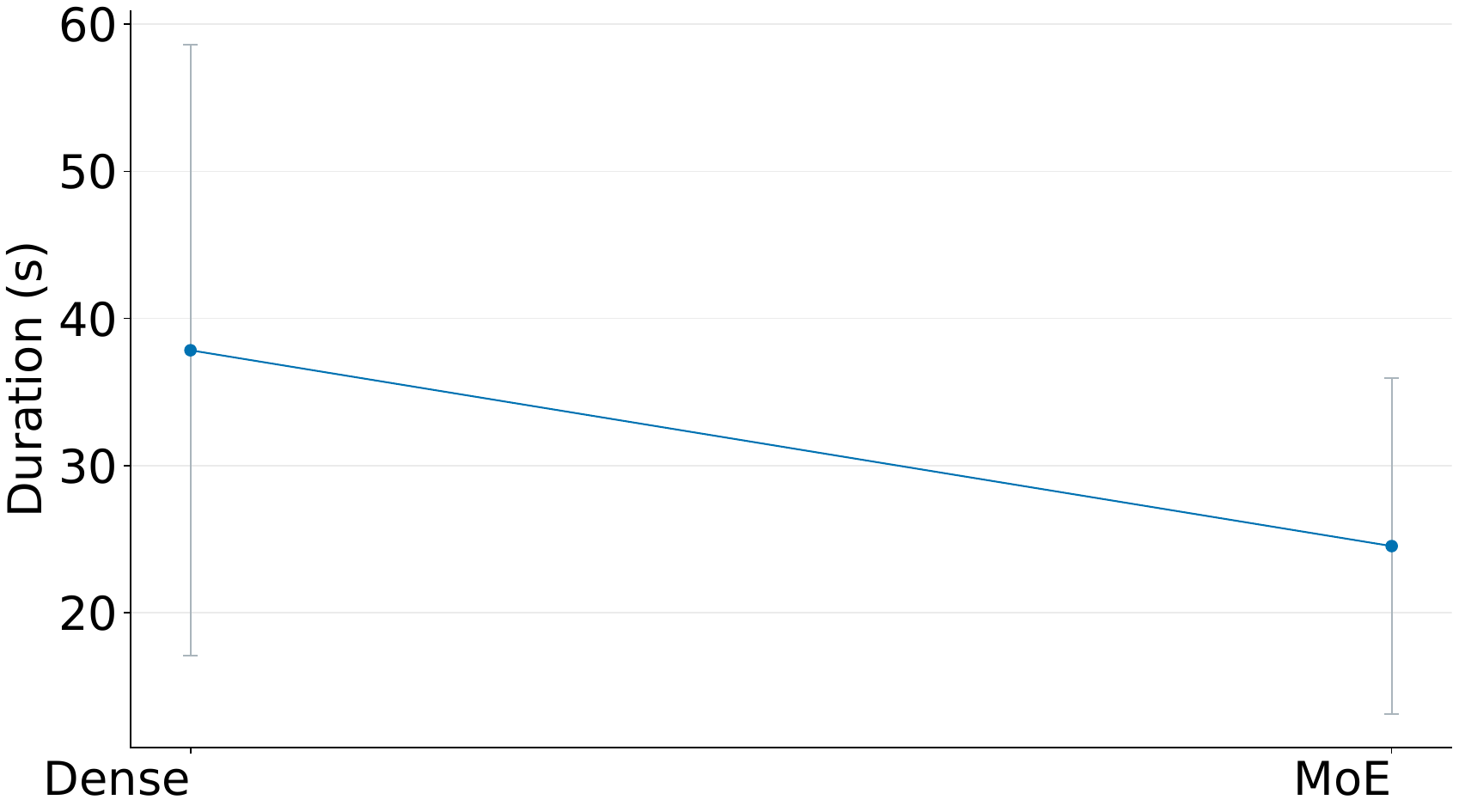}
        \caption{Iteration duration}
        \label{fig:quant_moe_dense_duration}
    \end{subfigure}
    \hfill
    \begin{subfigure}[t]{0.48\textwidth}
        \centering
        \includegraphics[width=\linewidth]
        {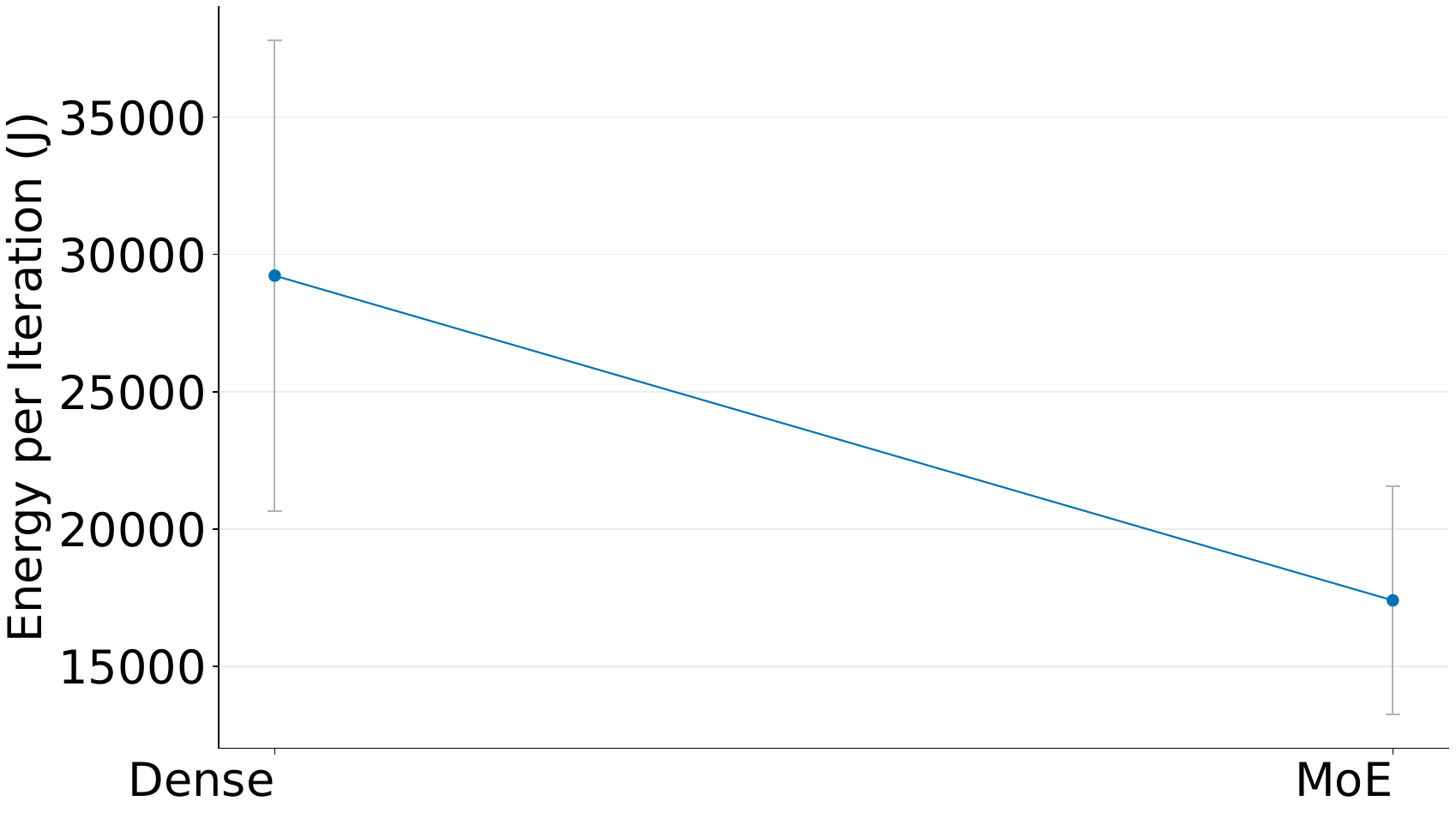}
        \caption{Energy per iteration}
        \label{fig:quant_moe_dense_energy_iteration}
    \end{subfigure}

    \vspace{0.5em}

    \begin{subfigure}[t]{0.48\textwidth}
        \centering
        \includegraphics[width=\linewidth]
        {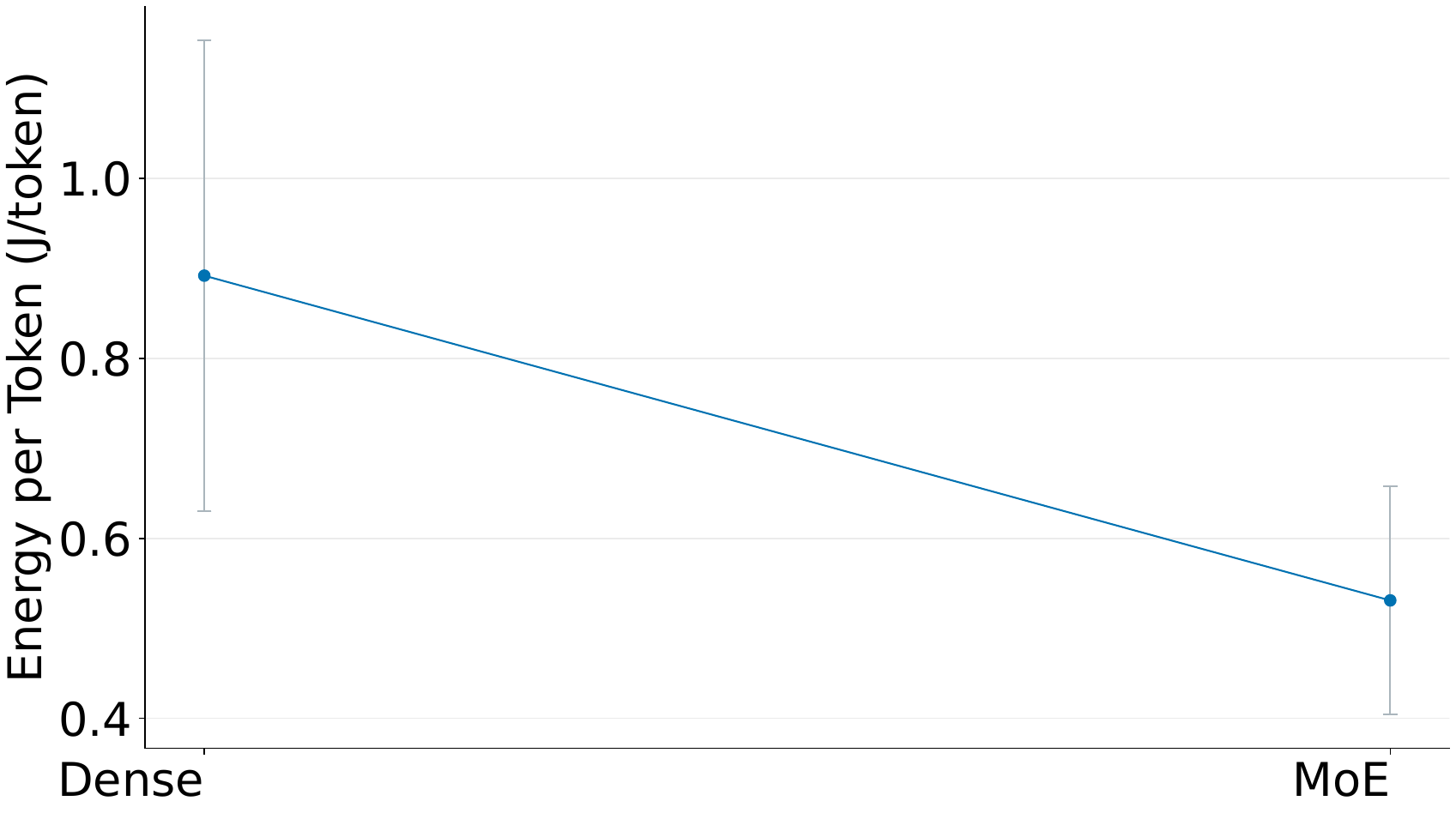}
        \caption{Energy per token}
        \label{fig:quant_moe_dense_energy_token}
    \end{subfigure}
    \hfill
    \begin{subfigure}[t]{0.48\textwidth}
        \centering
        \includegraphics[width=\linewidth]
        {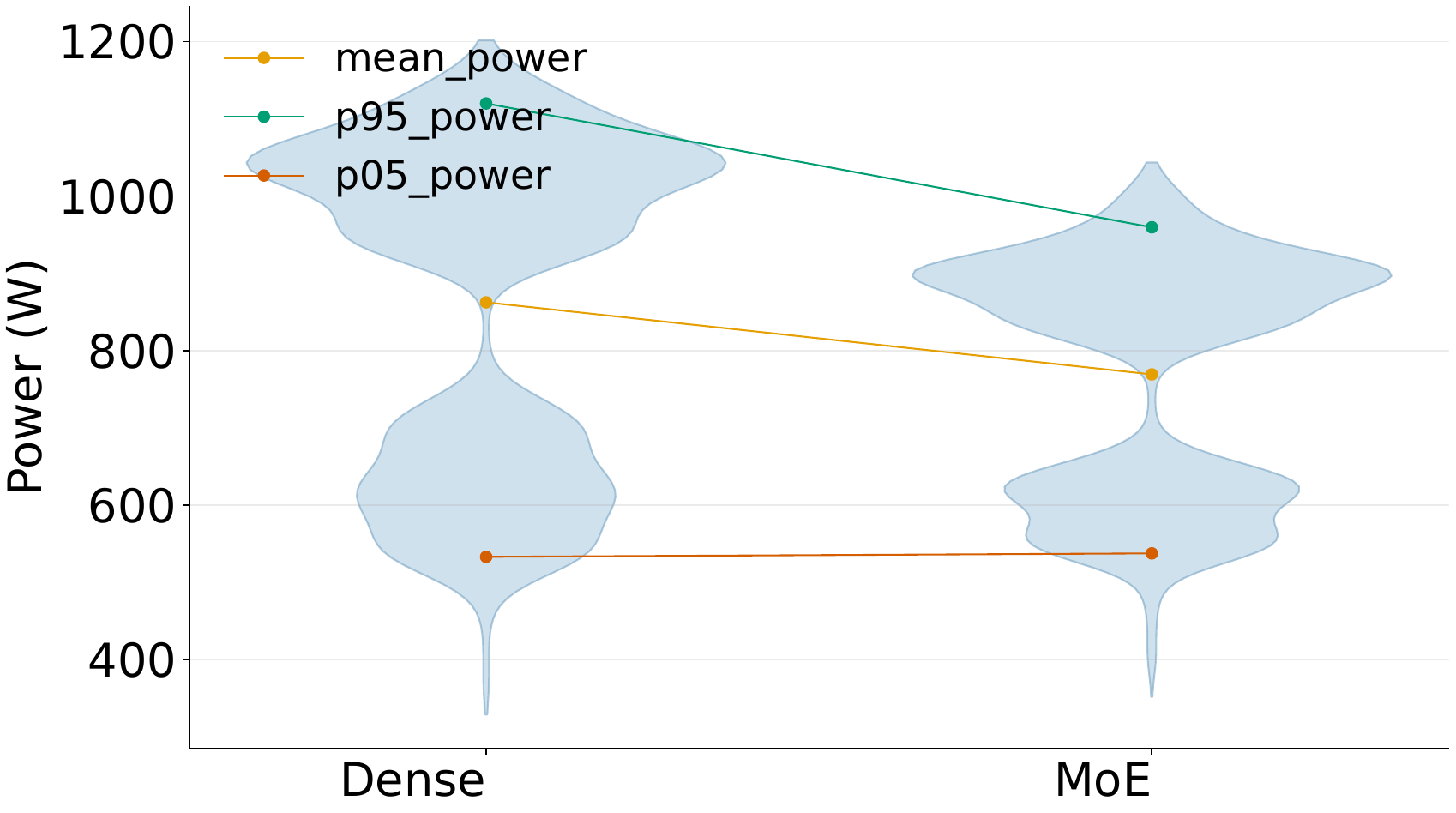}
        \caption{Power distribution}
        \label{fig:quant_moe_dense_power}
    \end{subfigure}

    \vspace{0.5em}

    \begin{subfigure}[t]{0.48\textwidth}
        \centering
        \includegraphics[width=\linewidth]
        {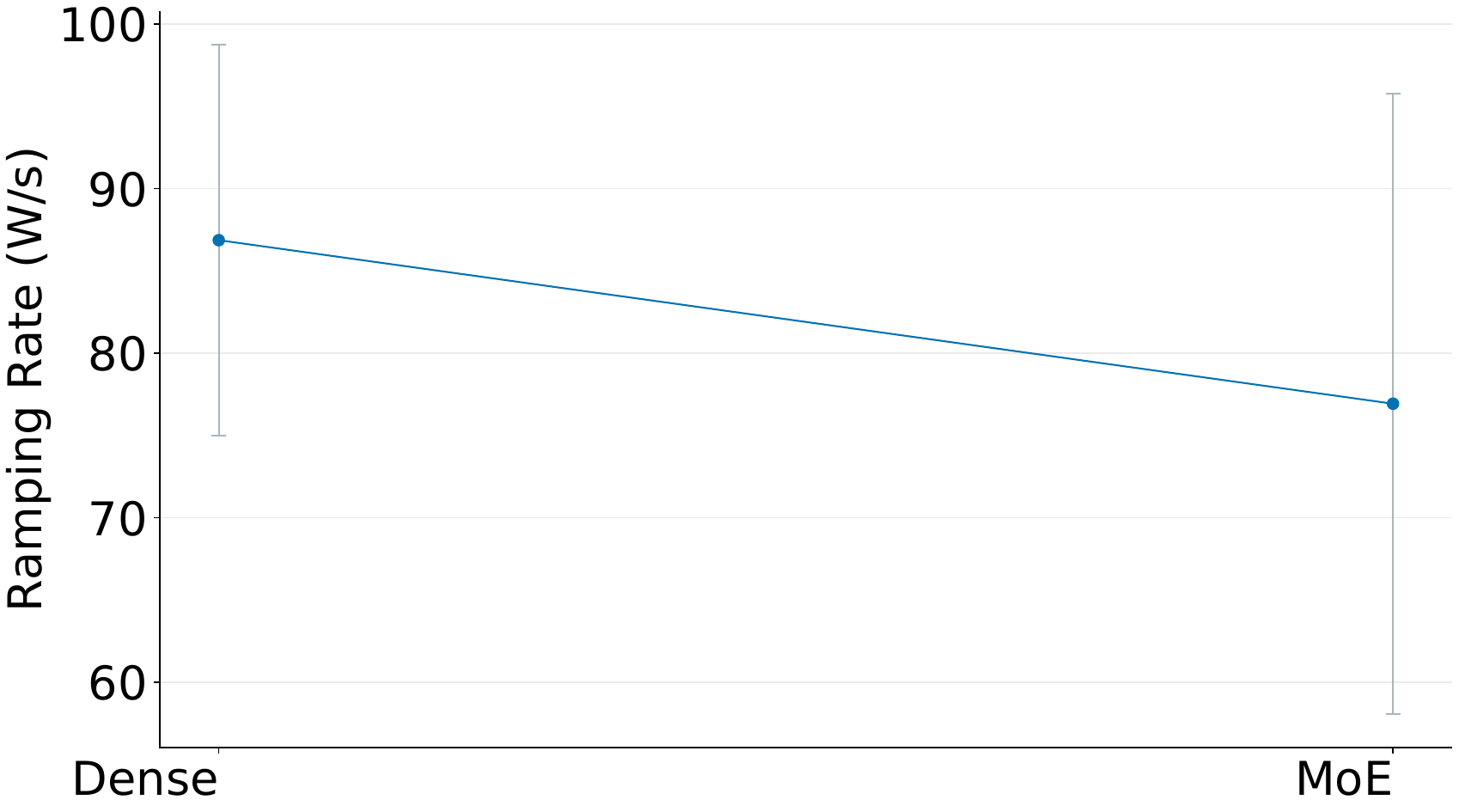}
        \caption{Ramping rate}
        \label{fig:quant_moe_dense_ramp}
    \end{subfigure}
    \hfill
    \begin{subfigure}[t]{0.48\textwidth}
        \centering
        \includegraphics[width=\linewidth]
        {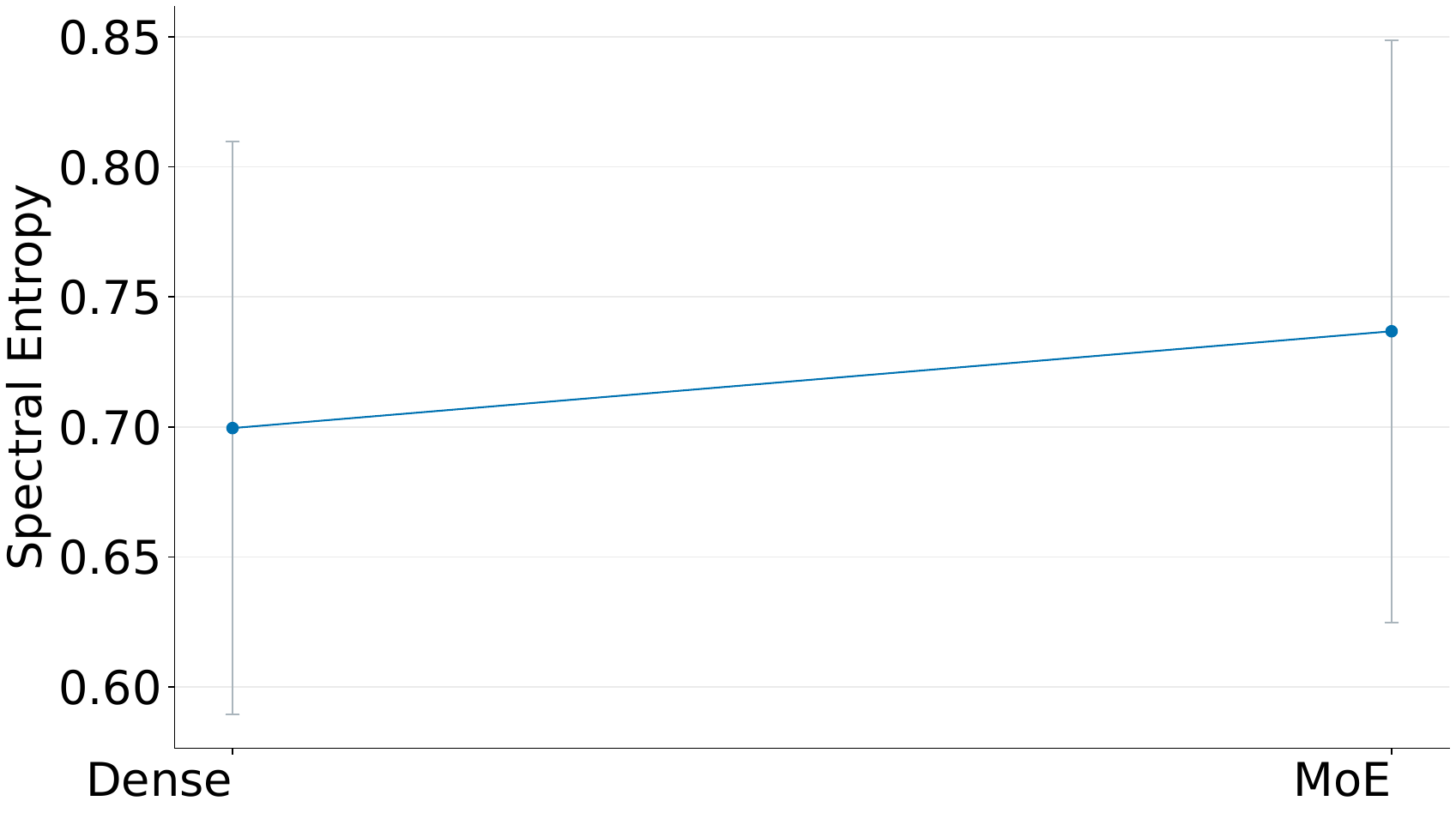}
        \caption{Spectral entropy}
        \label{fig:quant_moe_dense_entropy}
    \end{subfigure}

    \caption{[Fine-tuning] \textbf{Model architecture.} Complete quantitative comparison between mixture-of-experts (MoE)
    and dense model architectures. The panels report iteration duration,
    energy per iteration, energy per token, power distribution, ramping rate,
    and spectral entropy. It compares dense and mixture-of-experts (MoE)
architectures across the six metrics. Compared with dense models, MoE models
have shorter iteration duration and lower energy per iteration and per token
(Figs.~\ref{fig:quant_moe_dense_duration},
\ref{fig:quant_moe_dense_energy_iteration},
\ref{fig:quant_moe_dense_energy_token}). Mean and 95th-percentile power are
also lower, while 5th-percentile power remains similar
(Fig.~\ref{fig:quant_moe_dense_power}). MoE models show slightly lower ramping
rate and slightly higher spectral entropy, although both exhibit substantial
overlap in the error bars (Figs.~\ref{fig:quant_moe_dense_ramp},
\ref{fig:quant_moe_dense_entropy}).}
    \label{fig:quant_moe_dense}
\end{figure*}


\begin{figure*}[!t]
    \centering

    \begin{subfigure}[t]{0.48\textwidth}
        \centering
        \includegraphics[width=\linewidth]
        {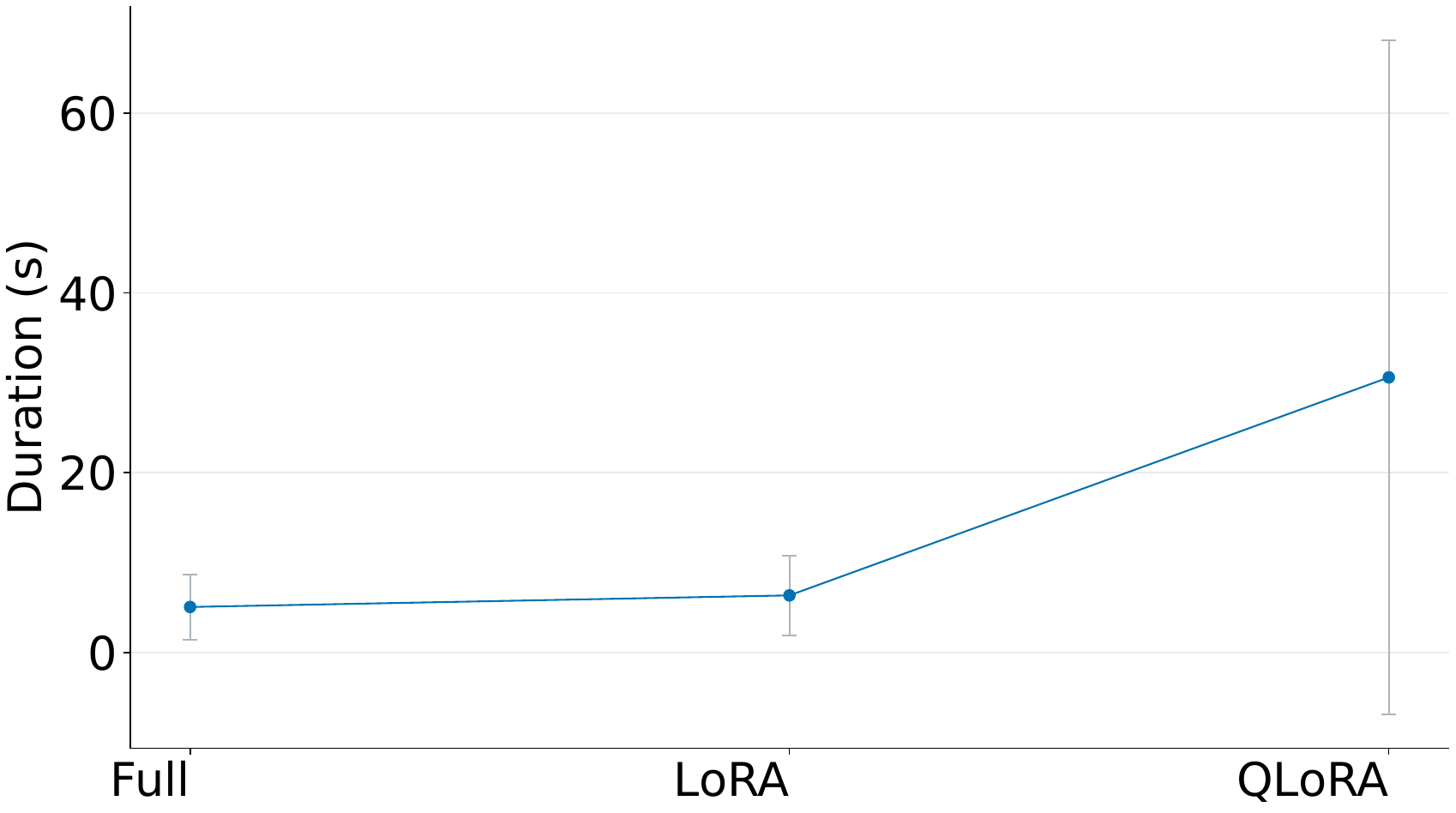}
        \caption{Iteration duration}
        \label{fig:quant_training_method_duration}
    \end{subfigure}
    \hfill
    \begin{subfigure}[t]{0.48\textwidth}
        \centering
        \includegraphics[width=\linewidth]
        {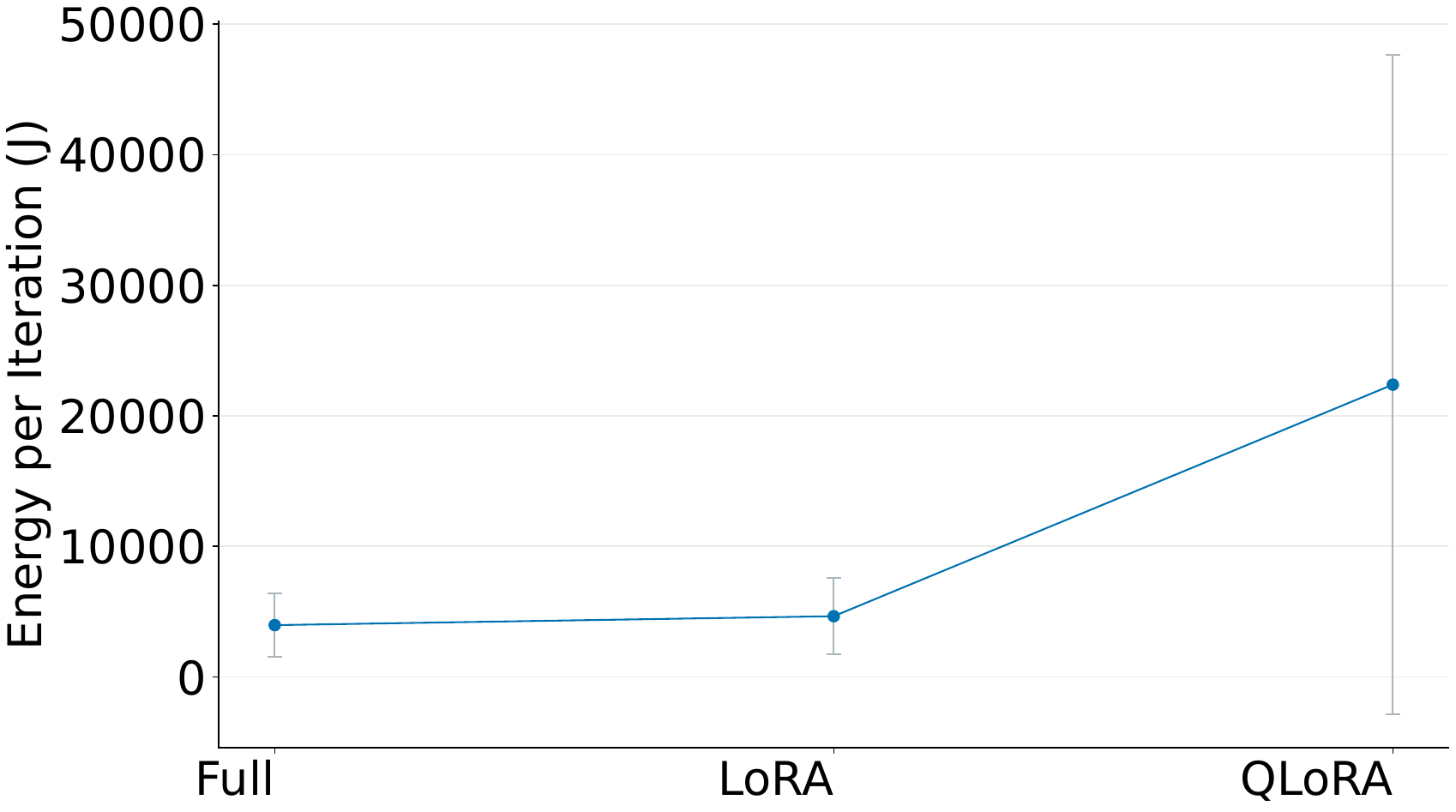}
        \caption{Energy per iteration}
        \label{fig:quant_training_method_energy_iteration}
    \end{subfigure}

    \vspace{0.5em}

    \begin{subfigure}[t]{0.48\textwidth}
        \centering
        \includegraphics[width=\linewidth]
        {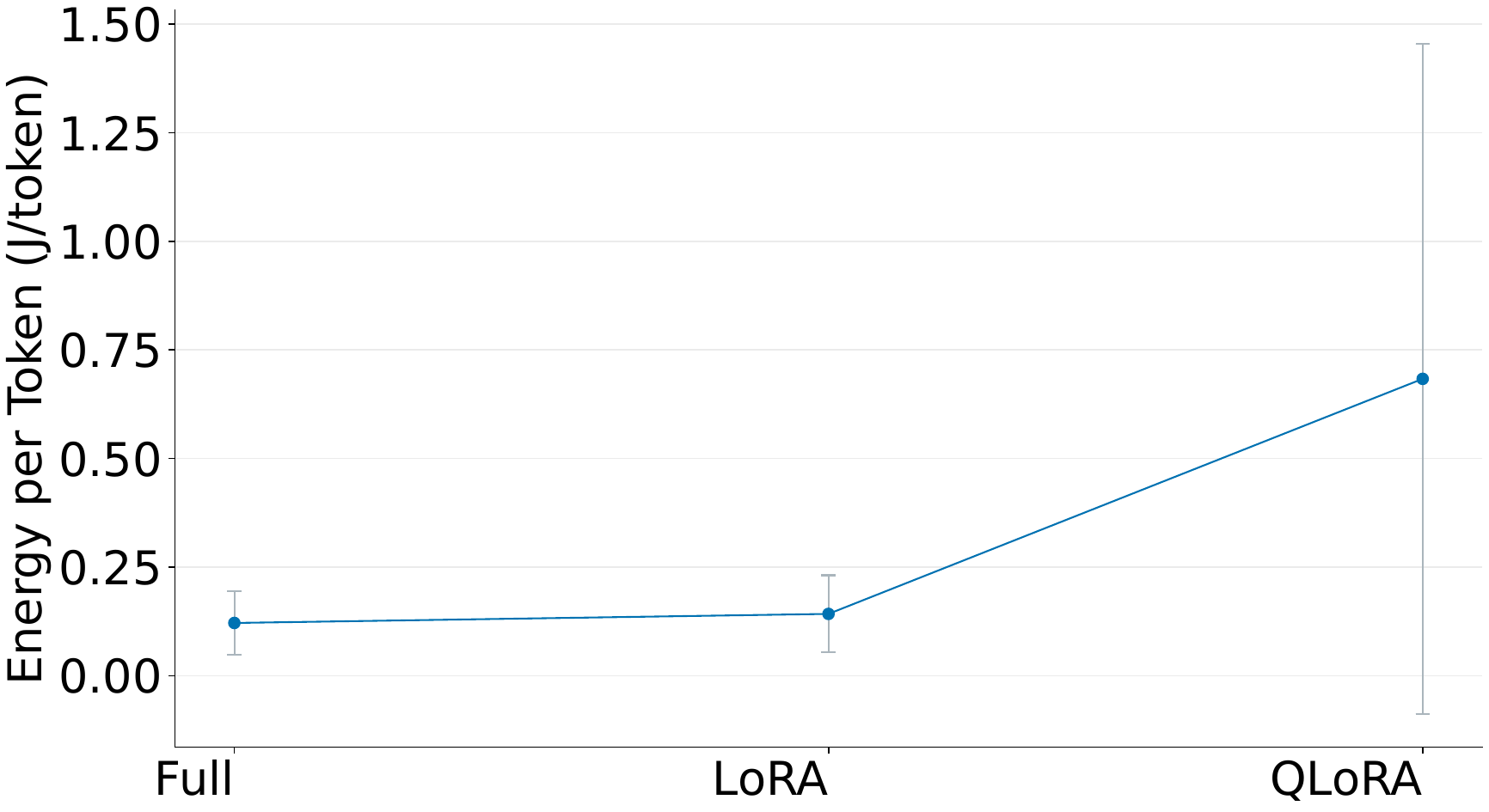}
        \caption{Energy per token}
        \label{fig:quant_training_method_energy_token}
    \end{subfigure}
    \hfill
    \begin{subfigure}[t]{0.48\textwidth}
        \centering
        \includegraphics[width=\linewidth]
        {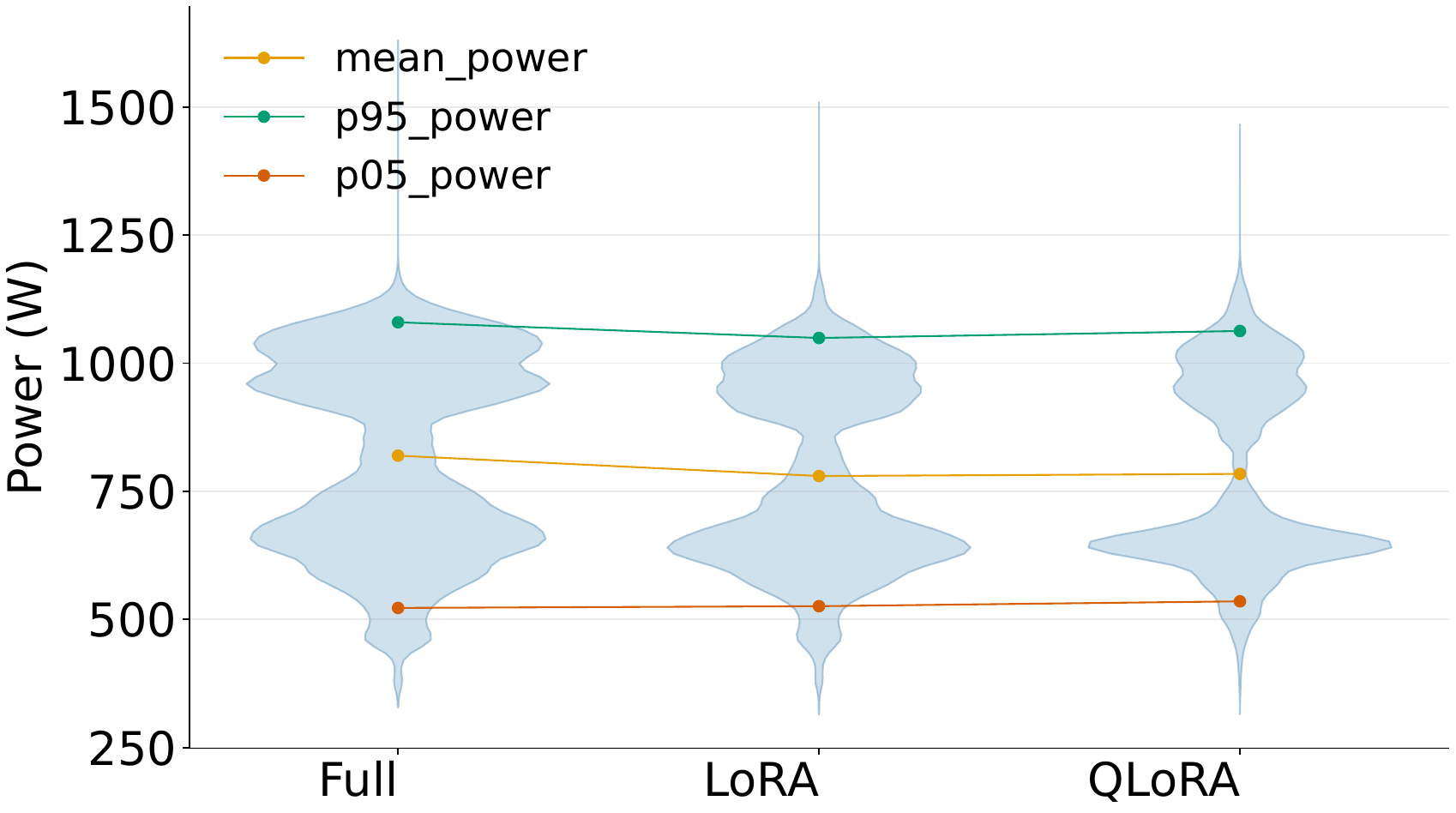}
        \caption{Power distribution}
        \label{fig:quant_training_method_power}
    \end{subfigure}

    \vspace{0.5em}

    \begin{subfigure}[t]{0.48\textwidth}
        \centering
        \includegraphics[width=\linewidth]
        {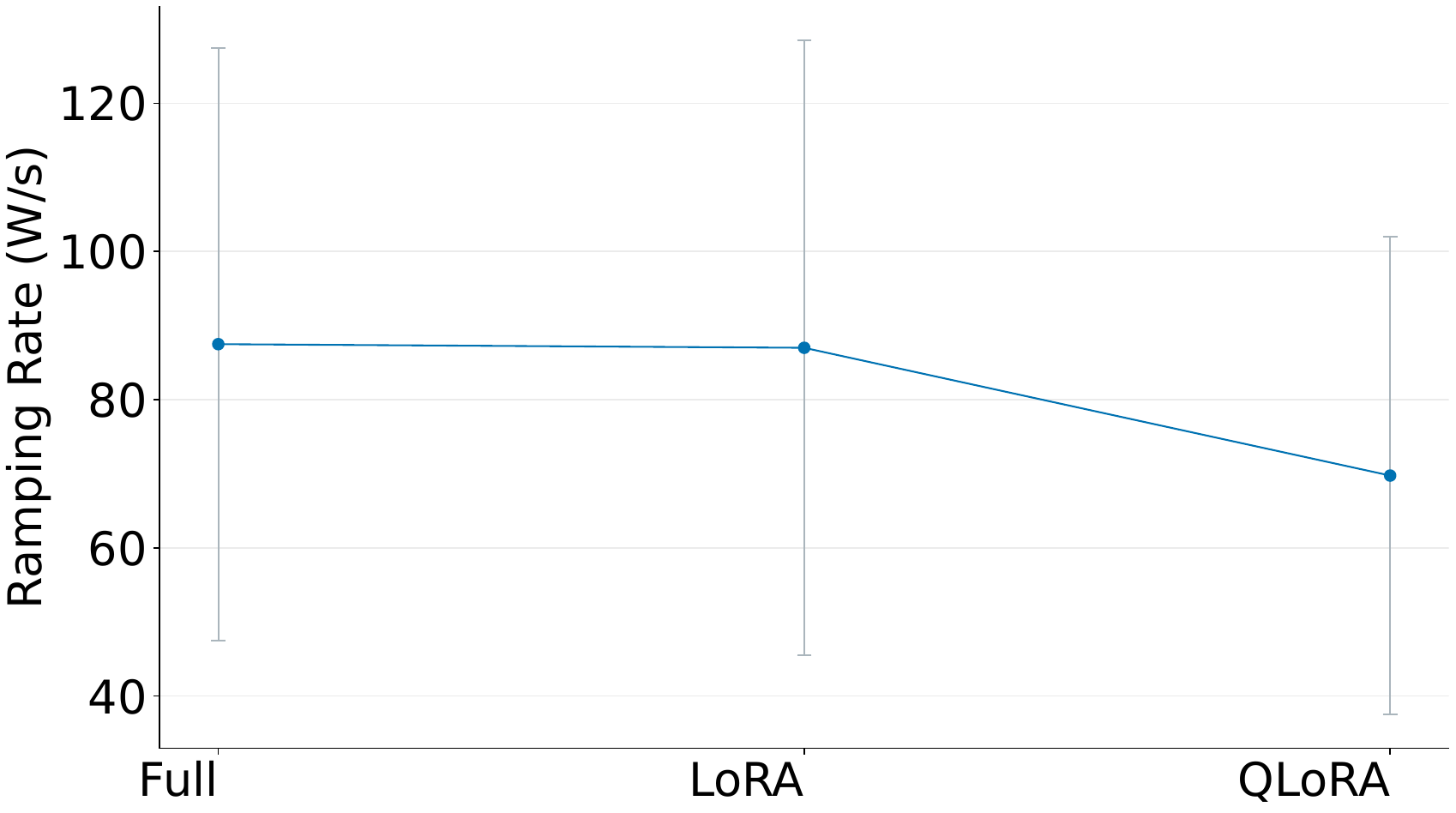}
        \caption{Ramping rate}
        \label{fig:quant_training_method_ramp}
    \end{subfigure}
    \hfill
    \begin{subfigure}[t]{0.48\textwidth}
        \centering
        \includegraphics[width=\linewidth]
        {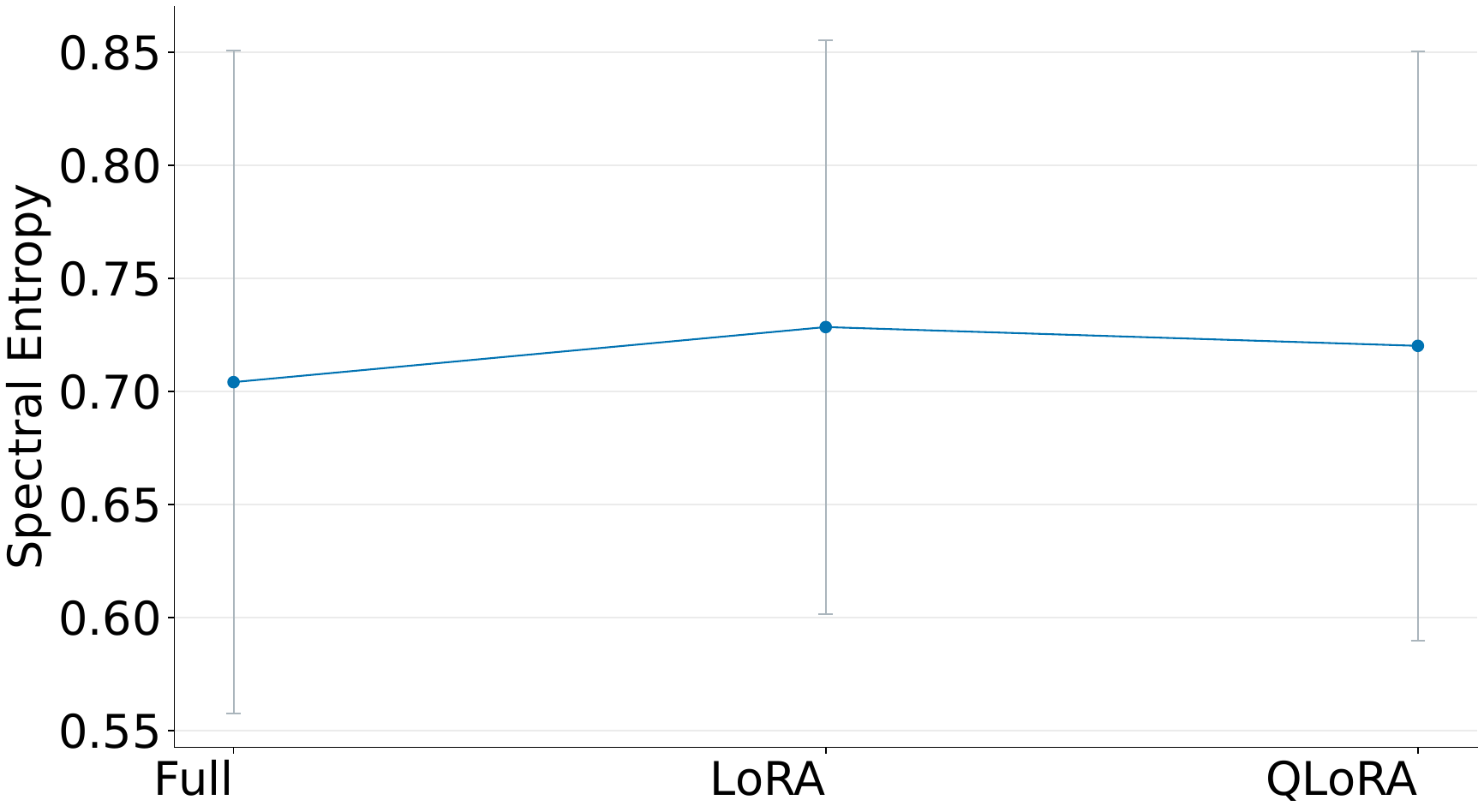}
        \caption{Spectral entropy}
        \label{fig:quant_training_method_entropy}
    \end{subfigure}

    \caption{[Fine-tuning]  \textbf{Fine-tuning method.} Complete quantitative comparison across different fine-tuning
    methods. The panels report iteration duration, energy per iteration,
    energy per token, power distribution, ramping rate, and spectral entropy. It compares full fine-tuning, LoRA, and
QLoRA across the six metrics. Full fine-tuning and LoRA have similar mean
iteration durations, while QLoRA is substantially slower and more variable
(Fig.~\ref{fig:quant_training_method_duration}). The same pattern appears in
energy per iteration and per token, with QLoRA showing the highest energy
consumption (Figs.~\ref{fig:quant_training_method_energy_iteration},
\ref{fig:quant_training_method_energy_token}). In contrast, power
distributions are broadly similar across the three methods, with only modest
differences in mean and percentile values
(Fig.~\ref{fig:quant_training_method_power}). Full fine-tuning and LoRA also
have similar ramping rates, while QLoRA is lower; spectral entropy varies only
slightly across methods (Figs.~\ref{fig:quant_training_method_ramp},
\ref{fig:quant_training_method_entropy}).}
    \label{fig:quant_training_method}
\end{figure*}


\begin{figure*}[!t]
    \centering

    \begin{subfigure}[t]{0.48\textwidth}
        \centering
        \includegraphics[width=\linewidth]
        {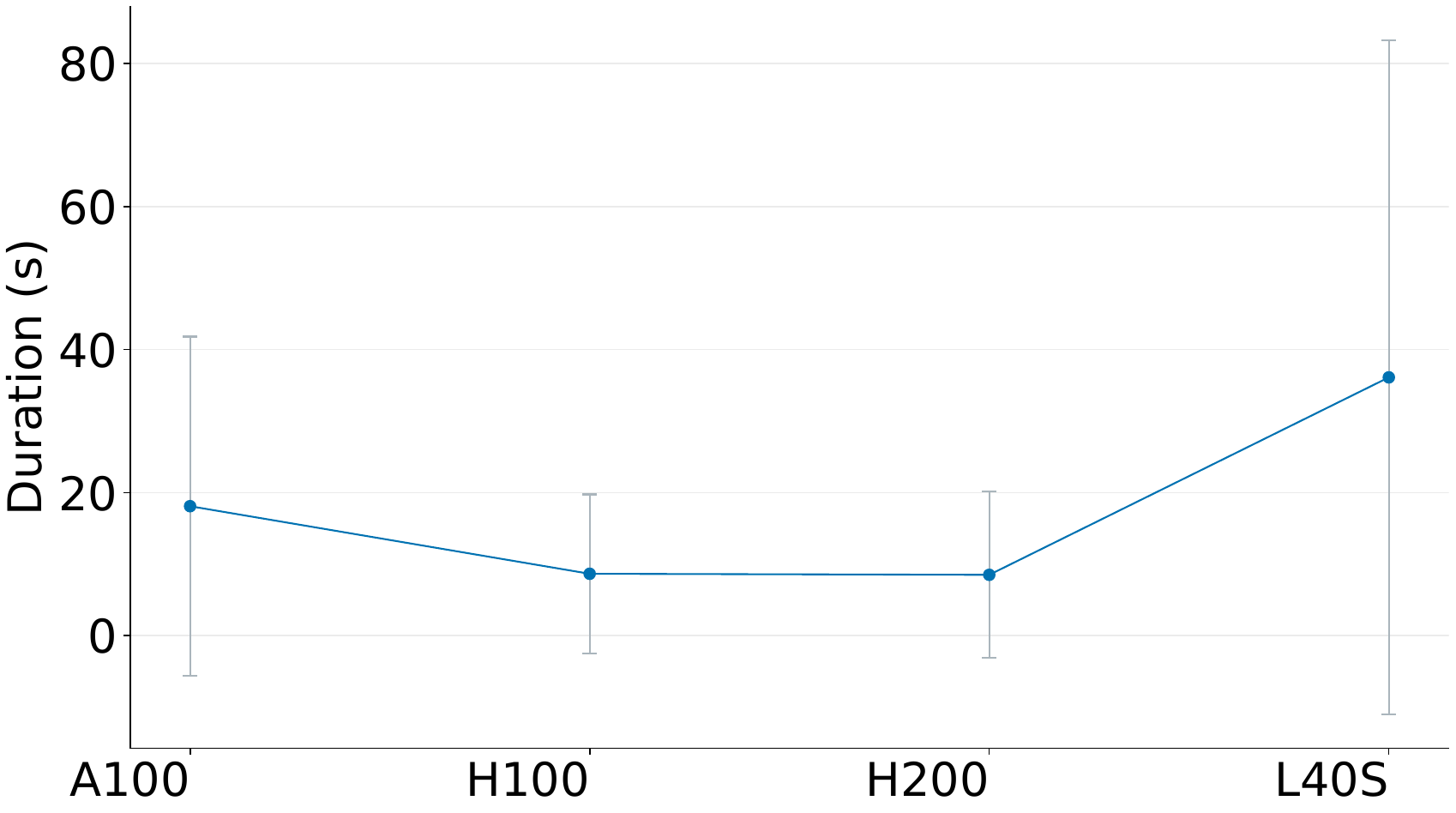}
        \caption{Iteration duration}
        \label{fig:quant_gpu_duration}
    \end{subfigure}
    \hfill
    \begin{subfigure}[t]{0.48\textwidth}
        \centering
        \includegraphics[width=\linewidth]
        {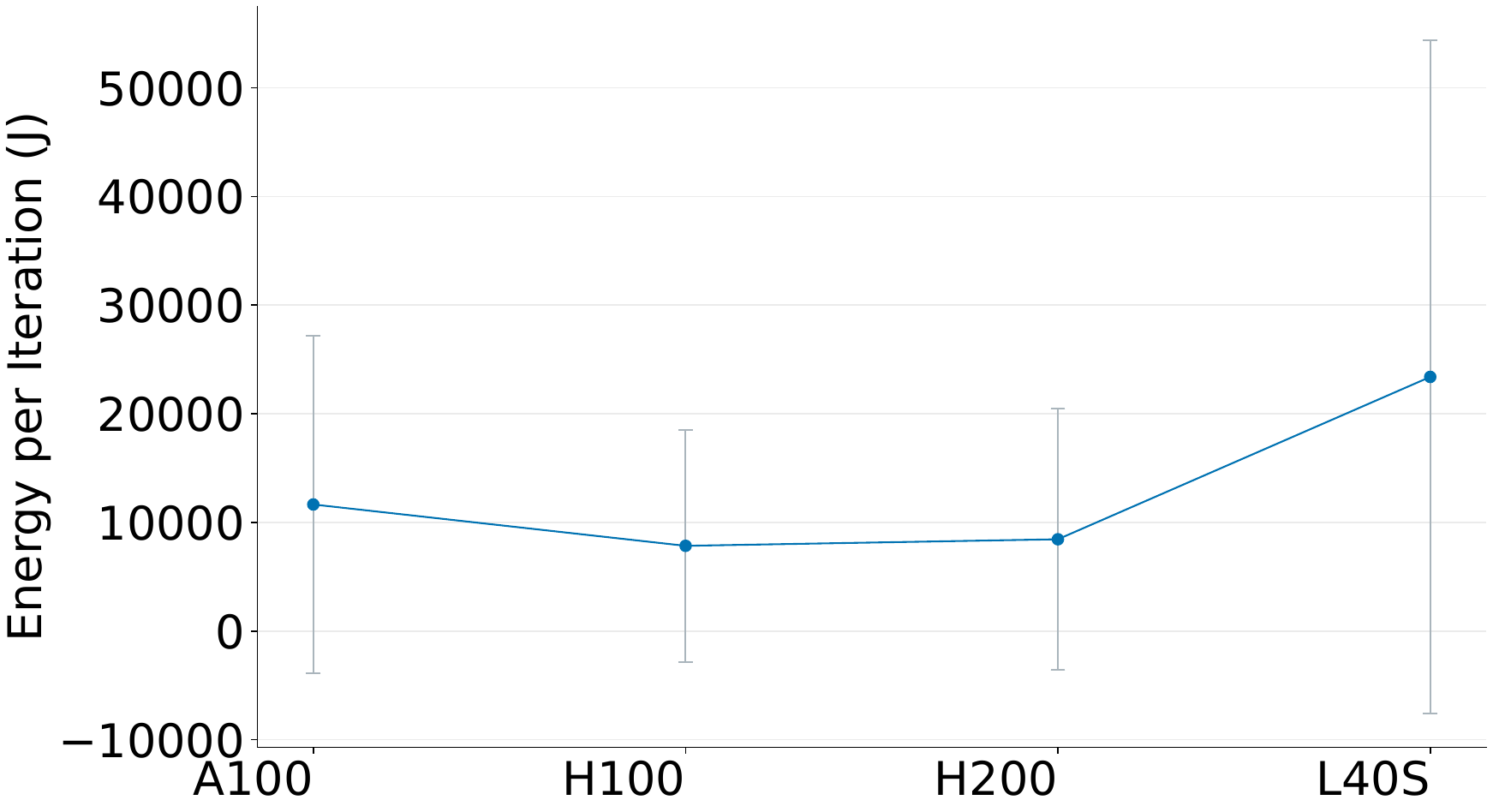}
        \caption{Energy per iteration}
        \label{fig:quant_gpu_energy_iteration}
    \end{subfigure}

    \vspace{0.5em}

    \begin{subfigure}[t]{0.48\textwidth}
        \centering
        \includegraphics[width=\linewidth]
        {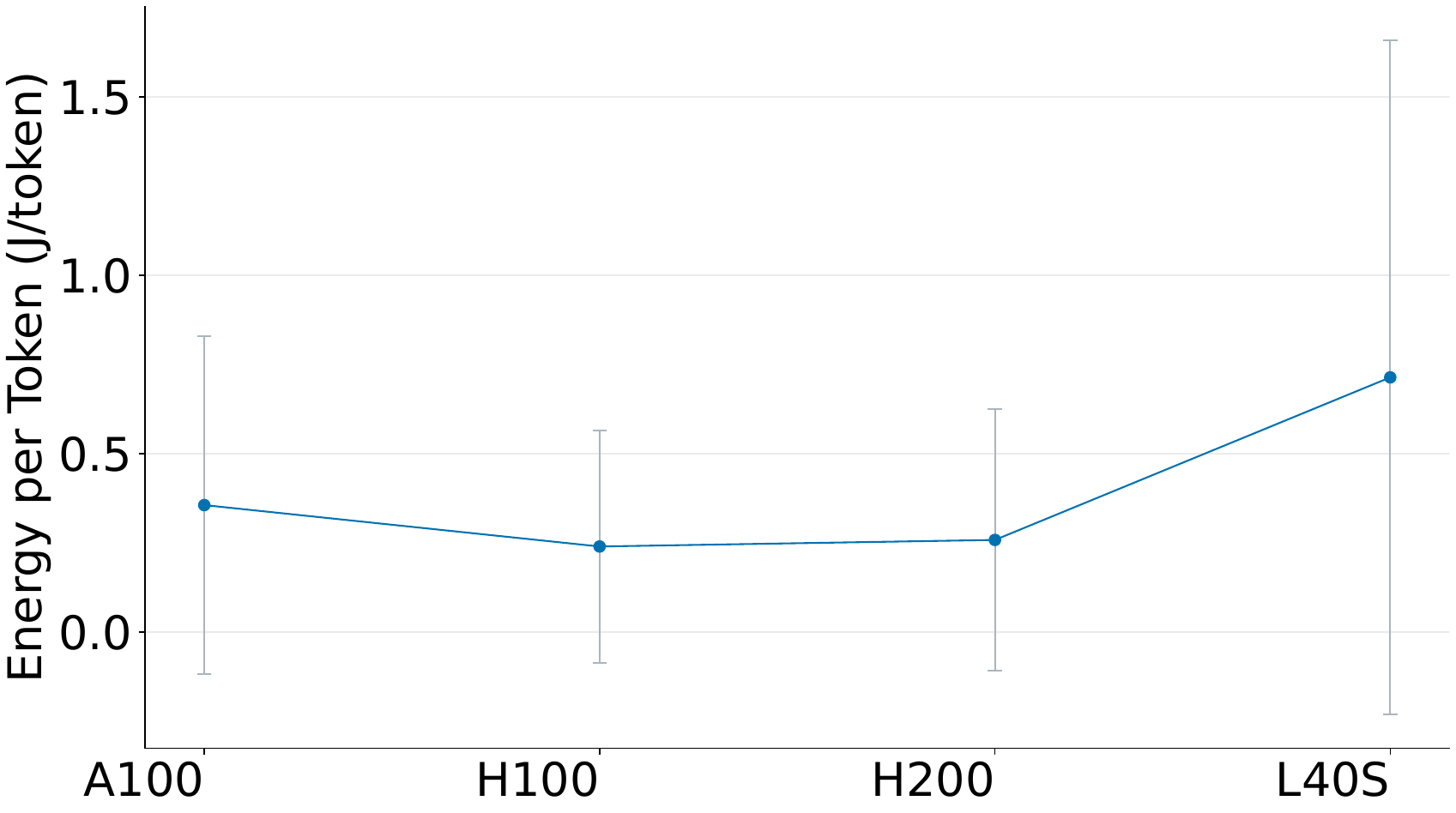}
        \caption{Energy per token}
        \label{fig:quant_gpu_energy_token}
    \end{subfigure}
    \hfill
    \begin{subfigure}[t]{0.48\textwidth}
        \centering
        \includegraphics[width=\linewidth]
        {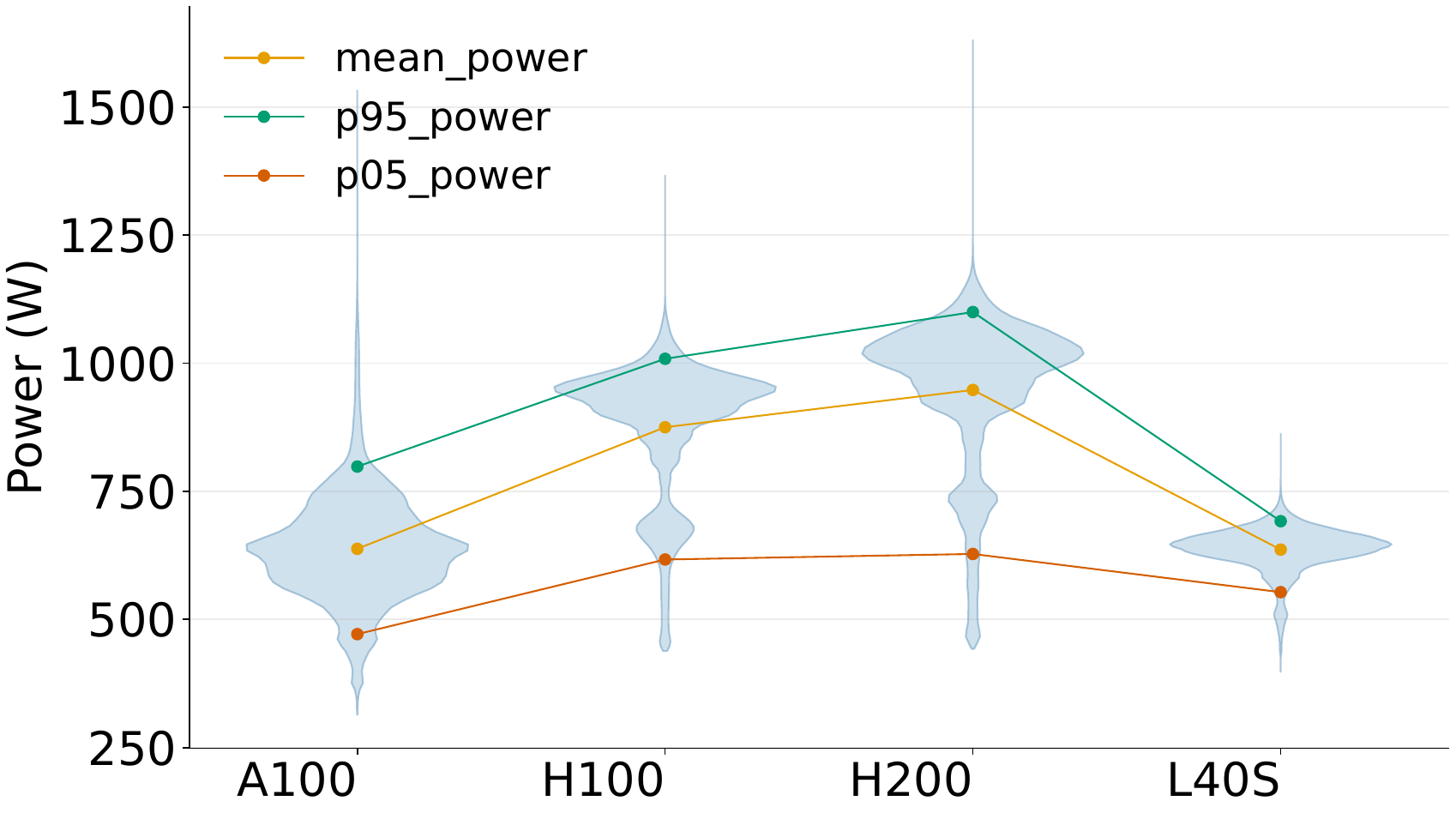}
        \caption{Power distribution}
        \label{fig:quant_gpu_power}
    \end{subfigure}

    \vspace{0.5em}

    \begin{subfigure}[t]{0.48\textwidth}
        \centering
        \includegraphics[width=\linewidth]
        {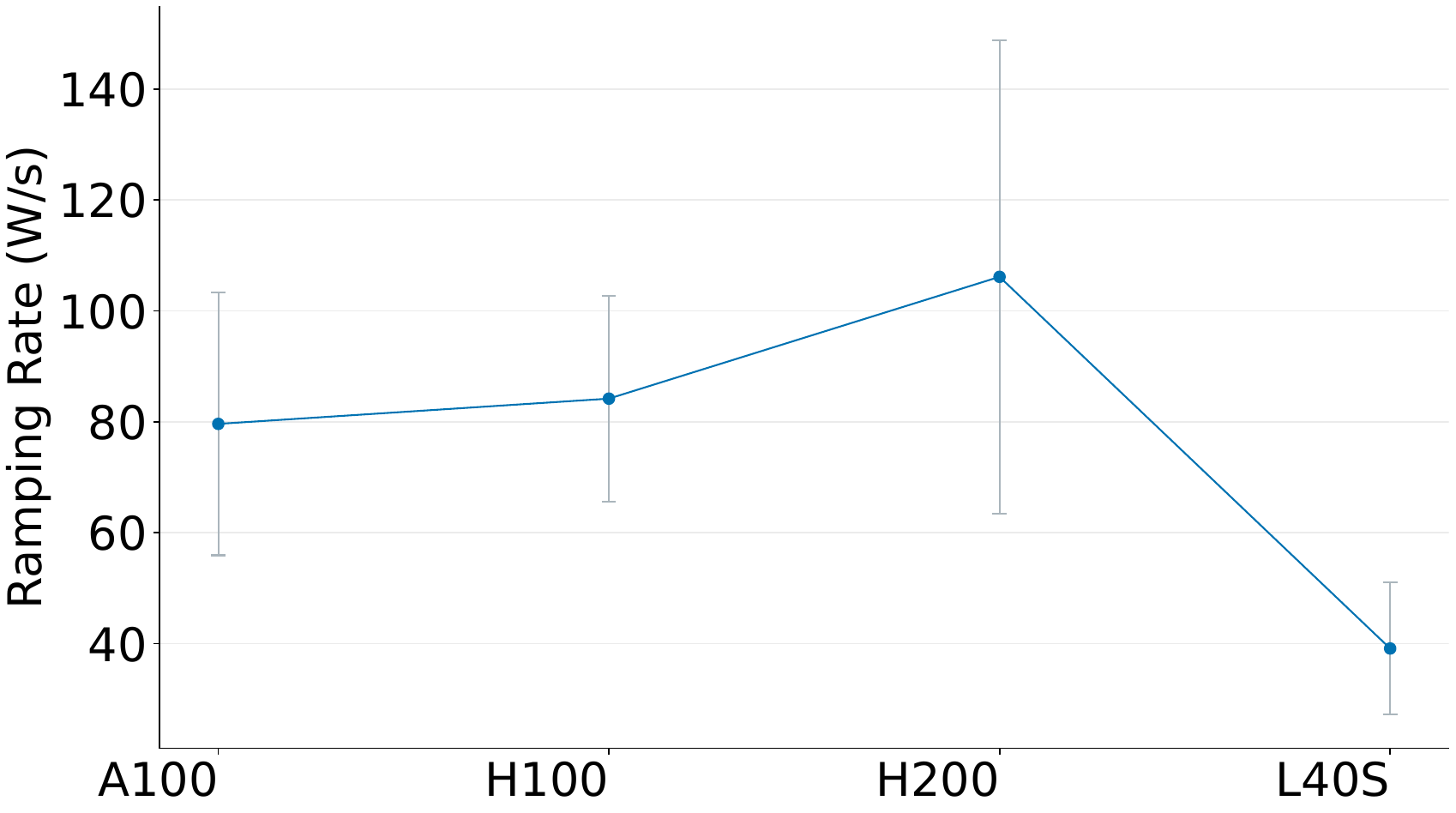}
        \caption{Ramping rate}
        \label{fig:quant_gpu_ramp}
    \end{subfigure}
    \hfill
    \begin{subfigure}[t]{0.48\textwidth}
        \centering
        \includegraphics[width=\linewidth]
        {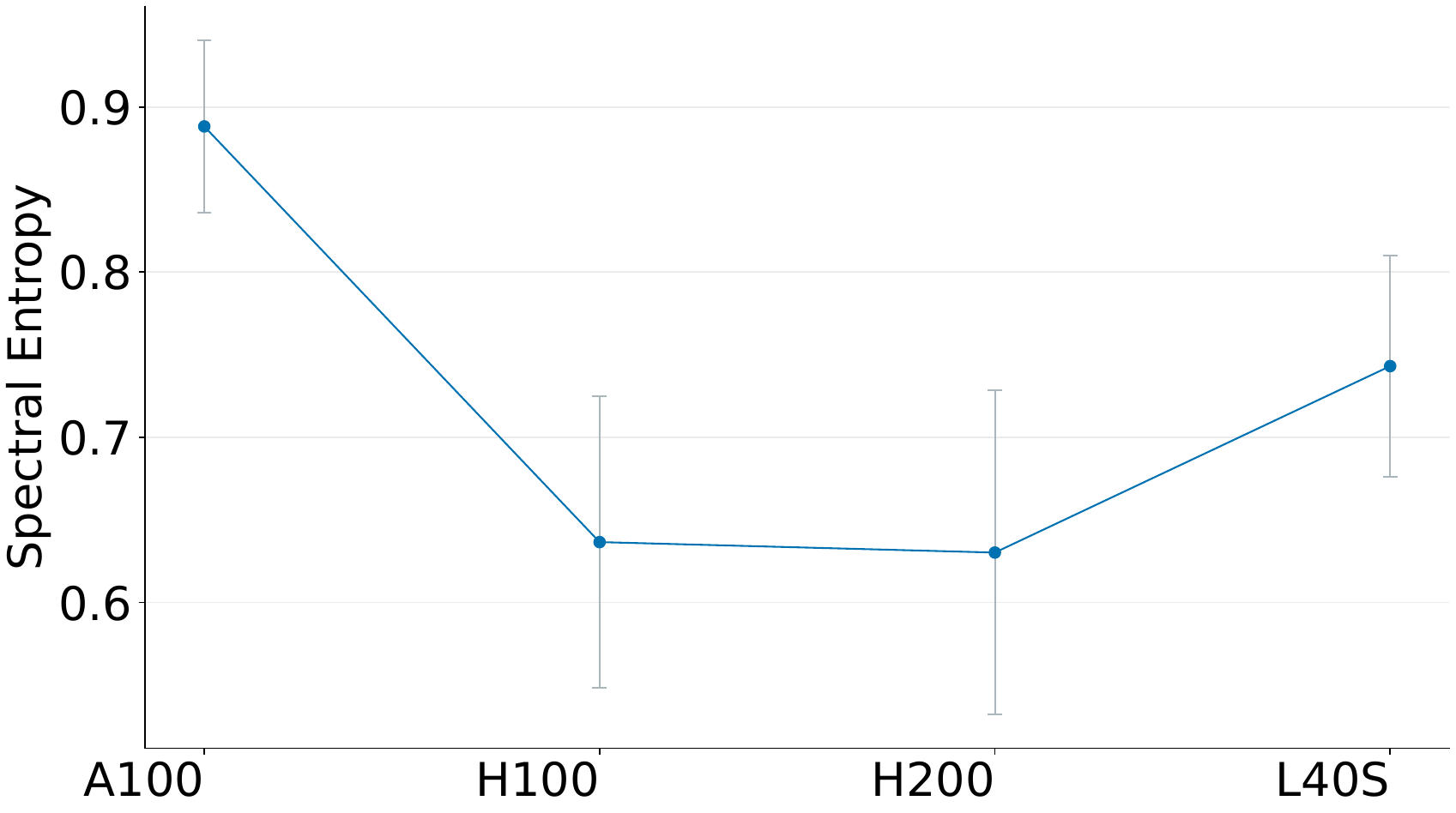}
        \caption{Spectral entropy}
        \label{fig:quant_gpu_entropy}
    \end{subfigure}

    \caption{[Fine-tuning] \textbf{GPU platform.} Complete quantitative comparison across different GPU platforms.
    The panels report iteration duration, energy per iteration, energy per token,
    power distribution, ramping rate, and spectral entropy. It compares A100, H100, H200, and L40S GPUs across the
six metrics. L40S has the longest mean iteration duration, followed by A100,
while H100 and H200 complete iterations substantially faster
(Fig.~\ref{fig:quant_gpu_duration}). Energy per iteration and per token follow
the same broad ordering, with H100 and H200 consuming less energy than A100
and L40S (Figs.~\ref{fig:quant_gpu_energy_iteration},
\ref{fig:quant_gpu_energy_token}). In contrast, H200 and H100 operate at
higher mean and upper-tail power, with H200 showing the highest values
(Fig.~\ref{fig:quant_gpu_power}). H200 also has the highest ramping rate,
whereas L40S has the lowest (Fig.~\ref{fig:quant_gpu_ramp}). Spectral entropy
is lower on H100 and H200 than on A100 and L40S
(Fig.~\ref{fig:quant_gpu_entropy}), indicating more regular power traces.}
    \label{fig:quant_gpu}
\end{figure*}


\begin{figure*}[!t]
    \centering

    \begin{subfigure}[t]{0.48\textwidth}
        \centering
        \includegraphics[width=\linewidth]
        {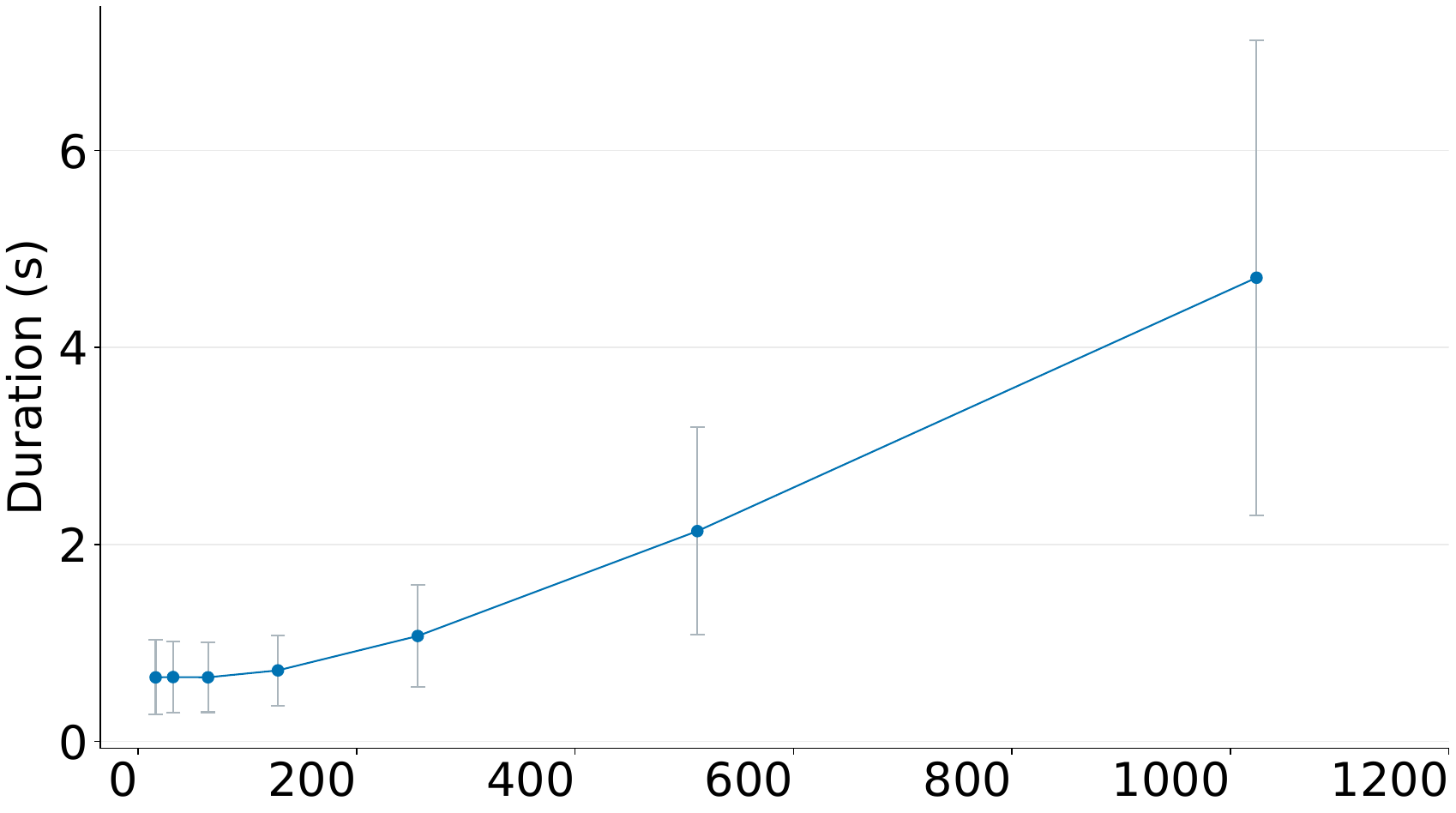}
        \caption{Iteration duration}
        \label{fig:quant_seq_length_duration}
    \end{subfigure}
    \hfill
    \begin{subfigure}[t]{0.48\textwidth}
        \centering
        \includegraphics[width=\linewidth]
        {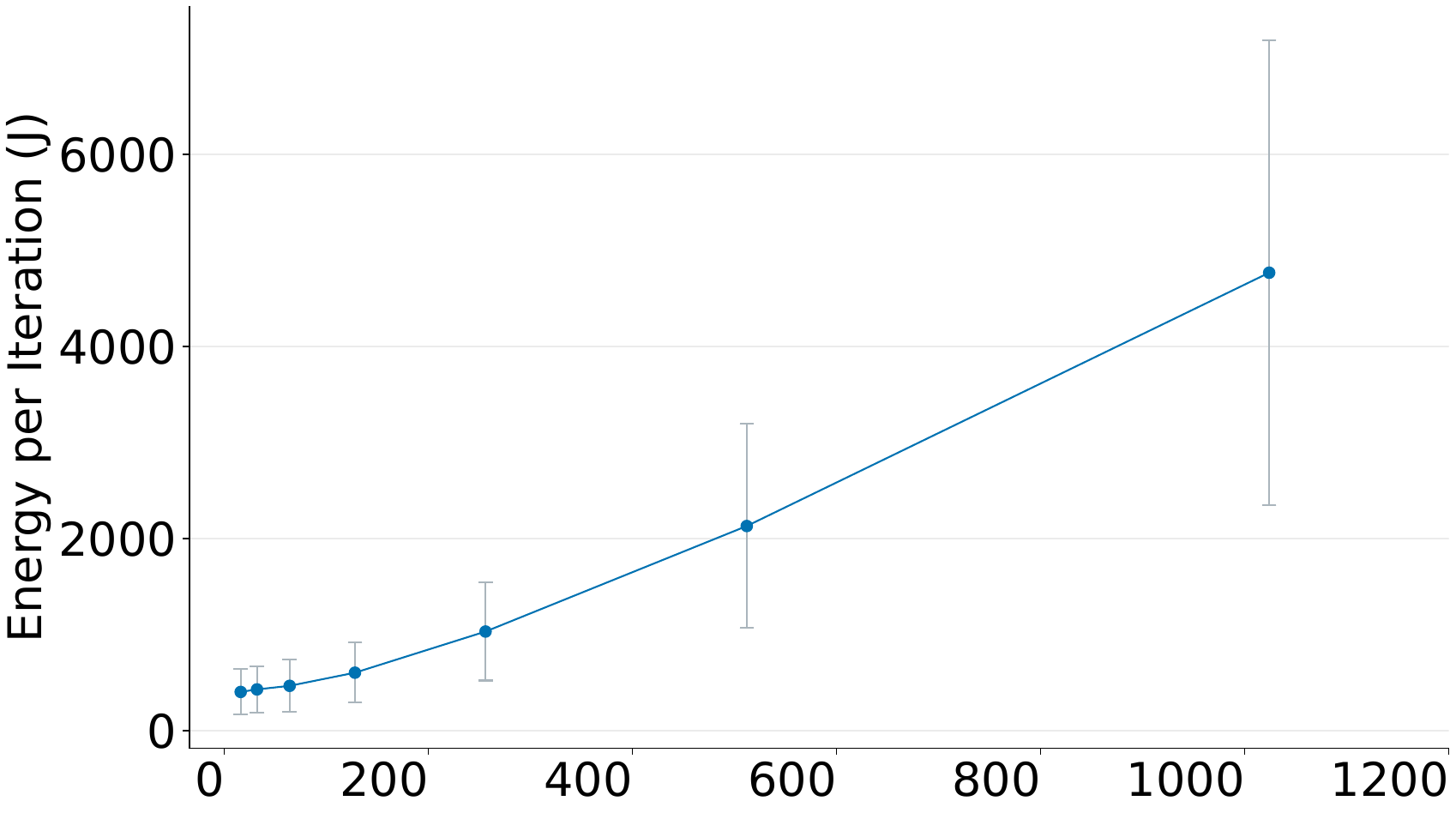}
        \caption{Energy per iteration}
        \label{fig:quant_seq_length_energy_iteration}
    \end{subfigure}

    \vspace{0.5em}

    \begin{subfigure}[t]{0.48\textwidth}
        \centering
        \includegraphics[width=\linewidth]
        {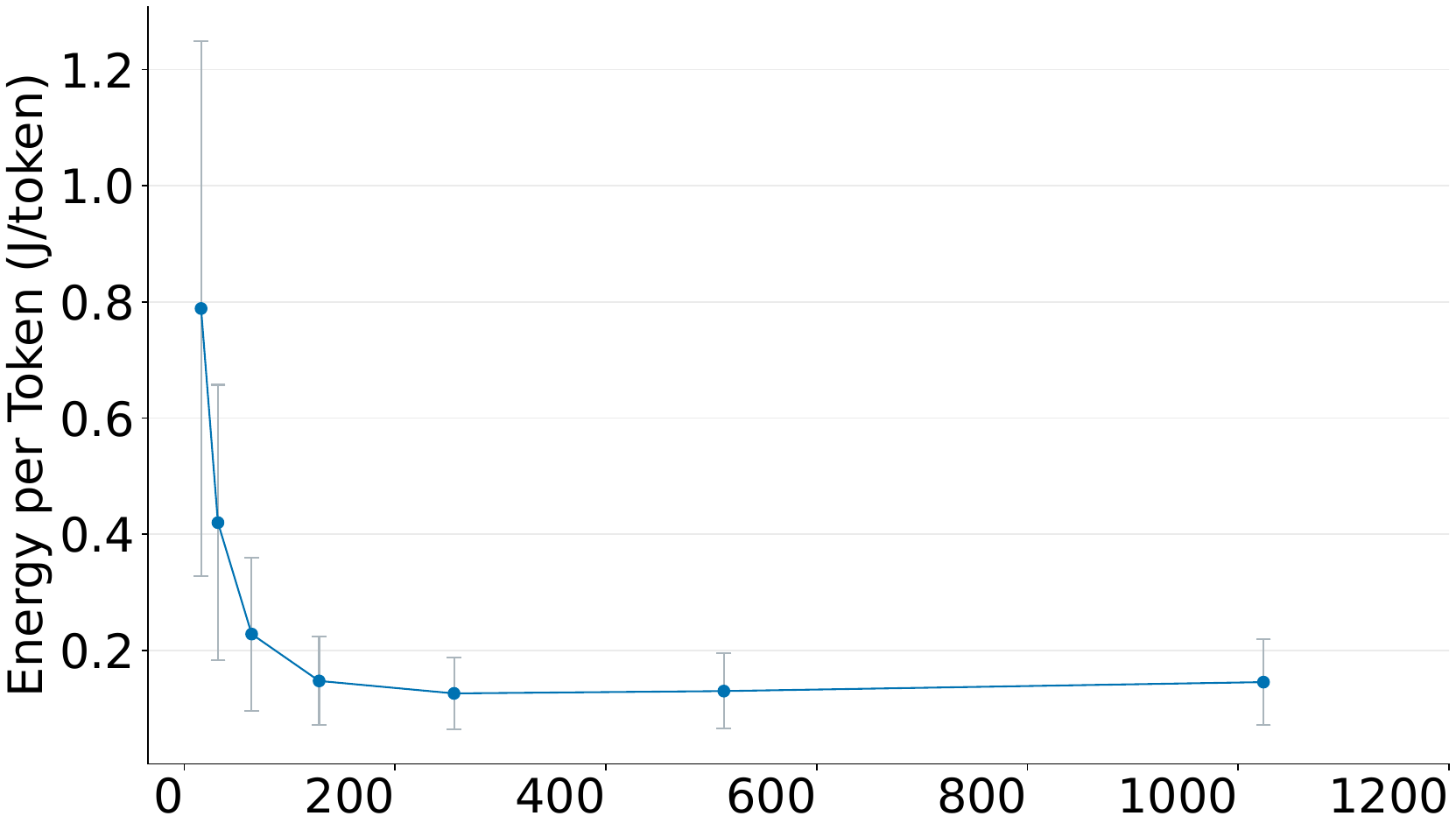}
        \caption{Energy per token}
        \label{fig:quant_seq_length_energy_token}
    \end{subfigure}
    \hfill
    \begin{subfigure}[t]{0.48\textwidth}
        \centering
        \includegraphics[width=\linewidth]
        {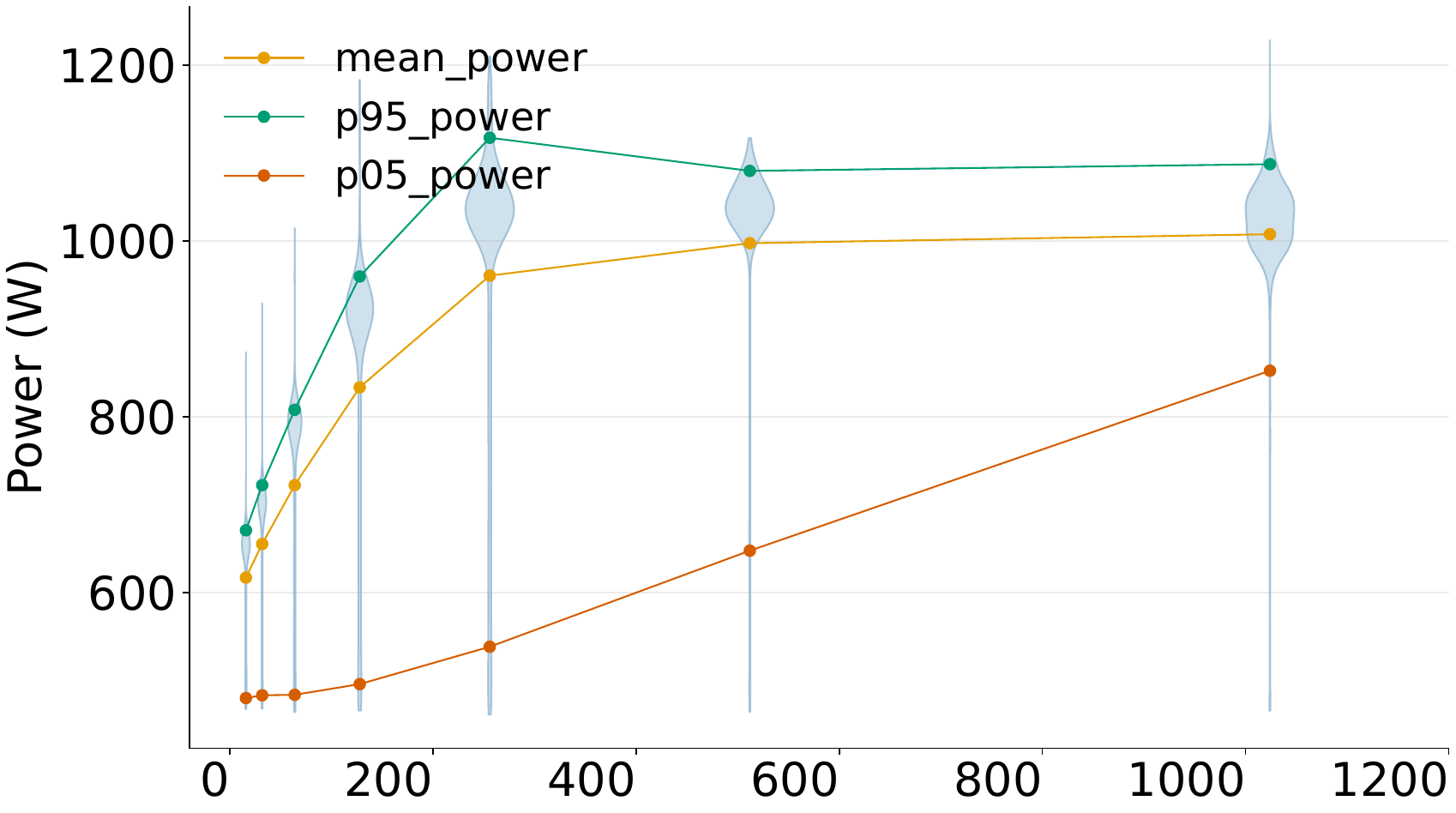}
        \caption{Power distribution}
        \label{fig:quant_seq_length_power}
    \end{subfigure}

    \vspace{0.5em}

    \begin{subfigure}[t]{0.48\textwidth}
        \centering
        \includegraphics[width=\linewidth]
        {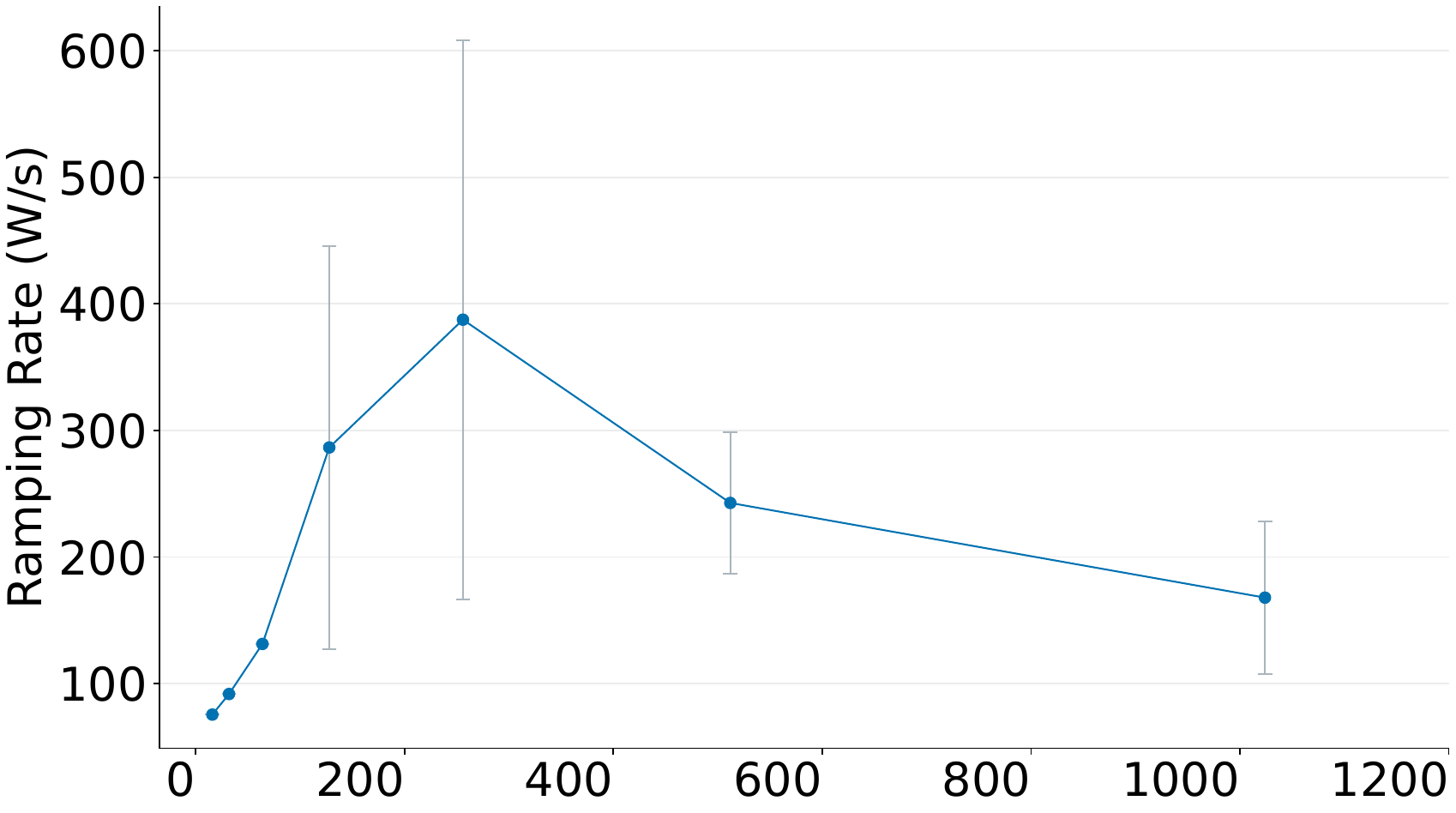}
        \caption{Ramping rate}
        \label{fig:quant_seq_length_ramp}
    \end{subfigure}
    \hfill
    \begin{subfigure}[t]{0.48\textwidth}
        \centering
        \includegraphics[width=\linewidth]
        {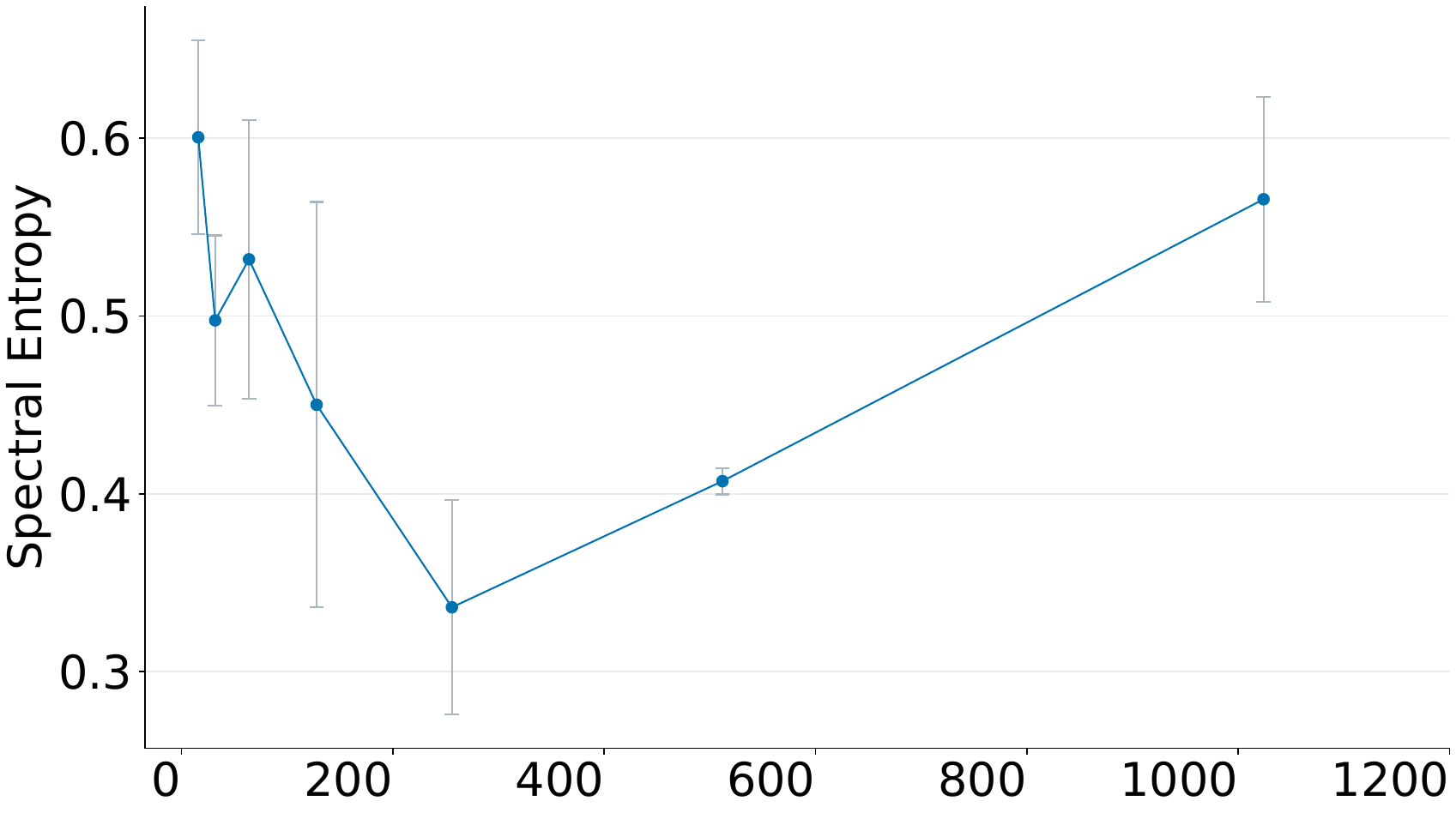}
        \caption{Spectral entropy}
        \label{fig:quant_seq_length_entropy}
    \end{subfigure}

    \caption{[Fine-tuning] \textbf{Sequence length.} Quantitative comparison across different sequence lengths.
    The panels report iteration duration, energy per iteration, energy per token,
    power distribution, ramping rate, and spectral entropy. It summarizes the effects of sequence length on the six fine-tuning metrics. Iteration duration and energy per iteration increase
sharply beyond 256 tokens (Figs.~\ref{fig:quant_seq_length_duration},
\ref{fig:quant_seq_length_energy_iteration}), while energy per token decreases
from 16 to 256 tokens before approaching a plateau
(Fig.~\ref{fig:quant_seq_length_energy_token}). Mean, 5th-percentile, and
95th-percentile power increase with sequence length, although upper-tail power
begins to saturate at the longest sequences
(Fig.~\ref{fig:quant_seq_length_power}). Ramping rate shows no clear monotonic
trend, while spectral entropy decreases toward 256 tokens before increasing
again at longer sequences (Figs.~\ref{fig:quant_seq_length_ramp},
\ref{fig:quant_seq_length_entropy}).}
    \label{fig:quant_seq_length}
\end{figure*}

\clearpage

\newpage
\section{Additional Inference Results}
\label{app:more-inf}


\subsection{Additional Inference Serving Traces}
\label{app:inf-traces}

We provide additional representative serving power traces to illustrate how the arrival pattern,
workload, execution configuration, and hardware reshape the LLM serving's power profile. Unless noted, runs are closed loop with 32 requests
in flight and at most 512 output tokens, shown over 60~s taken 120~s into the measurement window.

\begin{table}[h]
\centering
\caption{Guide to the inference figures in the appendix.}
\label{tab:inference_figure_guide}
\footnotesize
\setlength{\tabcolsep}{3.5pt}
\renewcommand{\arraystretch}{1.08}
\begin{tabularx}{\linewidth}{
@{}p{0.13\linewidth}
p{0.19\linewidth}
p{0.25\linewidth}
X@{}}
\toprule
\textbf{Figure} & \textbf{Analysis} &
\textbf{Factor / comparison} & \textbf{Content} \\
\midrule

Fig.~\ref{fig:inf-tp}
& Representative traces
& Tensor-parallel degree
& Per-GPU and node power of Qwen3-14B at TP~1, 2, and 4 on H100. \\

Fig.~\ref{fig:inf-hardware}
& Representative traces
& GPU platform
& Qwen3-14B at TP~4 on A100, H100, and H200. \\

Fig.~\ref{fig:inf-arrival}
& Representative traces
& Arrival pattern
& Steady and bursty BurstGPT replays of Qwen3-32B on two H200s. \\

Fig.~\ref{fig:inf-sweeps}
& Representative traces
& Concurrency; input/output length
& Segments across concurrency limits 1--256 (A100) and input/output pairs (H200). \\

Fig.~\ref{fig:inf-size}
& Representative traces
& Model size; dense vs.\ MoE
& Qwen3-8B, 14B, 32B, and 30B-A3B on two H200s at TP~2. \\

Fig.~\ref{fig:inf-precision}
& Representative traces
& Weight and KV-cache precision
& BF16, FP8, and INT4 weights and an FP8 KV cache for Qwen3-14B on H100. \\

Tab.~\ref{tab:inf-tp-period}
& Turnover period
& Tensor-parallel degree
& Batch-turnover period for every dense model and GPU at TP~1, 2, and 4. \\

Fig.~\ref{fig:inf_quant_model_size}
& Quantitative comparison
& Model size
& Latency, energy, power distribution, ramping rate, and spectral entropy. \\

Fig.~\ref{fig:inf_quant_architecture}
& Quantitative comparison
& Dense vs.\ MoE
& Six inference metrics across model architectures. \\

Fig.~\ref{fig:inf_quant_gpu_platform}
& Quantitative comparison
& GPU platform
& Six inference metrics across A100, H100, and H200. \\

Fig.~\ref{fig:inf_quant_concurrency}
& Quantitative comparison
& Concurrency limit
& Six inference metrics across 1--256 requests in flight. \\

Fig.~\ref{fig:inf_quant_arrival_rate}
& Quantitative comparison
& Poisson arrival rate
& Six inference metrics across 0.25--8 requests/s. \\

Fig.~\ref{fig:inf_quant_isl_osl}
& Quantitative comparison
& Input/output length
& Six inference metrics across forced input/output pairs. \\

Fig.~\ref{fig:inf_quant_weight_precision}
& Quantitative comparison
& Weight precision
& Six inference metrics for BF16, FP8, and INT4 weights. \\

Fig.~\ref{fig:inf_quant_kv_precision}
& Quantitative comparison
& KV-cache precision
& Six inference metrics for BF16 and FP8 KV caches. \\

Fig.~\ref{fig:inf_quant_tensor_parallelism}
& Quantitative comparison
& Tensor-parallel degree
& Six inference metrics at TP~1, 2, and 4. \\

Fig.~\ref{fig:inf_quant_sm_clock}
& Quantitative comparison
& SM clock
& Six inference metrics across locked clocks on H100. \\

\bottomrule
\end{tabularx}
\vspace{1mm}
\begin{minipage}{0.98\linewidth}
\footnotesize
\textit{Notes:} Representative-trace figures illustrate waveform structure under controlled serving configurations, while quantitative figures summarize how each serving factor affects request latency, energy, power, ramping, and temporal regularity.
\end{minipage}
\end{table}

\begin{figure}[h]
  \centering
  \includegraphics[width=0.85\linewidth]{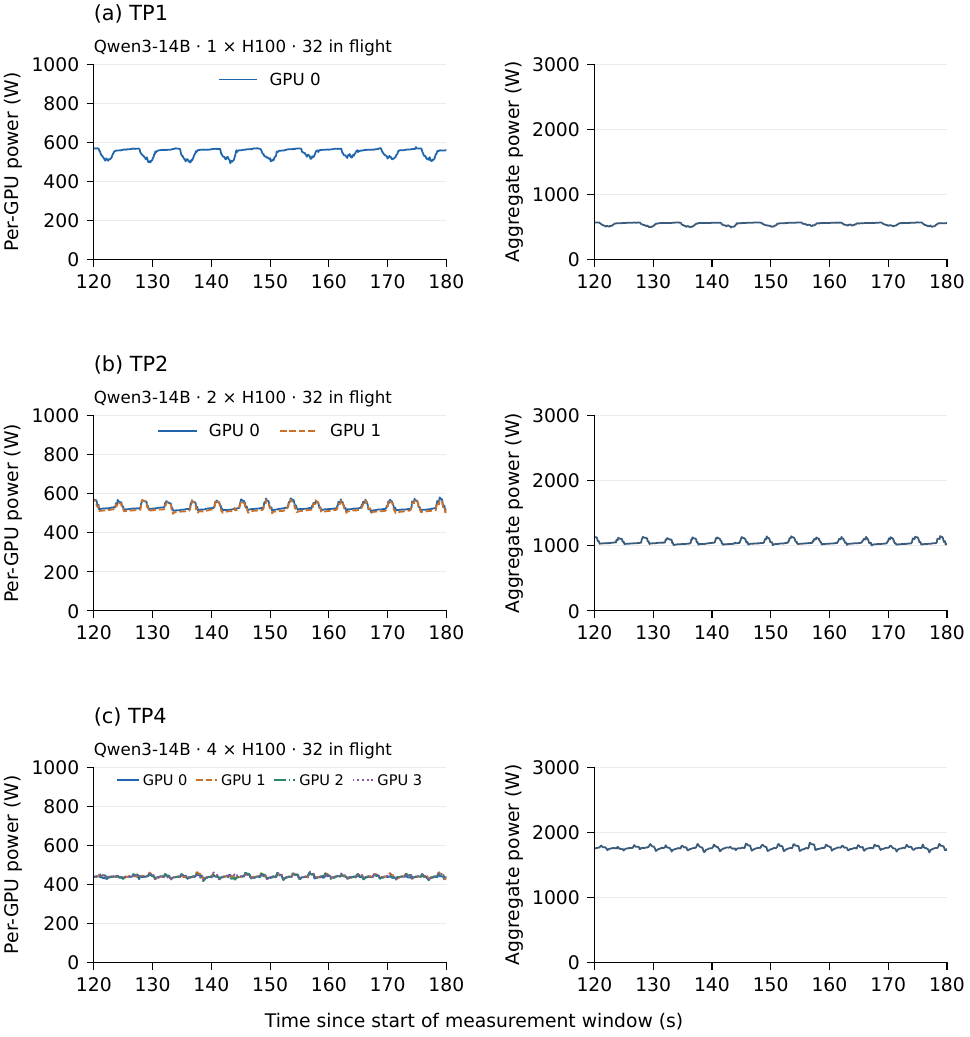}
  \caption{[Inference] Tensor-parallel degree: Qwen3-14B on (a) one, (b) two, and (c) four H100 GPUs. Each row
  shows per-GPU power (left) and their sum (right) over 60~s taken 120~s into the measurement window.
  The TP~1 run ran on a host with the GPU power limit lowered to 575~W; TP~2 and TP~4 ran at 700~W.}
  \label{fig:inf-tp}
\end{figure}

\begin{figure}[h]
  \centering
  \includegraphics[width=0.85\linewidth]{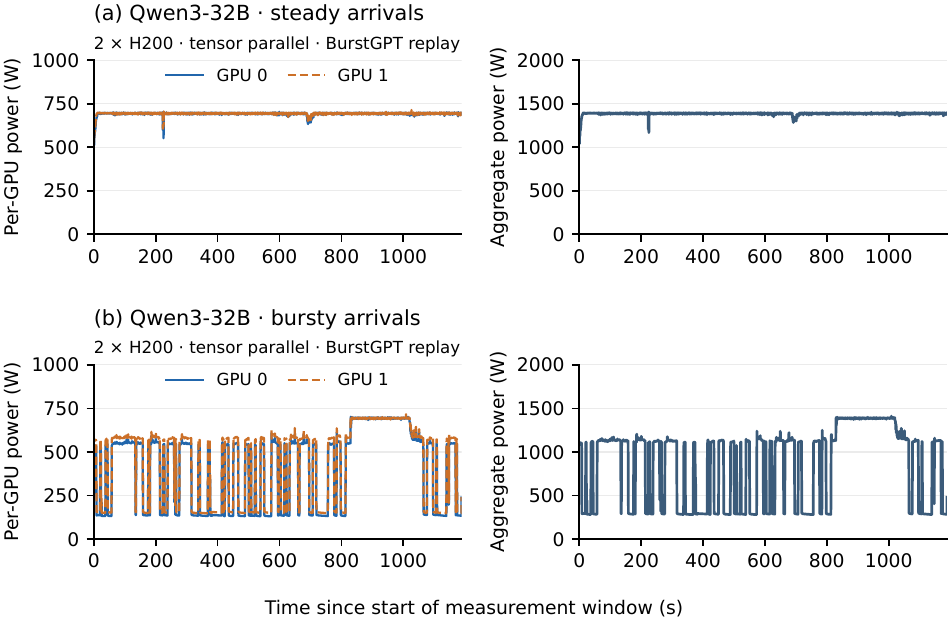}
  \caption{[Inference] Representative power traces of Qwen3-32B served on two H200 GPUs with tensor
  parallelism, replaying (a) a steady and (b) a bursty segment of the BurstGPT trace. Each row shows per-GPU power (left) and their sum (right) over the full measurement window. Model,
  hardware, and serving settings are identical; only the arrival pattern differs.}
  \label{fig:inf-arrival}
\end{figure}

\textbf{Arrival pattern.} Fig.~\ref{fig:inf-arrival} replays a steady and a bursty segment of the
BurstGPT trace on Qwen3-32B served on two H200 GPUs. Under steady arrivals, both GPUs stay near
their power limit for the whole window, with only a few brief dips. Under bursty arrivals, the node
switches repeatedly between an idle floor of about 280~W and a serving plateau of about 1.1~kW, with
one sustained stretch at the limit during the heaviest burst. Both GPUs switch together, so the
node swings by the full sum of their changes, and the 99th-percentile 1-s ramp rises from 35 to
738~W/s.

\begin{figure}[h]
  \centering
  \includegraphics[width=0.85\linewidth]{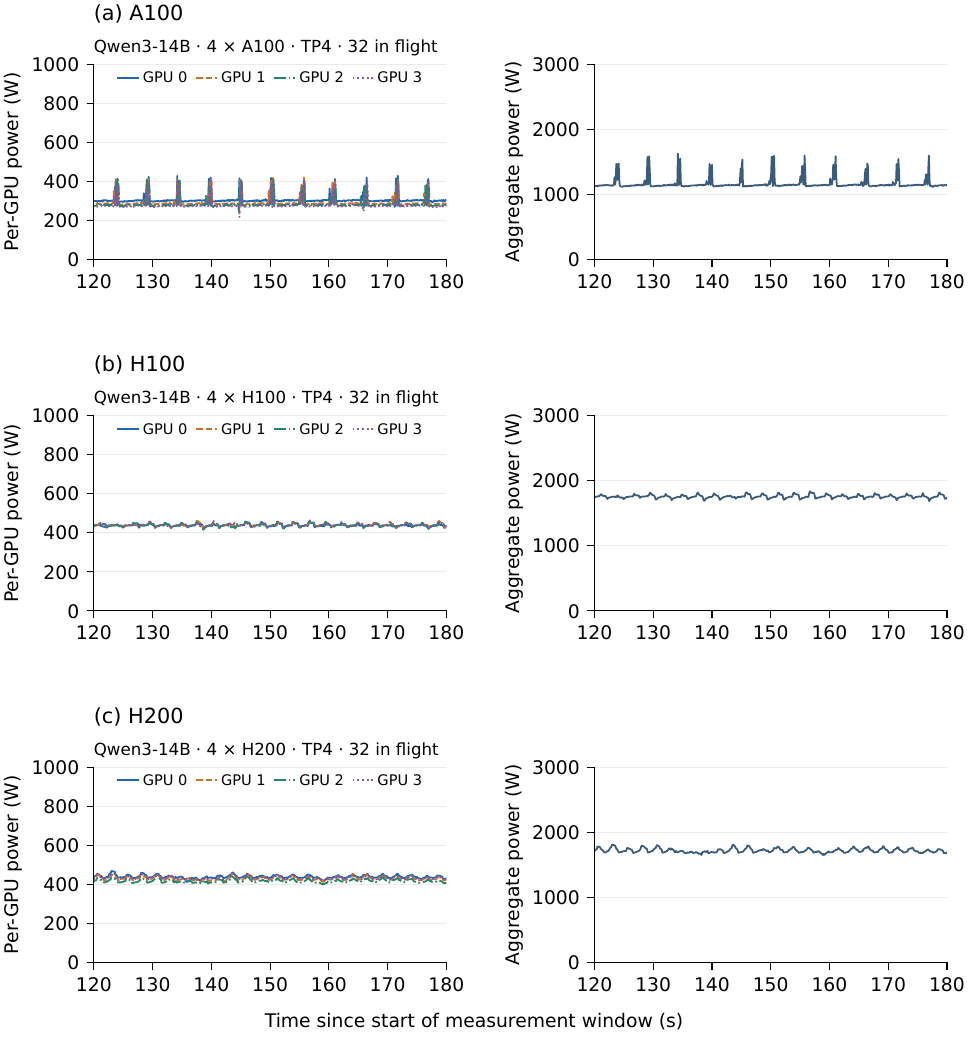}
  \caption{[Inference] GPU platform: Qwen3-14B at TP~4 on (a) A100, (b) H100, and (c) H200. Each row shows
  per-GPU power (left) and their sum (right) over 60~s taken 120~s into the measurement window.}
  \label{fig:inf-hardware}
\end{figure}

\begin{figure}[h]
  \centering
  \includegraphics[width=0.7\linewidth]{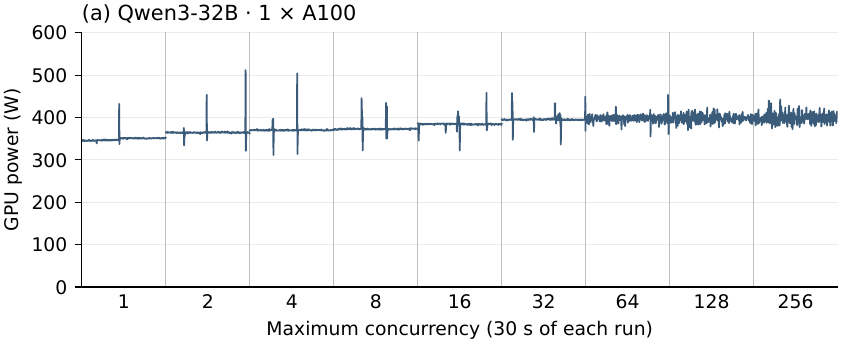}\\[4pt]
  \includegraphics[width=0.7\linewidth]{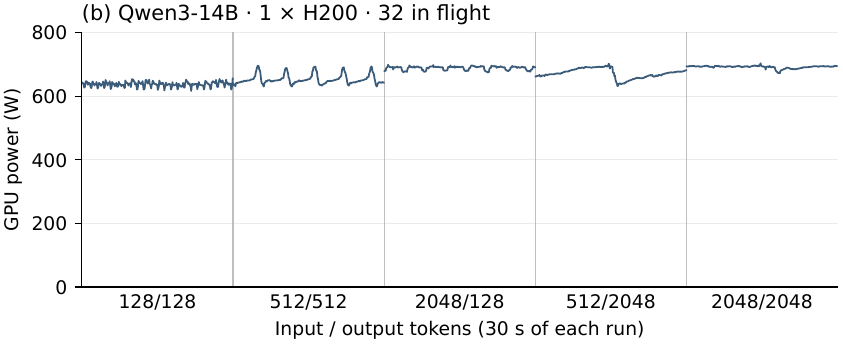}
  \caption{[Inference] Workload sweeps: (a) maximum concurrency from 1 to 256 for Qwen3-32B on one A100;
  (b) input/output length from 128/128 to 2048/2048 tokens for Qwen3-14B on one H200 with 32
  requests in flight. Each setting is a separate run; 30~s of each run are shown side by side,
  separated by grey lines, so the segments are not contiguous in time.}
  \label{fig:inf-sweeps}
\end{figure}

\textbf{Concurrency and input/output length.} Fig.~\ref{fig:inf-sweeps} places 30-s segments of
separate runs side by side. For Qwen3-32B on one A100, power rises only from about 350 to 400~W as
the concurrency limit grows from 1 to 256; at low concurrency the trace is flat with isolated
spikes at request turnover, and from 64 requests on it becomes a dense fluctuation near the power
limit. For Qwen3-14B on one H200, 2048-token prompts hold power within a few percent of the 700~W
limit with small ramps, whereas the 512/512 and 512/2048 workloads show regular turnover dips spaced
several seconds apart, which produce the largest ramps.

\begin{figure}[h]
  \centering
  \includegraphics[width=\linewidth]{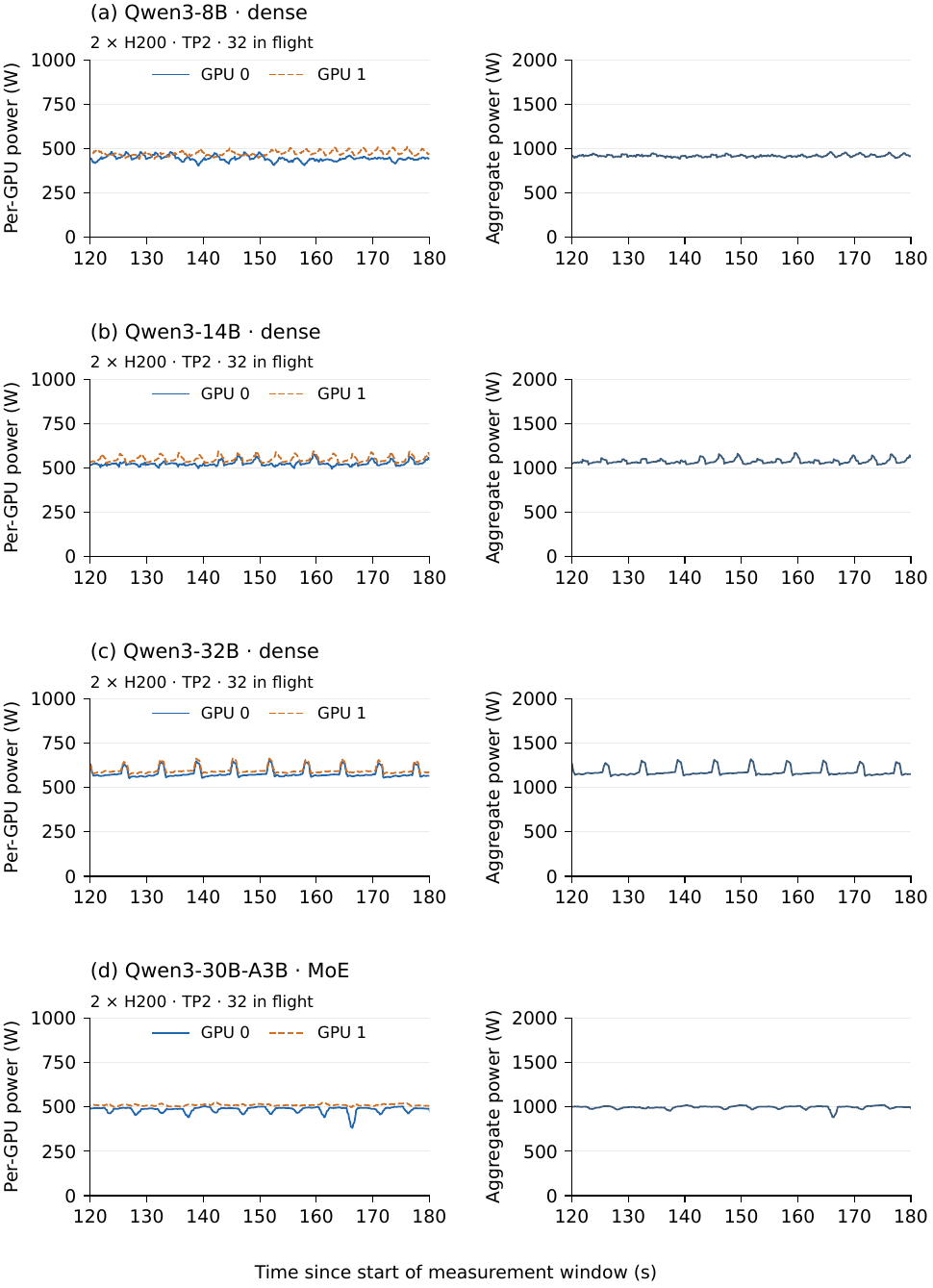}
  \caption{[Inference] Model-size and architecture examples on two H200 GPUs at TP~2: (a) Qwen3-8B,
  (b) Qwen3-14B, (c) Qwen3-32B (dense), and (d) Qwen3-30B-A3B (MoE). Each row shows per-GPU power
  (left) and their sum (right) over 60~s taken 120~s into the measurement window. The dense/MoE
  pair differs in both total and active parameters.}
  \label{fig:inf-size}
\end{figure}

\textbf{Model size and architecture.} Fig.~\ref{fig:inf-size} compares four Qwen3 models on two
H200 GPUs at TP~2. From Qwen3-8B to 14B to 32B, node power rises from 0.91 to 1.06 to 1.17~kW and
the batch-turnover period lengthens from 2.6 to 3.4 to 6.5~s, with the turnover spikes becoming more
distinct. The MoE Qwen3-30B-A3B draws less power than the dense Qwen3-32B (0.98~kW) and turns over
faster (5.1~s), but its dips appear mainly on one of its two GPUs.

\begin{figure}[h]
  \centering
  \includegraphics[width=\linewidth]{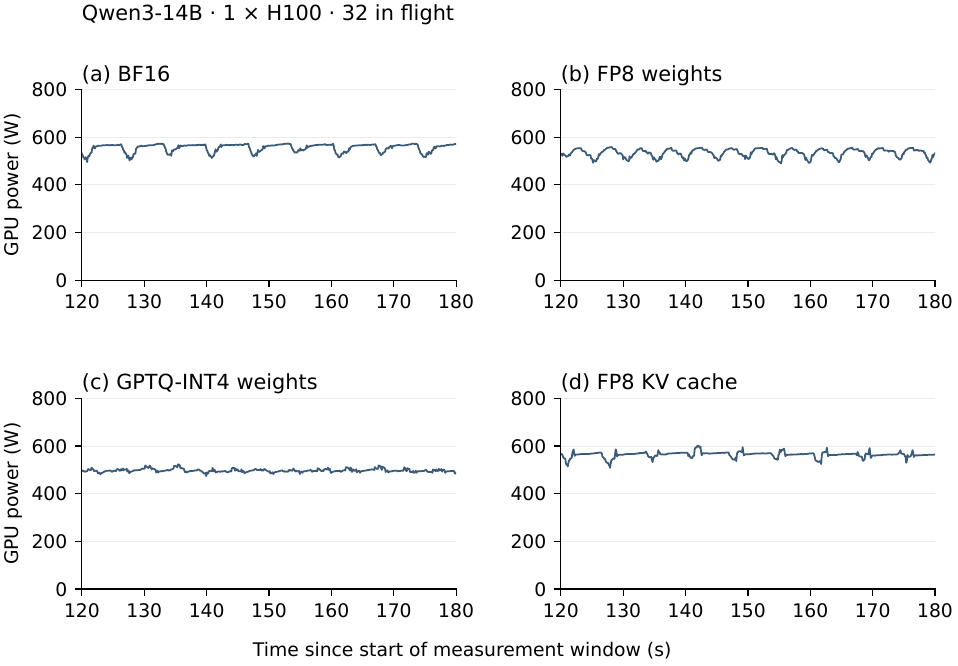}
  \caption{[Inference] Numerical-precision examples for Qwen3-14B on one H100: (a) BF16, (b) FP8 weights,
  (c) GPTQ-INT4 weights, and (d) FP8 KV cache with BF16 weights, over 60~s taken 120~s into the
  measurement window. Runs (a)--(c) ran on a host with the GPU power limit lowered to 575~W.}
  \label{fig:inf-precision}
\end{figure}

\textbf{Numerical precision.} Fig.~\ref{fig:inf-precision} serves Qwen3-14B on one H100. FP8 and
GPTQ-INT4 weights lower power (554, 532, and 498~W) and shorten the turnover period from 6.9 to 4.8
and 4.6~s, so the dips become more frequent and, for INT4, nearly vanish. An FP8 KV cache leaves the
period unchanged (6.9~s) and changes power little.

\textbf{Tensor parallelism.} Fig.~\ref{fig:inf-tp} serves Qwen3-14B on one, two, and four H100
GPUs. Per-GPU power falls from 551 to 527 and 439~W while node power rises from 0.55 to 1.05 and
1.75~kW, and the GPUs of each group rise and fall together. The batch-turnover period shortens from
6.9 to 4.2 and 2.8~s; Table~\ref{tab:inf-tp-period} gives the period for every dense model and GPU.

\textbf{GPU platform.} Fig.~\ref{fig:inf-hardware} serves Qwen3-14B at TP~4 on A100, H100, and
H200. The A100 node holds a flat floor of about 1.15~kW and spikes by about 400~W at each
turnover, every 5.3~s, as each GPU briefly reaches its 400~W limit. The H100 and H200 nodes draw
about 1.75 and 1.70~kW with only small ripples and turn over every 2.8 and 2.5~s. The turnover spikes
give the A100 node a 99th-percentile 1-s ramp of 432~W/s, against 99 and 111~W/s on H100 and H200.
We compare at TP~4 because the only two-A100 host held its application clock at 1155~MHz
(App.~\ref{app:inf}).

\begin{table}[h]
  \centering
  \small
  \caption{Batch-turnover period (s) versus tensor-parallel degree, closed-loop load with 32
  requests in flight. $^\dagger$Host with its application clock held at 1155~MHz (maximum
  1410~MHz). $^\ddagger$Host with the GPU power limit lowered to 575~W. Both are excluded
  from the scaling-law fit and the pooled comparisons.}
  \label{tab:inf-tp-period}
  \begin{tabular}{llrrr}
    \toprule
    GPU & Model & TP1 & TP2 & TP4 \\
    \midrule
    A100 & Qwen3-8B  & 7.4  & 5.3$^\dagger$ & -- \\
    A100 & Qwen3-14B & 12.9 & 8.2$^\dagger$ & 5.3 \\
    A100 & Qwen3-32B & 26.4 & 16.2$^\dagger$ & 9.7 \\
    H100 & Qwen3-14B & 6.9$^\ddagger$  & 4.2 & 2.8 \\
    H100 & Qwen3-32B & 14.3$^\ddagger$ & 8.3 & 5.0 \\
    H200 & Qwen3-8B  & 3.3  & 2.6 & -- \\
    H200 & Qwen3-14B & 5.2  & 3.4 & 2.5 \\
    H200 & Qwen3-32B & 11.1 & 6.5 & 4.2 \\
    \bottomrule
  \end{tabular}
\end{table}

\subsection{Factors Shaping Serving Power Traces}
\label{app:inf-factors}

\begin{figure*}[!t]
    \centering

    \begin{subfigure}[t]{0.48\textwidth}
        \centering
        \includegraphics[width=\linewidth]
        {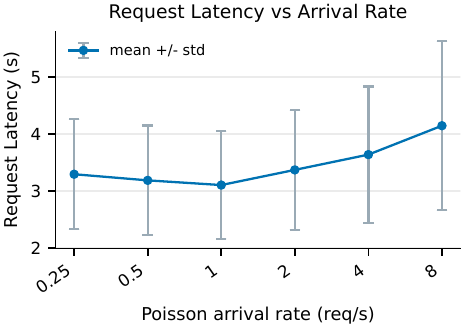}
        \caption{Request latency}
        \label{fig:inf_quant_arrival_rate_latency}
    \end{subfigure}
    \hfill
    \begin{subfigure}[t]{0.48\textwidth}
        \centering
        \includegraphics[width=\linewidth]
        {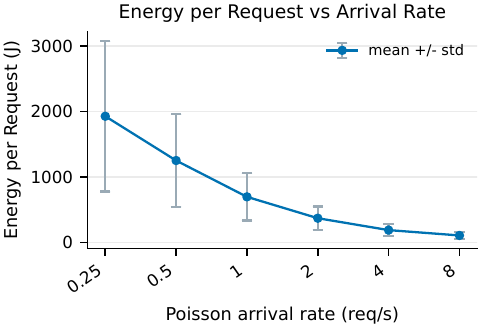}
        \caption{Energy per request}
        \label{fig:inf_quant_arrival_rate_energy_per_request}
    \end{subfigure}

    \vspace{0.5em}

    \begin{subfigure}[t]{0.48\textwidth}
        \centering
        \includegraphics[width=\linewidth]
        {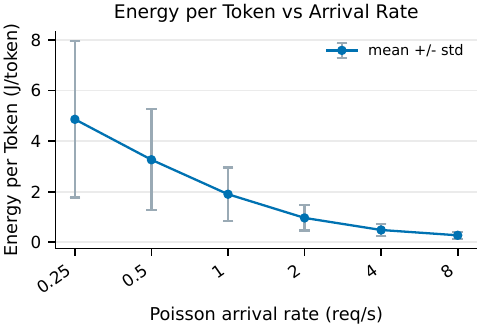}
        \caption{Energy per token}
        \label{fig:inf_quant_arrival_rate_energy_per_token}
    \end{subfigure}
    \hfill
    \begin{subfigure}[t]{0.48\textwidth}
        \centering
        \includegraphics[width=\linewidth]
        {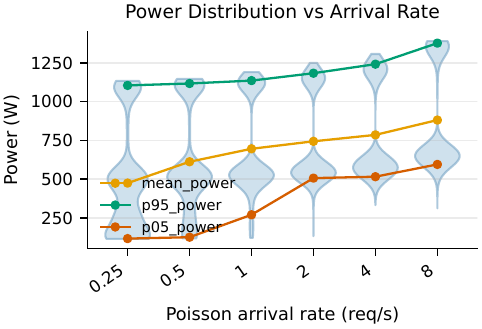}
        \caption{Power distribution}
        \label{fig:inf_quant_arrival_rate_power_distribution}
    \end{subfigure}

    \vspace{0.5em}

    \begin{subfigure}[t]{0.48\textwidth}
        \centering
        \includegraphics[width=\linewidth]
        {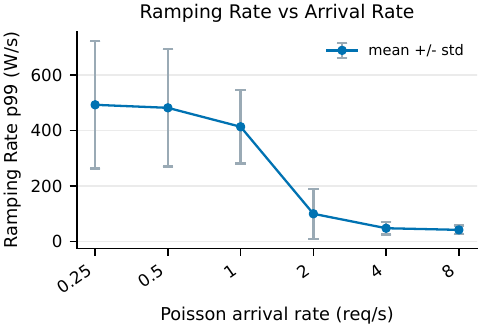}
        \caption{Ramping rate}
        \label{fig:inf_quant_arrival_rate_ramping_rate}
    \end{subfigure}
    \hfill
    \begin{subfigure}[t]{0.48\textwidth}
        \centering
        \includegraphics[width=\linewidth]
        {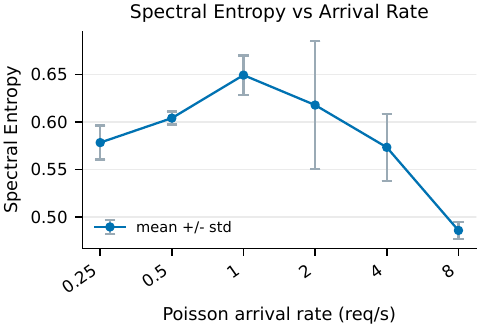}
        \caption{Spectral entropy}
        \label{fig:inf_quant_arrival_rate_spectral_entropy}
    \end{subfigure}

    \caption{[Inference] Complete quantitative comparison across different Poisson arrival rates. The panels report request latency, energy per request, energy per token, power distribution, ramping rate, and spectral entropy. It summarizes the effects of the arrival rate on the six inference metrics (Qwen3-8B, 14B, and 32B on H200, free-length generation; Qwen3-32B at TP~2). Request latency rises only slightly with rate, from 3.1 to 4.1~s, while energy per request falls from 1.9~kJ to 109~J (Figs.~\ref{fig:inf_quant_arrival_rate_latency}, \ref{fig:inf_quant_arrival_rate_energy_per_request}). Energy per token falls from 4.9 to 0.28~J as idle time is amortized over more requests (Fig.~\ref{fig:inf_quant_arrival_rate_energy_per_token}). The 5th-percentile power stays near idle up to 0.5 requests/s (117--128~W on one GPU and 261--270~W for the two-GPU Qwen3-32B node) and rises to 577--638~W on one GPU (1.28~kW for Qwen3-32B) at 8 requests/s, as the GPUs stop draining between requests (Fig.~\ref{fig:inf_quant_arrival_rate_power_distribution}). The ramping rate stays at 410--490~W/s up to 1 request/s and drops to 42--48~W/s from 4 requests/s, while spectral entropy peaks at 1 request/s (Figs.~\ref{fig:inf_quant_arrival_rate_ramping_rate}, \ref{fig:inf_quant_arrival_rate_spectral_entropy}).}
    \label{fig:inf_quant_arrival_rate}
\end{figure*}

\textbf{Arrival process.} How requests arrive changes the trace more than any other knob (Fig.~\ref{fig:inf-arrival}). A steady replay of the BurstGPT trace holds the GPU near its power limit, whereas a bursty segment on the same model and GPU switches between an idle floor and a serving plateau, widening the 5th-95th-percentile band from about 20 to 1100~W and raising the 99th-percentile 1-s ramp about twentyfold. Poisson arrivals show the same split (Fig.~\ref{fig:inf_quant_arrival_rate}): up to 0.5 requests per second the GPU drains to near its idle power between requests, and up to one per second the 99th-percentile ramp exceeds 330~W/s; once arrivals keep the GPU busy (from two per second for Qwen3-14B and 32B, from four for Qwen3-8B), ramps fall about tenfold, to 29--73~W/s.

\begin{figure*}[!t]
    \centering

    \begin{subfigure}[t]{0.48\textwidth}
        \centering
        \includegraphics[width=\linewidth]
        {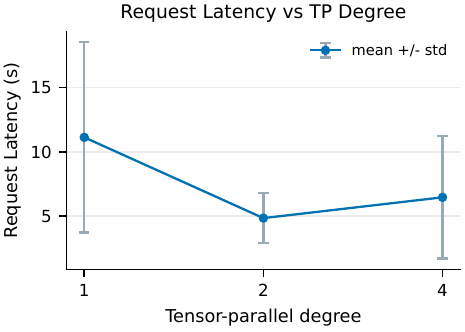}
        \caption{Request latency}
        \label{fig:inf_quant_tensor_parallelism_latency}
    \end{subfigure}
    \hfill
    \begin{subfigure}[t]{0.48\textwidth}
        \centering
        \includegraphics[width=\linewidth]
        {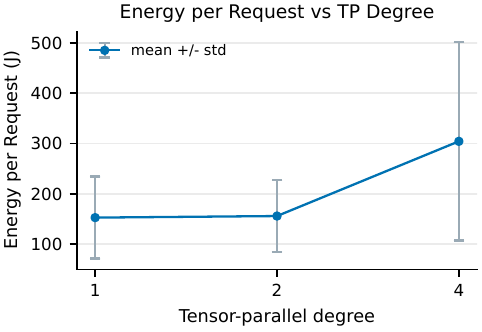}
        \caption{Energy per request}
        \label{fig:inf_quant_tensor_parallelism_energy_per_request}
    \end{subfigure}

    \vspace{0.5em}

    \begin{subfigure}[t]{0.48\textwidth}
        \centering
        \includegraphics[width=\linewidth]
        {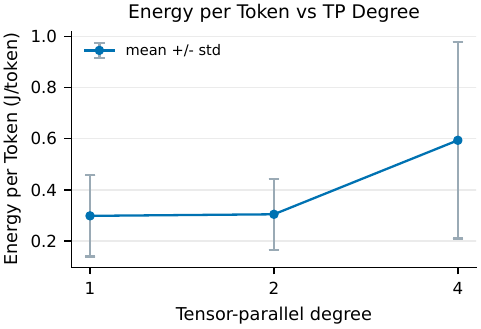}
        \caption{Energy per token}
        \label{fig:inf_quant_tensor_parallelism_energy_per_token}
    \end{subfigure}
    \hfill
    \begin{subfigure}[t]{0.48\textwidth}
        \centering
        \includegraphics[width=\linewidth]
        {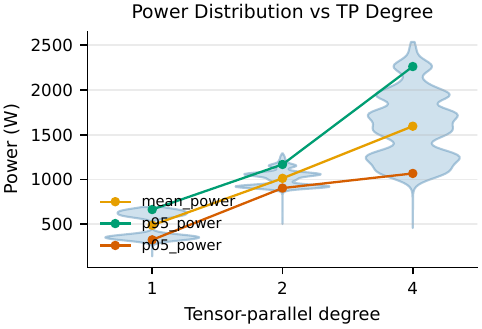}
        \caption{Power distribution}
        \label{fig:inf_quant_tensor_parallelism_power_distribution}
    \end{subfigure}

    \vspace{0.5em}

    \begin{subfigure}[t]{0.48\textwidth}
        \centering
        \includegraphics[width=\linewidth]
        {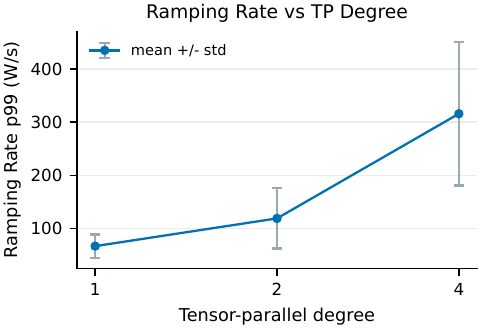}
        \caption{Ramping rate}
        \label{fig:inf_quant_tensor_parallelism_ramping_rate}
    \end{subfigure}
    \hfill
    \begin{subfigure}[t]{0.48\textwidth}
        \centering
        \includegraphics[width=\linewidth]
        {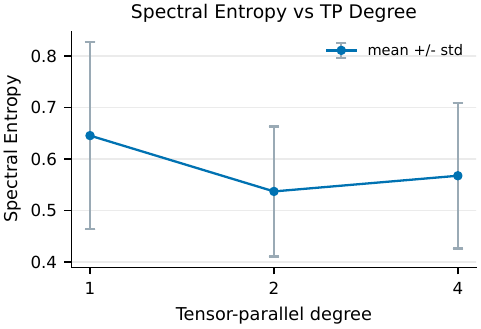}
        \caption{Spectral entropy}
        \label{fig:inf_quant_tensor_parallelism_spectral_entropy}
    \end{subfigure}

    \caption{[Inference] Complete quantitative comparison across different tensor-parallel degrees. The panels report request latency, energy per request, energy per token, power distribution, ramping rate, and spectral entropy. It summarizes the effects of the tensor-parallel degree on the six inference metrics (five models on A100, H100, and H200; node power and energy summed over the GPUs of the group; Llama-3.1-70B runs only at TP~4; with capped hosts excluded, TP~1 pools A100 and H200 runs, TP~2 pools H100 and H200 runs, and TP~4 pools all three GPUs, so the pooled trend mixes the TP degree with the platform). Request latency falls from 11.1~s at TP~1 to 4.8~s at TP~2 and is 6.4~s at TP~4, whose group adds the A100 and Llama-3.1-70B, while energy per request and per token stay flat from TP~1 to 2 and double at TP~4, partly because the TP~4 group includes Llama-3.1-70B (Figs.~\ref{fig:inf_quant_tensor_parallelism_latency}, \ref{fig:inf_quant_tensor_parallelism_energy_per_request}, \ref{fig:inf_quant_tensor_parallelism_energy_per_token}). Node power rises from 489 to 1{,}013 and 1{,}594~W (Fig.~\ref{fig:inf_quant_tensor_parallelism_power_distribution}). The ramping rate grows with the degree, from 67 to 119 and 316~W/s, as more GPUs ramp together, while spectral entropy falls from 0.65 to 0.54--0.57 (Figs.~\ref{fig:inf_quant_tensor_parallelism_ramping_rate}, \ref{fig:inf_quant_tensor_parallelism_spectral_entropy}).}
    \label{fig:inf_quant_tensor_parallelism}
\end{figure*}

\textbf{Parallelism.} Tensor parallelism splits every layer, so all GPUs run each decode step together and rise and fall at the same time (Fig.~\ref{fig:inf-tp}): the 1-s synchronization factor (Sec.~\ref{sec:metrics}) is $\Gamma_{1\mathrm{s}}=0.76$--$1.01$ (median 0.93 over 19 closed-loop TP runs), so the node's 99th-percentile ramp is close to the sum of the per-GPU ones. More GPUs lower per-GPU power but raise node power, and they shorten the batch-turnover period, each doubling of the TP degree cutting it to 0.58--0.76 of its previous value for the dense models (Table~\ref{tab:inf-tp-period}, Fig.~\ref{fig:inf_quant_tensor_parallelism}).

\begin{figure*}[!t]
    \centering

    \begin{subfigure}[t]{0.48\textwidth}
        \centering
        \includegraphics[width=\linewidth]
        {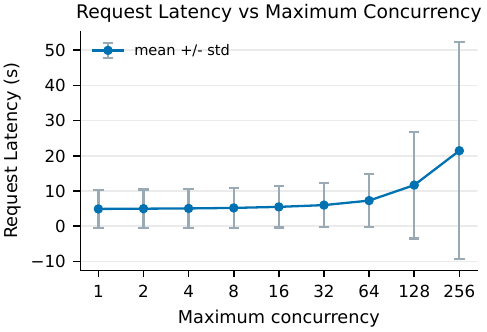}
        \caption{Request latency}
        \label{fig:inf_quant_concurrency_latency}
    \end{subfigure}
    \hfill
    \begin{subfigure}[t]{0.48\textwidth}
        \centering
        \includegraphics[width=\linewidth]
        {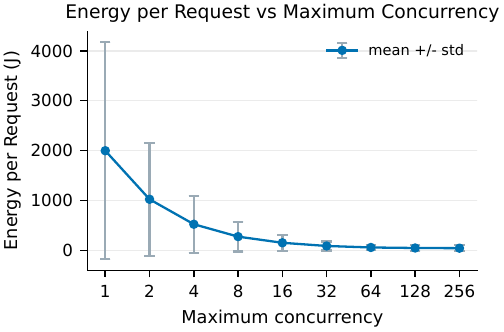}
        \caption{Energy per request}
        \label{fig:inf_quant_concurrency_energy_per_request}
    \end{subfigure}

    \vspace{0.5em}

    \begin{subfigure}[t]{0.48\textwidth}
        \centering
        \includegraphics[width=\linewidth]
        {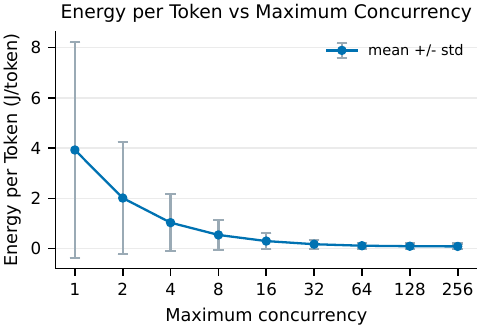}
        \caption{Energy per token}
        \label{fig:inf_quant_concurrency_energy_per_token}
    \end{subfigure}
    \hfill
    \begin{subfigure}[t]{0.48\textwidth}
        \centering
        \includegraphics[width=\linewidth]
        {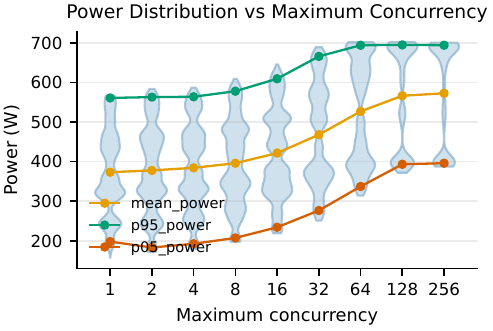}
        \caption{Power distribution}
        \label{fig:inf_quant_concurrency_power_distribution}
    \end{subfigure}

    \vspace{0.5em}

    \begin{subfigure}[t]{0.48\textwidth}
        \centering
        \includegraphics[width=\linewidth]
        {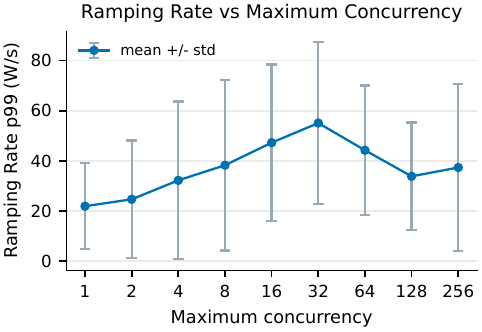}
        \caption{Ramping rate}
        \label{fig:inf_quant_concurrency_ramping_rate}
    \end{subfigure}
    \hfill
    \begin{subfigure}[t]{0.48\textwidth}
        \centering
        \includegraphics[width=\linewidth]
        {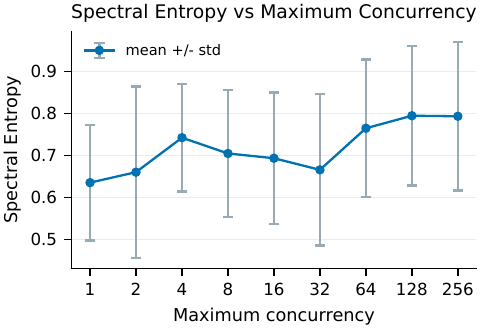}
        \caption{Spectral entropy}
        \label{fig:inf_quant_concurrency_spectral_entropy}
    \end{subfigure}

    \caption{[Inference] Complete quantitative comparison across different maximum concurrency levels. The panels report request latency, energy per request, energy per token, power distribution, ramping rate, and spectral entropy. It summarizes the effects of the concurrency limit on the six inference metrics (six dense Qwen3 models from 0.6B to 32B on A100, H100, and H200). Request latency stays near 5--6~s up to 32 concurrent requests and then rises to 21~s at 256, while energy per request falls from 2.0~kJ to 45~J (Figs.~\ref{fig:inf_quant_concurrency_latency}, \ref{fig:inf_quant_concurrency_energy_per_request}). Energy per token falls about 45-fold, from 3.9 to 0.088~J, as more requests share each decode step (Fig.~\ref{fig:inf_quant_concurrency_energy_per_token}). Mean power rises from 373 to 573~W, and the pooled band spans the three GPUs: its 95th percentile (near 694~W from 64 requests on) reflects the H100 and H200 runs and its 5th percentile (198 to 396~W) the A100 runs, which reach their 400~W limit at high concurrency (Fig.~\ref{fig:inf_quant_concurrency_power_distribution}). The ramping rate peaks at 32 requests (55~W/s) and falls once the GPUs saturate, while spectral entropy rises at high concurrency (Figs.~\ref{fig:inf_quant_concurrency_ramping_rate}, \ref{fig:inf_quant_concurrency_spectral_entropy}).}
    \label{fig:inf_quant_concurrency}
\end{figure*}

\textbf{Concurrency.} Raising the concurrency limit from 1 to 256 cuts energy per output token by 17--72$\times$, while mean power rises until it reaches the device limit (about 400~W on A100 and up to 690~W on H100 and H200) (Figs.~\ref{fig:inf-sweeps}, \ref{fig:inf_quant_concurrency}). In 12 of the 18 GPU--model series, ramps peak at intermediate concurrency (8--64 requests) and fall once the device is saturated: saturation compresses the power envelope and turns additional work into longer execution rather than higher power.

\begin{figure*}[!t]
    \centering

    \begin{subfigure}[t]{0.48\textwidth}
        \centering
        \includegraphics[width=\linewidth]{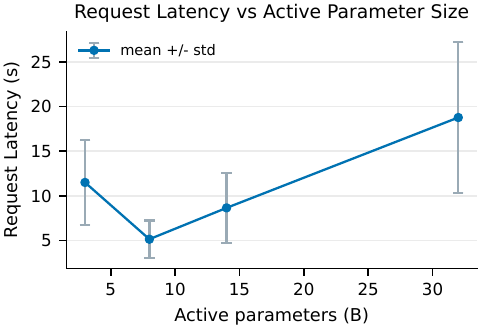}
        \caption{Request latency}
        \label{fig:inf_quant_model_size_latency}
    \end{subfigure}
    \hfill
    \begin{subfigure}[t]{0.48\textwidth}
        \centering
        \includegraphics[width=\linewidth]{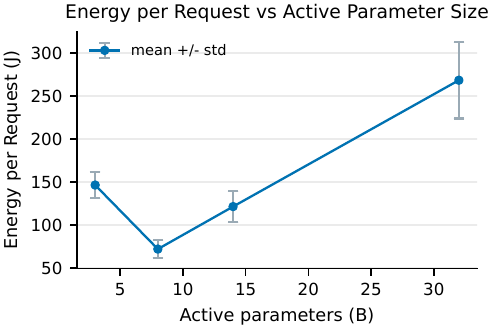}
        \caption{Energy per request}
        \label{fig:inf_quant_model_size_energy_per_request}
    \end{subfigure}

    \vspace{0.5em}

    \begin{subfigure}[t]{0.48\textwidth}
        \centering
        \includegraphics[width=\linewidth]{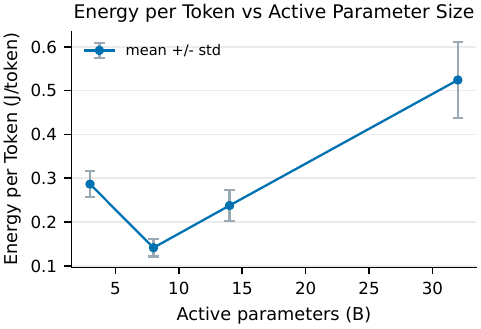}
        \caption{Energy per token}
        \label{fig:inf_quant_model_size_energy_per_token}
    \end{subfigure}
    \hfill
    \begin{subfigure}[t]{0.48\textwidth}
        \centering
        \includegraphics[width=\linewidth]{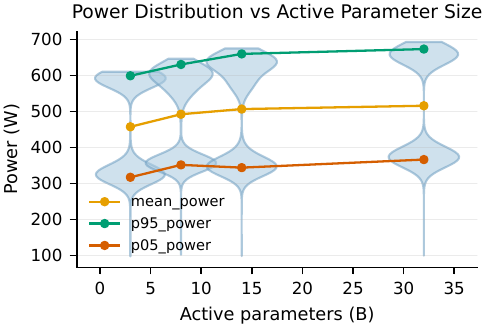}
        \caption{Power distribution}
        \label{fig:inf_quant_model_size_power_distribution}
    \end{subfigure}

    \vspace{0.5em}

    \begin{subfigure}[t]{0.48\textwidth}
        \centering
        \includegraphics[width=\linewidth]{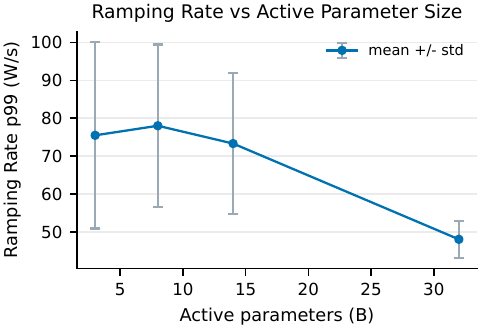}
        \caption{Ramping rate}
        \label{fig:inf_quant_model_size_ramping_rate}
    \end{subfigure}
    \hfill
    \begin{subfigure}[t]{0.48\textwidth}
        \centering
        \includegraphics[width=\linewidth]{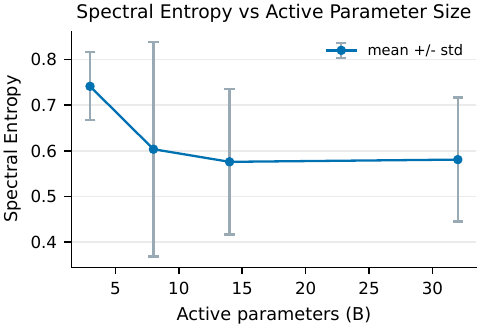}
        \caption{Spectral entropy}
        \label{fig:inf_quant_model_size_spectral_entropy}
    \end{subfigure}

    \caption{[Inference] Complete quantitative comparison across different model sizes. The panels report request latency, energy per request, energy per token, power distribution, ramping rate, and spectral entropy. It summarizes how the six inference metrics vary with active parameter count (BF16, TP~1, closed loop, pooled over A100, H100, and H200). Request latency and energy per request rise from 5.2~s and 72~J for Qwen3-8B to 18.8~s and 269~J for Qwen3-32B, while the MoE Qwen3-30B-A3B (3B active) is slower and costlier than the dense 8B model (Figs.~\ref{fig:inf_quant_model_size_latency}, \ref{fig:inf_quant_model_size_energy_per_request}). Energy per token follows the same order, from 0.14~J for Qwen3-8B to 0.52~J for Qwen3-32B (Fig.~\ref{fig:inf_quant_model_size_energy_per_token}). Mean and 95th-percentile power rise only modestly with size, from 493 to 516~W on average (Fig.~\ref{fig:inf_quant_model_size_power_distribution}). The ramping rate falls from 78 to 48~W/s between Qwen3-8B and Qwen3-32B, with the MoE model (76~W/s) close to Qwen3-8B, while spectral entropy is highest for the MoE model (0.74 against 0.58--0.60) (Figs.~\ref{fig:inf_quant_model_size_ramping_rate}, \ref{fig:inf_quant_model_size_spectral_entropy}).}
    \label{fig:inf_quant_model_size}
\end{figure*}

\begin{figure*}[!t]
    \centering

    \begin{subfigure}[t]{0.48\textwidth}
        \centering
        \includegraphics[width=\linewidth]
        {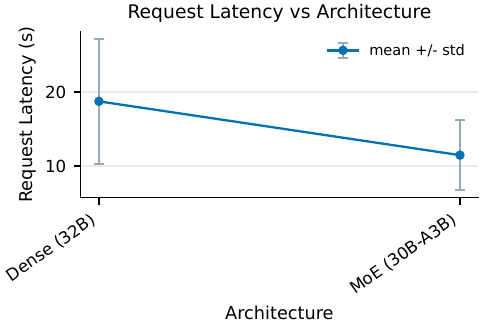}
        \caption{Request latency}
        \label{fig:inf_quant_architecture_latency}
    \end{subfigure}
    \hfill
    \begin{subfigure}[t]{0.48\textwidth}
        \centering
        \includegraphics[width=\linewidth]
        {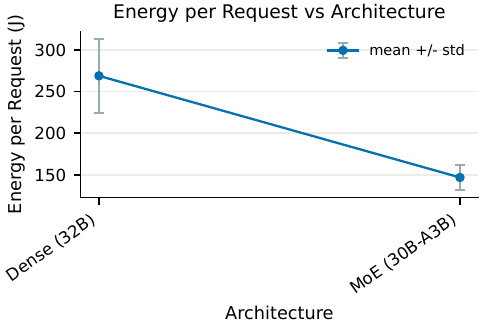}
        \caption{Energy per request}
        \label{fig:inf_quant_architecture_energy_per_request}
    \end{subfigure}

    \vspace{0.5em}

    \begin{subfigure}[t]{0.48\textwidth}
        \centering
        \includegraphics[width=\linewidth]
        {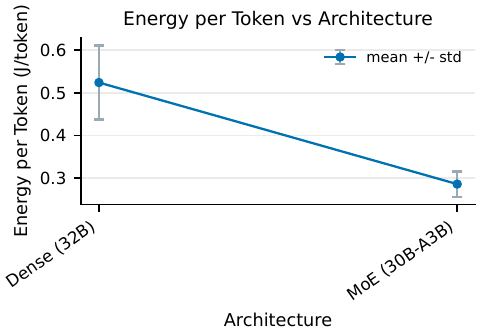}
        \caption{Energy per token}
        \label{fig:inf_quant_architecture_energy_per_token}
    \end{subfigure}
    \hfill
    \begin{subfigure}[t]{0.48\textwidth}
        \centering
        \includegraphics[width=\linewidth]
        {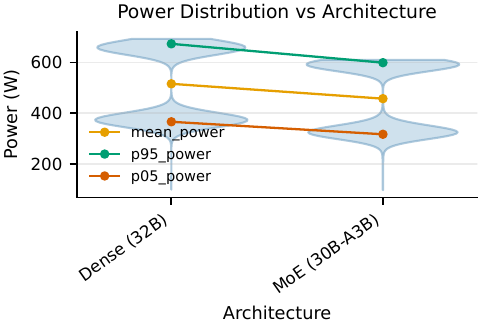}
        \caption{Power distribution}
        \label{fig:inf_quant_architecture_power_distribution}
    \end{subfigure}

    \vspace{0.5em}

    \begin{subfigure}[t]{0.48\textwidth}
        \centering
        \includegraphics[width=\linewidth]
        {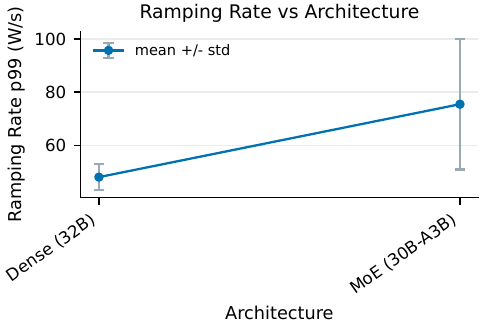}
        \caption{Ramping rate}
        \label{fig:inf_quant_architecture_ramping_rate}
    \end{subfigure}
    \hfill
    \begin{subfigure}[t]{0.48\textwidth}
        \centering
        \includegraphics[width=\linewidth]
        {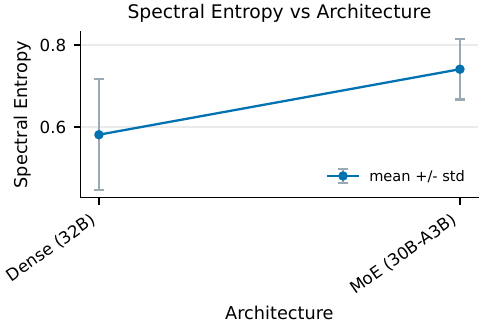}
        \caption{Spectral entropy}
        \label{fig:inf_quant_architecture_spectral_entropy}
    \end{subfigure}

    \caption{[Inference] Complete quantitative comparison across different model architectures. The panels report request latency, energy per request, energy per token, power distribution, ramping rate, and spectral entropy. It compares the dense Qwen3-32B and the MoE Qwen3-30B-A3B (BF16, TP~1, closed loop, pooled over A100, H100, and H200). The MoE model serves requests faster (11.5 versus 18.8~s) at lower energy per request (147 versus 269~J) and per token (0.29 versus 0.52~J) (Figs.~\ref{fig:inf_quant_architecture_latency}, \ref{fig:inf_quant_architecture_energy_per_request}, \ref{fig:inf_quant_architecture_energy_per_token}). It also draws less power, 458 versus 516~W on average and 599 versus 674~W at the 95th percentile (Fig.~\ref{fig:inf_quant_architecture_power_distribution}). Its ramping rate is about 1.6 times that of the dense model (76 versus 48~W/s), and its spectral entropy is higher (0.74 versus 0.58) (Figs.~\ref{fig:inf_quant_architecture_ramping_rate}, \ref{fig:inf_quant_architecture_spectral_entropy}).}
    \label{fig:inf_quant_architecture}
\end{figure*}

\begin{figure*}[!t]
    \centering

    \begin{subfigure}[t]{0.48\textwidth}
        \centering
        \includegraphics[width=\linewidth]
        {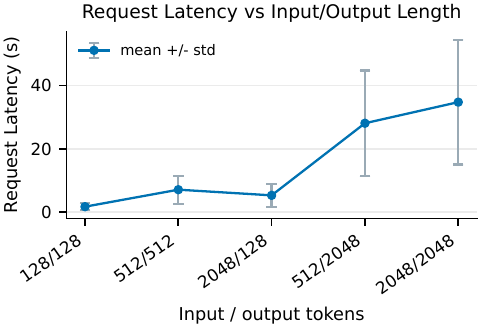}
        \caption{Request latency}
        \label{fig:inf_quant_isl_osl_latency}
    \end{subfigure}
    \hfill
    \begin{subfigure}[t]{0.48\textwidth}
        \centering
        \includegraphics[width=\linewidth]
        {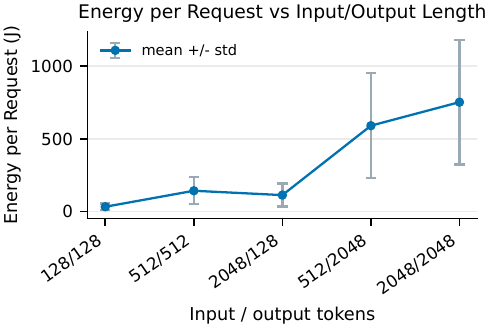}
        \caption{Energy per request}
        \label{fig:inf_quant_isl_osl_energy_per_request}
    \end{subfigure}

    \vspace{0.5em}

    \begin{subfigure}[t]{0.48\textwidth}
        \centering
        \includegraphics[width=\linewidth]
        {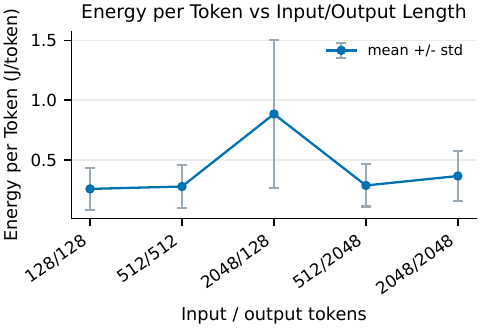}
        \caption{Energy per token}
        \label{fig:inf_quant_isl_osl_energy_per_token}
    \end{subfigure}
    \hfill
    \begin{subfigure}[t]{0.48\textwidth}
        \centering
        \includegraphics[width=\linewidth]
        {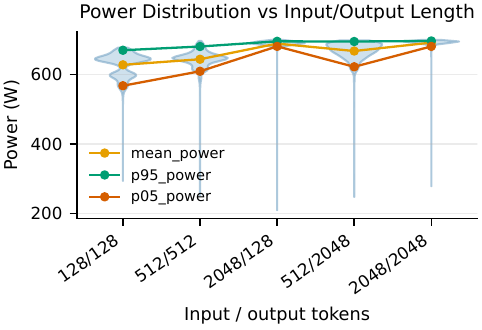}
        \caption{Power distribution}
        \label{fig:inf_quant_isl_osl_power_distribution}
    \end{subfigure}

    \vspace{0.5em}

    \begin{subfigure}[t]{0.48\textwidth}
        \centering
        \includegraphics[width=\linewidth]
        {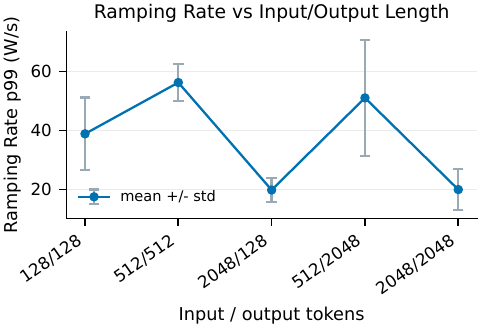}
        \caption{Ramping rate}
        \label{fig:inf_quant_isl_osl_ramping_rate}
    \end{subfigure}
    \hfill
    \begin{subfigure}[t]{0.48\textwidth}
        \centering
        \includegraphics[width=\linewidth]
        {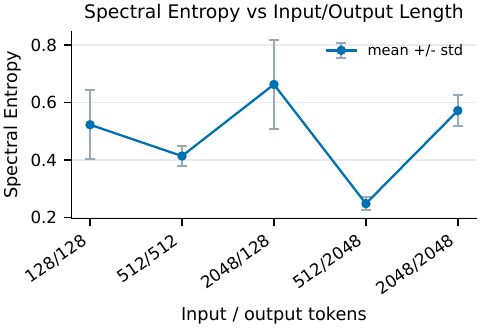}
        \caption{Spectral entropy}
        \label{fig:inf_quant_isl_osl_spectral_entropy}
    \end{subfigure}

    \caption{[Inference] Complete quantitative comparison across different input/output lengths. The panels report request latency, energy per request, energy per token, power distribution, ramping rate, and spectral entropy. It summarizes the effects of the forced input/output length on the six inference metrics (Qwen3-8B, 14B, and 32B on one H200 with 32 requests in flight). Request latency and energy per request grow mainly with the output length, from 1.7~s and 33~J at 128/128 to 35~s and 753~J at 2048/2048 (Figs.~\ref{fig:inf_quant_isl_osl_latency}, \ref{fig:inf_quant_isl_osl_energy_per_request}). Energy per output token is highest for long prompts with short outputs (0.89~J at 2048/128) and 0.26--0.37~J otherwise (Fig.~\ref{fig:inf_quant_isl_osl_energy_per_token}). The 2048-token prompts hold power within a few percent of the 700~W limit, with the 5th percentile at 681~W (Fig.~\ref{fig:inf_quant_isl_osl_power_distribution}). The same workloads have the smallest ramps (20~W/s against 39--56~W/s) and the highest spectral entropy (Figs.~\ref{fig:inf_quant_isl_osl_ramping_rate}, \ref{fig:inf_quant_isl_osl_spectral_entropy}).}
    \label{fig:inf_quant_isl_osl}
\end{figure*}

\textbf{Workload characteristics.} Larger models raise power and lengthen the turnover period (2.6, 3.4, and 6.5~s for Qwen3-8B, 14B, and 32B on two H200s). How ramps respond depends on the headroom: on two H200s, where each GPU stays well below its limit, the 99th-percentile ramp rises with size (62, 78, and 157~W/s; Fig.~\ref{fig:inf-size}), whereas on a single GPU, which the larger models push closer to its limit, it falls (from 96 to 50~W/s on A100 and from 55 to 47~W/s on H200; Fig.~\ref{fig:inf_quant_model_size}); the MoE Qwen3-30B-A3B draws less power than the dense Qwen3-32B but ramps about 1.6 times as fast (Fig.~\ref{fig:inf_quant_architecture}). Long prompts hold the GPU within a few percent of its limit with small ramps, while shorter prompts leave headroom and widely spaced turnover events produce the largest ramps (Figs.~\ref{fig:inf-sweeps}, \ref{fig:inf_quant_isl_osl}).

\begin{figure*}[!t]
    \centering

    \begin{subfigure}[t]{0.48\textwidth}
        \centering
        \includegraphics[width=\linewidth]
        {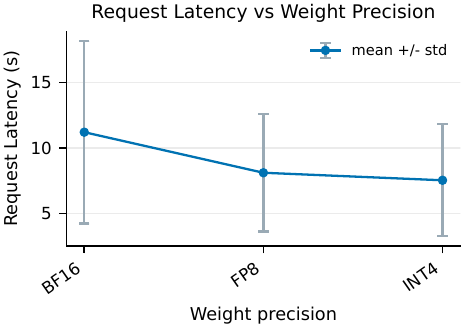}
        \caption{Request latency}
        \label{fig:inf_quant_weight_precision_latency}
    \end{subfigure}
    \hfill
    \begin{subfigure}[t]{0.48\textwidth}
        \centering
        \includegraphics[width=\linewidth]
        {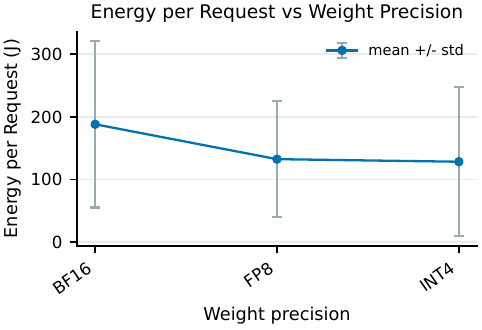}
        \caption{Energy per request}
        \label{fig:inf_quant_weight_precision_energy_per_request}
    \end{subfigure}

    \vspace{0.5em}

    \begin{subfigure}[t]{0.48\textwidth}
        \centering
        \includegraphics[width=\linewidth]
        {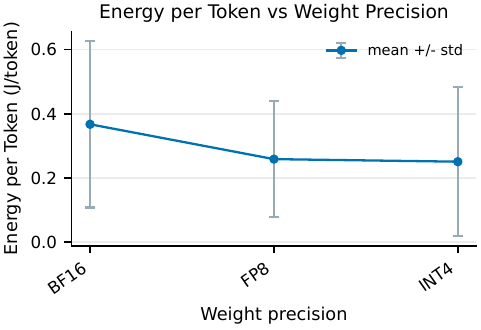}
        \caption{Energy per token}
        \label{fig:inf_quant_weight_precision_energy_per_token}
    \end{subfigure}
    \hfill
    \begin{subfigure}[t]{0.48\textwidth}
        \centering
        \includegraphics[width=\linewidth]
        {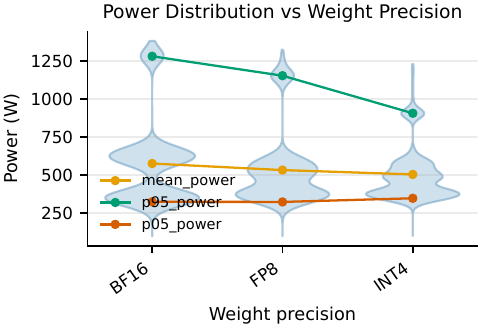}
        \caption{Power distribution}
        \label{fig:inf_quant_weight_precision_power_distribution}
    \end{subfigure}

    \vspace{0.5em}

    \begin{subfigure}[t]{0.48\textwidth}
        \centering
        \includegraphics[width=\linewidth]
        {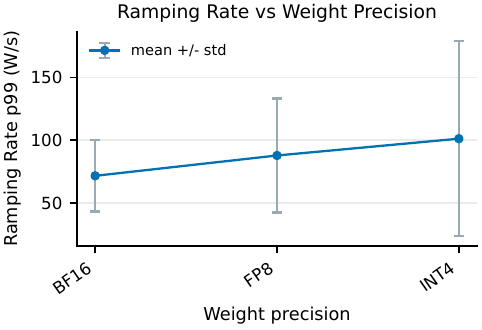}
        \caption{Ramping rate}
        \label{fig:inf_quant_weight_precision_ramping_rate}
    \end{subfigure}
    \hfill
    \begin{subfigure}[t]{0.48\textwidth}
        \centering
        \includegraphics[width=\linewidth]
        {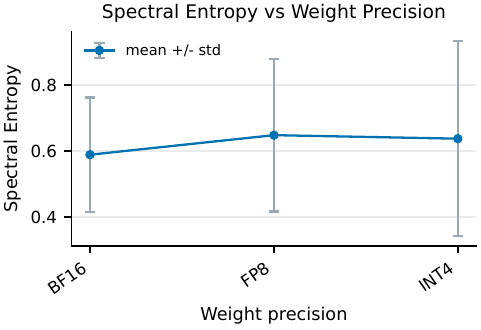}
        \caption{Spectral entropy}
        \label{fig:inf_quant_weight_precision_spectral_entropy}
    \end{subfigure}

    \caption{[Inference] Complete quantitative comparison across different weight precisions. The panels report request latency, energy per request, energy per token, power distribution, ramping rate, and spectral entropy. It compares BF16, FP8, and GPTQ-INT4 weights across the six metrics (five models on A100 and H200; all H100 weight-precision runs ran with a lowered power limit and are excluded). Quantized weights shorten request latency from 11.2 to 8.1 and 7.5~s and cut energy per request and per token by about 30\% (Figs.~\ref{fig:inf_quant_weight_precision_latency}, \ref{fig:inf_quant_weight_precision_energy_per_request}, \ref{fig:inf_quant_weight_precision_energy_per_token}). Mean power falls from 576 to 533 and 505~W, and 95th-percentile power falls more (Fig.~\ref{fig:inf_quant_weight_precision_power_distribution}). The ramping rate rises from 71 to 88 and 101~W/s, while spectral entropy rises slightly, from 0.59 to 0.65 and 0.64 (Figs.~\ref{fig:inf_quant_weight_precision_ramping_rate}, \ref{fig:inf_quant_weight_precision_spectral_entropy}).}
    \label{fig:inf_quant_weight_precision}
\end{figure*}

\begin{figure*}[!t]
    \centering

    \begin{subfigure}[t]{0.48\textwidth}
        \centering
        \includegraphics[width=\linewidth]
        {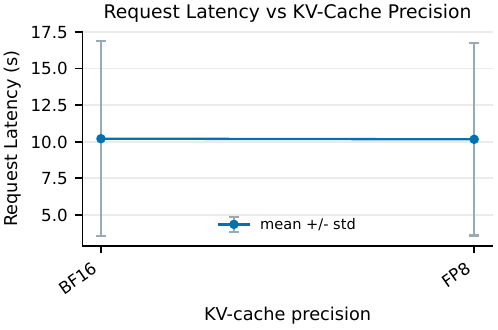}
        \caption{Request latency}
        \label{fig:inf_quant_kv_precision_latency}
    \end{subfigure}
    \hfill
    \begin{subfigure}[t]{0.48\textwidth}
        \centering
        \includegraphics[width=\linewidth]
        {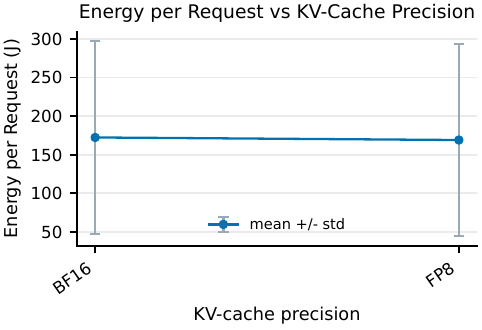}
        \caption{Energy per request}
        \label{fig:inf_quant_kv_precision_energy_per_request}
    \end{subfigure}

    \vspace{0.5em}

    \begin{subfigure}[t]{0.48\textwidth}
        \centering
        \includegraphics[width=\linewidth]
        {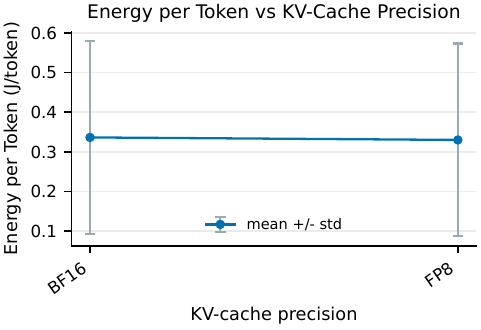}
        \caption{Energy per token}
        \label{fig:inf_quant_kv_precision_energy_per_token}
    \end{subfigure}
    \hfill
    \begin{subfigure}[t]{0.48\textwidth}
        \centering
        \includegraphics[width=\linewidth]
        {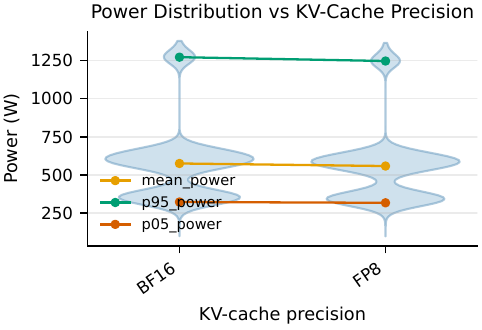}
        \caption{Power distribution}
        \label{fig:inf_quant_kv_precision_power_distribution}
    \end{subfigure}

    \vspace{0.5em}

    \begin{subfigure}[t]{0.48\textwidth}
        \centering
        \includegraphics[width=\linewidth]
        {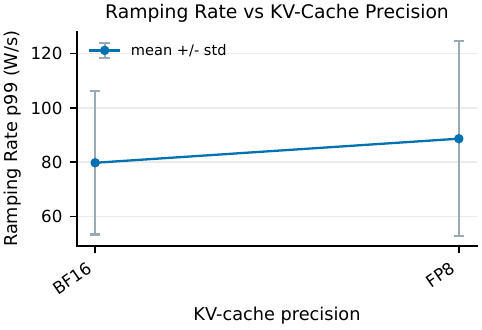}
        \caption{Ramping rate}
        \label{fig:inf_quant_kv_precision_ramping_rate}
    \end{subfigure}
    \hfill
    \begin{subfigure}[t]{0.48\textwidth}
        \centering
        \includegraphics[width=\linewidth]
        {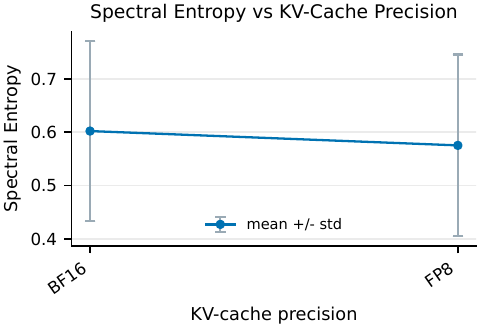}
        \caption{Spectral entropy}
        \label{fig:inf_quant_kv_precision_spectral_entropy}
    \end{subfigure}

    \caption{[Inference] Complete quantitative comparison across different KV-cache precisions. The panels report request latency, energy per request, energy per token, power distribution, ramping rate, and spectral entropy. It compares BF16 and FP8 (E4M3) KV caches across the six metrics (five models on A100, H100, and H200, BF16 weights). Request latency changes by less than 1\%, and energy per request and per token fall by about 2\% on average (a median 2.7\% and at most 5\% across the 13 same-host pairs) (Figs.~\ref{fig:inf_quant_kv_precision_latency}, \ref{fig:inf_quant_kv_precision_energy_per_request}, \ref{fig:inf_quant_kv_precision_energy_per_token}). Mean power falls slightly, from 574 to 559~W (Fig.~\ref{fig:inf_quant_kv_precision_power_distribution}). The ramping rate rises from 80 to 89~W/s, less than its run-to-run spread, and spectral entropy falls from 0.60 to 0.58 (Figs.~\ref{fig:inf_quant_kv_precision_ramping_rate}, \ref{fig:inf_quant_kv_precision_spectral_entropy}).}
    \label{fig:inf_quant_kv_precision}
\end{figure*}

\textbf{Weight precision.} 
Weight precision is the inference-side analogue of the fine-tuning method: it changes how much data each decode step moves. It shortens the batch-turnover period in every model and GPU pair and cuts energy per token by a median 30\% (FP8) and 39\% (GPTQ-INT4). Its effect on power depends on the platform: mean power falls on H100 and H200 (by up to 20\%), and ramps change inconsistently (Figs.~\ref{fig:inf-precision}, \ref{fig:inf_quant_weight_precision}). Weight quantization therefore shows up mainly as faster, cheaper execution rather than as a different waveform. An FP8 KV cache changes even less: energy per token falls by a median 2.7\% (at most 5\%) and the ramp changes by less than its run-to-run spread (Fig.~\ref{fig:inf_quant_kv_precision}).

\begin{figure*}[!t]
    \centering

    \begin{subfigure}[t]{0.48\textwidth}
        \centering
        \includegraphics[width=\linewidth]
        {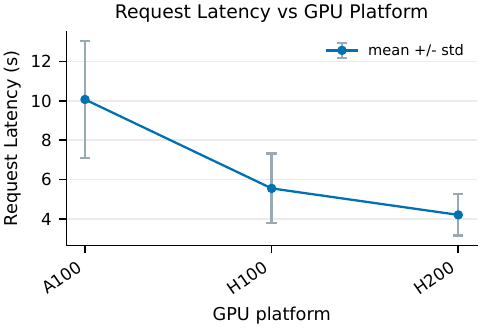}
        \caption{Request latency}
        \label{fig:inf_quant_gpu_platform_latency}
    \end{subfigure}
    \hfill
    \begin{subfigure}[t]{0.48\textwidth}
        \centering
        \includegraphics[width=\linewidth]
        {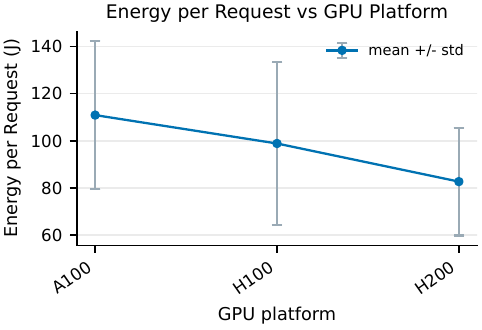}
        \caption{Energy per request}
        \label{fig:inf_quant_gpu_platform_energy_per_request}
    \end{subfigure}

    \vspace{0.5em}

    \begin{subfigure}[t]{0.48\textwidth}
        \centering
        \includegraphics[width=\linewidth]
        {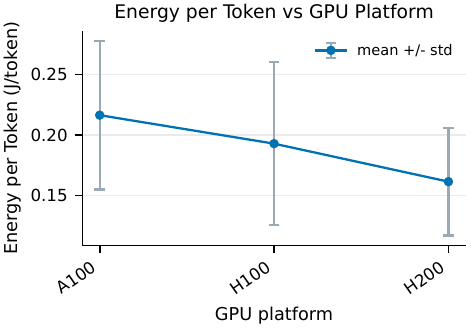}
        \caption{Energy per token}
        \label{fig:inf_quant_gpu_platform_energy_per_token}
    \end{subfigure}
    \hfill
    \begin{subfigure}[t]{0.48\textwidth}
        \centering
        \includegraphics[width=\linewidth]
        {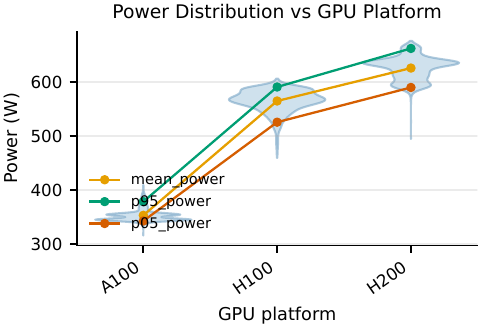}
        \caption{Power distribution}
        \label{fig:inf_quant_gpu_platform_power_distribution}
    \end{subfigure}

    \vspace{0.5em}

    \begin{subfigure}[t]{0.48\textwidth}
        \centering
        \includegraphics[width=\linewidth]
        {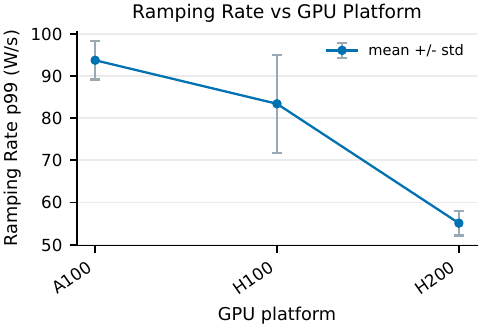}
        \caption{Ramping rate}
        \label{fig:inf_quant_gpu_platform_ramping_rate}
    \end{subfigure}
    \hfill
    \begin{subfigure}[t]{0.48\textwidth}
        \centering
        \includegraphics[width=\linewidth]
        {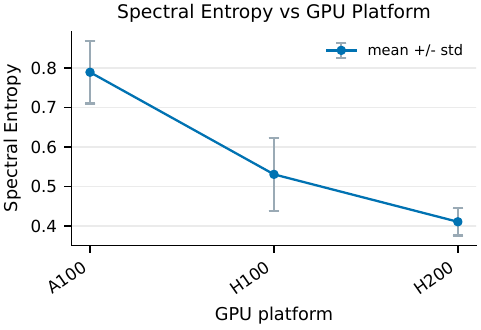}
        \caption{Spectral entropy}
        \label{fig:inf_quant_gpu_platform_spectral_entropy}
    \end{subfigure}

    \caption{[Inference] Complete quantitative comparison across different GPU platforms. The panels report request latency, energy per request, energy per token, power distribution, ramping rate, and spectral entropy. It compares A100, H100, and H200 (BF16, TP~1, closed loop, pooled over Qwen3-8B and 14B, the only models with an H100 run at the standard power limit; 6, 2, and 6 runs). Request latency falls from 10.1~s on A100 to 5.5 and 4.2~s on H100 and H200, and energy per request and per token fall in the same order (111, 99, and 83~J; 0.22, 0.19, and 0.16~J) (Figs.~\ref{fig:inf_quant_gpu_platform_latency}, \ref{fig:inf_quant_gpu_platform_energy_per_request}, \ref{fig:inf_quant_gpu_platform_energy_per_token}). Mean power rises from 353~W on A100 to 565 and 625~W on H100 and H200, each within a narrow band (Fig.~\ref{fig:inf_quant_gpu_platform_power_distribution}). The ramping rate falls from 94 to 83 and 55~W/s, and spectral entropy from 0.79 to 0.53 and 0.41 (Figs.~\ref{fig:inf_quant_gpu_platform_ramping_rate}, \ref{fig:inf_quant_gpu_platform_spectral_entropy}).}
    \label{fig:inf_quant_gpu_platform}
\end{figure*}

\begin{figure*}[!t]
    \centering

    \begin{subfigure}[t]{0.48\textwidth}
        \centering
        \includegraphics[width=\linewidth]
        {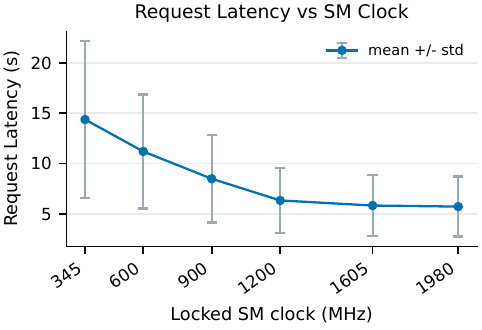}
        \caption{Request latency}
        \label{fig:inf_quant_sm_clock_latency}
    \end{subfigure}
    \hfill
    \begin{subfigure}[t]{0.48\textwidth}
        \centering
        \includegraphics[width=\linewidth]
        {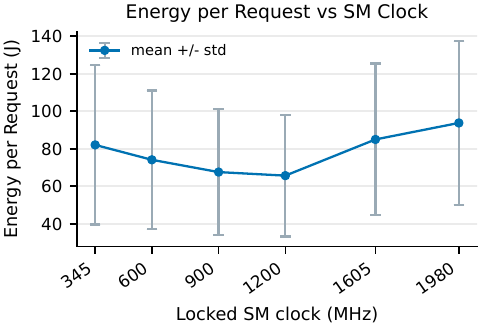}
        \caption{Energy per request}
        \label{fig:inf_quant_sm_clock_energy_per_request}
    \end{subfigure}

    \vspace{0.5em}

    \begin{subfigure}[t]{0.48\textwidth}
        \centering
        \includegraphics[width=\linewidth]
        {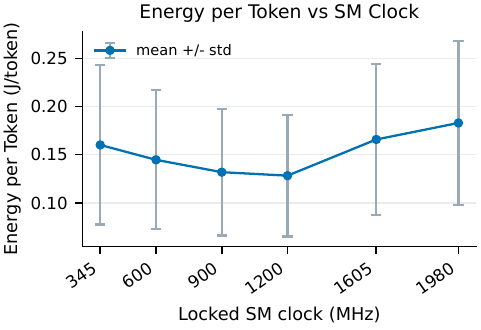}
        \caption{Energy per token}
        \label{fig:inf_quant_sm_clock_energy_per_token}
    \end{subfigure}
    \hfill
    \begin{subfigure}[t]{0.48\textwidth}
        \centering
        \includegraphics[width=\linewidth]
        {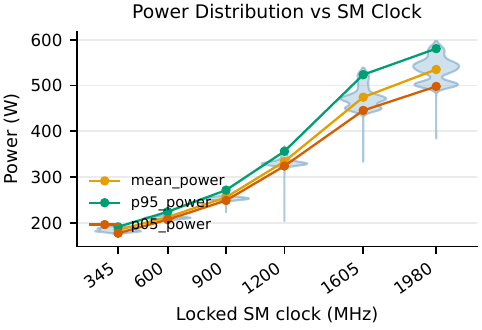}
        \caption{Power distribution}
        \label{fig:inf_quant_sm_clock_power_distribution}
    \end{subfigure}

    \vspace{0.5em}

    \begin{subfigure}[t]{0.48\textwidth}
        \centering
        \includegraphics[width=\linewidth]
        {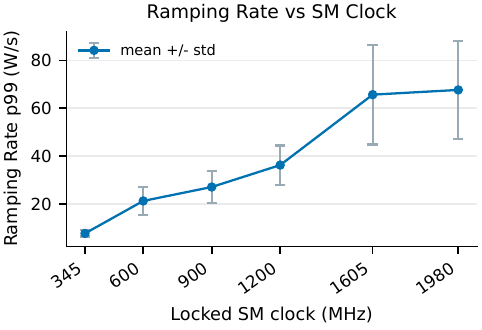}
        \caption{Ramping rate}
        \label{fig:inf_quant_sm_clock_ramping_rate}
    \end{subfigure}
    \hfill
    \begin{subfigure}[t]{0.48\textwidth}
        \centering
        \includegraphics[width=\linewidth]
        {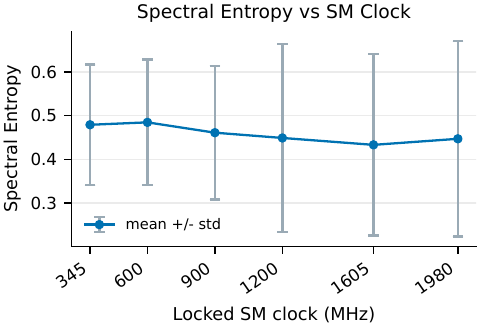}
        \caption{Spectral entropy}
        \label{fig:inf_quant_sm_clock_spectral_entropy}
    \end{subfigure}

    \caption{[Inference] Complete quantitative comparison across different SM clock frequencies. The panels report request latency, energy per request, energy per token, power distribution, ramping rate, and spectral entropy. It summarizes the effects of the locked SM clock on the six inference metrics (Qwen3-8B, Qwen3-30B-A3B, and Llama-3.1-8B on one H100, closed loop; the 1980~MHz setting throttles to a median of 1830~MHz). Request latency falls from 14.4~s at 345~MHz to 5.8 and 5.7~s at 1605 and 1980~MHz, while energy per request and per token are lowest at 1200~MHz (66~J and 0.13~J) and rise again at higher clocks (Figs.~\ref{fig:inf_quant_sm_clock_latency}, \ref{fig:inf_quant_sm_clock_energy_per_request}, \ref{fig:inf_quant_sm_clock_energy_per_token}). Mean power rises from 184 to 535~W, with a narrow band at every clock (Fig.~\ref{fig:inf_quant_sm_clock_power_distribution}). The ramping rate grows at every clock step, from 8 to 68~W/s, while spectral entropy falls from 0.48 to 0.43--0.45 (Figs.~\ref{fig:inf_quant_sm_clock_ramping_rate}, \ref{fig:inf_quant_sm_clock_spectral_entropy}).}
    \label{fig:inf_quant_sm_clock}
\end{figure*}

\textbf{Hardware.} The GPU platform sets the power range and the timescale (Figs.~\ref{fig:inf-hardware}, \ref{fig:inf_quant_gpu_platform}): the same configuration (Qwen3-14B at TP~4) turns over every 5.3~s on A100 and every 2.8 and 2.5~s on H100 and H200, at about 1.5 times the node power. Within the H100 platform, the SM clock exposes a two-objective trade-off (Fig.~\ref{fig:inf_quant_sm_clock}): energy per token is lowest near 1200~MHz, whereas the 99th-percentile ramp keeps growing with clock. The clock that minimizes energy is not the one that minimizes grid impact.

\subsection{Complete Inference Comparisons}
\label{app:inf-quantitative}

We next provide the complete quantitative comparisons across the major inference factors considered in our experiments, in the same format as Appendix~\ref{app:training_quantitative}.
For each factor, runs at the same factor level are pooled over the models and
GPU platforms represented in that comparison, and error bars show one standard
deviation.
These results quantify how each factor affects request latency, energy per request, energy per token, power distribution, ramping behavior, and temporal regularity. Runs on hosts with a lowered GPU power limit or a capped application clock are excluded.

\end{document}